\documentclass{LMCS}                                                                                   
\def\doi{8 (1:28) 2012}                                                  
\lmcsheading%
{\doi}                                                                      
{1--44}                                                                    
{}                                                                         
{}                                                                        
{Oct.~\phantom04, 2011}                                                    
{Mar.~26, 2012}                                                            
{}                                                                                                      

\usepackage[usenames,dvipsnames]{color}                                                                
\usepackage{amssymb}                                                                                   
\usepackage{amsmath}                                                                                   
\usepackage{stmaryrd}                                                                                  
\usepackage{bussproofs}                                                                                
\usepackage{url}                                                                                       
\usepackage{proof}                                                                                     
\usepackage{fancybox}                                                                                  
\usepackage{ulem}                                                                                     
\usepackage{enumerate,hyperref}

\newcommand{\ShaneIgnore}[1]{}

\def\itmath#1{\leavevmode\ifmmode{\mbox{\it#1} }\else{\it#1 }\fi}
\def\sfmath#1{\leavevmode\ifmmode{\mbox{\sf#1} }\else{\sf#1 }\fi}
\def\condmath#1{\leavevmode\ifmmode{#1}\else{$#1$}\fi}

\newcommand{\lab}[2]{[ \! [ #1/ #2 ] \! ]}

\newcommand{\ie}{{\it  i.e.}~}
\newcommand{\cf}{{\it  cf.}~}
\newcommand{\eg}{{\it  e.g.}~}

  {\begin{enumerate}%
    \setlength{\itemsep}{0pt}%
    \setlength{\parskip}{0pt}%
    \setlength{\topsep}{0pt}
    \setlength{\partopsep}{0pt}}
  {\end{enumerate}}

  {\begin{description}%
    \setlength{\itemsep}{0.1pt}%
    \setlength{\parskip}{0pt}%
    \setlength{\topsep}{0pt}
    \setlength{\partopsep}{0pt}}
  {\end{description}}

\newcommand{\sep}{\hspace*{0.5cm}}

\def\lam{\lambda}

\def\sig{\sigma}
\def\Gam{\Gamma}

\newcommand{\Rew}[1]{\rightarrow_{#1}}
\newcommand{\rRew}[1]{\mapsto_{#1}}

\newcommand{\Rewn}[2][*]{\rightarrow^{#1}_{#2}}
\newcommand{\Rewnmod}[2]{\rightarrow^*_{#1/#2}}
\newcommand{\Rewplusmod}[2]{\rightarrow^+_{#1/#2}}
\newcommand{\Rewp}[1]{\rightarrow^{+}_{#1}}

\newcommand{\lx}{\lam{\tt x}}

\newcommand{\es}{{\tt es}}
\newcommand{\esw}{{\tt esw}}

\newcommand{\les}{\lam \es}

\newcommand{\lesw}{\lam \esw}

\newcommand{\isubs}[1]{ \{ #1  \} }

\newcommand{\SN}[1]{\mathcal{SN}_{#1}}
\newcommand{\WN}[1]{\mathcal{WN}_{#1}}

\newcommand{\multiset}[1]{ [ #1 ] }
\newcommand{\pos}[2]{{\tt pos}_{#1}(#2)}

\newcommand{\nat}{{\mathbb N}}

\newcommand{\lsigma}{\lam_{\sigma}}

\newcommand{\calA}{\mathcal{A}}
\newcommand{\calB}{\mathcal{B}}
\newcommand{\B}{{\tt dB}}
\newcommand{\ttE}{{\tt E}}
\newcommand{\ttF}{{\tt F}}

\newcommand{\ttv}{{\tt v}}

\newcommand{\ttR}{{\tt R}}

\newcommand{\shB}{{\tt B}}

\newcommand{\R}{\mathcal{R}}

\renewcommand{\S}{\mathcal{S}}
\newcommand{\Q}{\mathcal{Q}}

\newcommand{\dis}{{\tt j}}

\newcommand{\ldis}{\lam{\dis}}

\newcommand{\lj}{\lam{\dis}}

\newcommand{\ljdag}{\lam\jop\mbox{-dag}}

\newcommand{\pair}[2]{\langle #1, #2 \rangle}

\newcommand{\ren}[4]{R^{#2, #3}_{#4}(#1)}

\newcommand{\Rewplus}[1]{\rightarrow^{+}_{#1}}

\newcommand{\LRew}[1]{\mbox{}_{#1}{\leftarrow} }
\newcommand{\LRewn}[1]{\mbox{}^{*}_{#1}{\leftarrow}}

\newcommand{\Rewnumber}[2]{\stackrel{#1}{\rightarrow_{#2}}}

\newcommand{\fv}[1]{{\tt fv}(#1)}

\newcommand{\bv}[1]{\texttt{bv}(#1)}

\newcommand{\proj}{\wfc}

\newcommand{\Var}{{\tt d}}

\newcommand{\DSubs}{{\tt c}}

\newcommand{\out}{{\tt out}}
\newcommand{\outv}{{\tt out\overline{d}}}
\newcommand{\iinn}{{\tt in}}

\newcommand{\ldisout}{\lam \dis_{\out}}

\newcommand{\ldisin}{\lam \dis_{\iinn}}

\newcommand{\Gc}{{\tt w}}

\newcommand{\CS}{{\tt CS}}

\newcommand{\Beta}{{\tt \beta}}

\newcommand{\set}[1]{ \{ #1 \}}

\newcommand{\mul}[2]{{\tt P}_{#2}(#1)}

\newcommand{\ems}{\emptyset}

\newcommand{\paralp}[1]{\Rrightarrow_{#1}}

\newcommand{\paralpn}[1]{\Rrightarrow^{*}_{#1}}

\newcommand{\ih}{i.h.}

\renewcommand{\vec}[1]{\overline{#1}}

\newcommand{\terms}{\mathcal{T}}
\newcommand{\termsv}{\mathcal{T}_{\tt v}}

\newcommand{\termslambda}{\terms_{\lam}}

\newcommand{\dm}[1]{\dis {\tt m}(#1)}

\newcommand{\fo}[2]{\ttv_{#2}(#1)}
\newcommand{\fop}[1]{\ttv_{#1}}

\newcommand{\deft}[1]{{\bf #1}}

\newcommand{\iep}{{\bf IE}}

\newcommand{\giep}{{\bf GIE}}
\newcommand{\viep}{{\bf VIE}}

\newcommand{\fc}{\dis}
\newcommand{\wfc}{\Gc\dis}

\newcommand{\jop}{{\tt j}}

\newcommand{\unboxed}{{\tt u}}

\newcommand{\slist}{{\tt L}}

\newcommand{\modulo}[2]{#1/#2}

\renewcommand{\sp}[4]{\{#1_{i}/#2_{i}\}^{#3}_{#4}}
\newcommand{\esp}[4]{[{#1}_i/{#2}_i]^{#3}_{#4}}
\newcommand{\espv}[4]{[\void/{#2}_i]^{#3}_{#4}}

\newcommand{\ovl}[3]{\overline{#1}^{#2}_{#3}}
\newcommand{\jump}{jump}

\newcommand{\maxi}[2]{{\tt max}(#1,#2)}
\newcommand{\mmax}[1]{\maxi{1}{#1}\cdot}

\newcommand{\development}{development}

\newcommand{\supersuperdevelopment}{{\tt XL}-\development}

\newcommand{\weakmes}[1]{|#1|_{\nGc}} 
\newcommand{\nGc}{{\neg\Gc}}

\newcommand{\espp}[4]{[\cdot]^{#3}_{#4}}
\newcommand{\spp}[4]{\{\cdot\}^{#3}_{#4}}
 \newcommand{\ignore}[1]{}
\newcommand{\mellies}{Melli{\`e}s}

\newcommand{\eqw}[1]{\equiv_{#1}}
\newcommand{\eqo}{\equiv_\osym}
\newcommand{\eqttE}{\equiv_{\ttE}}
\newcommand{\preeq}{\sim}
\newcommand{\preeqw}[1]{\sim_{#1}}

\def\rsig{\hat{\sigma}}
\newcommand{\runboxed}{\hat{\unboxed}}

\newcommand{\osym}{{\tt o}}

\newcommand{\preeqsigu}{\preeq_{\sigma_1}}
\newcommand{\preeqsigt}{\preeq_{\sigma_2}}

\newcommand{\ldiso}{\ldis/{\osym}}
\newcommand{\ldisttE}{\ldis/{\ttE}}
\newcommand{\ldisf}{\lam \modulo{\dis}{\fsymb}}

\newcommand{\ldisfz}{\ldis/\fz}
\newcommand{\fsymb}{\osymb\boite}
\newcommand{\osymb}{{\tt o}}
\newcommand{\eqf}{\eqw{\fsymb}}
\newcommand{\eqfz}{\equiv_\fz}
\newcommand{\fz}{{\tt n}}
\newcommand{\aux}{{\tt void}}
\newcommand{\laux}{\lam\aux}
\newcommand{\lauxm}{\lam\modulo{ \aux }{ \osymb }}
\newcommand{\New}{{\tt h}}

\newcommand{\mlist}{{\tt M}}

\newcommand{\snsudd}[2]{\mathbb{SNSJ}_{#1}(#2)}

\newcommand{\ctx}[2]{#1 [ \! [#2] \! ]}
\newcommand{\bs}[1]{{\tt bs}(#1)}

\newcommand{\etamd}[2]{\mathbb{MSJ}_{#1}(#2)}

\newcommand{\eqcs}{\equiv_\CS}
\newcommand{\List}{{\tt L}}
\definecolor{LightGray}{gray}{.80}
\definecolor{DarkGray}{gray}{.60}
\newcommand{\grisar}[1]{\colorbox{LightGray}{\ensuremath{#1}}}
\newcommand{\grisarOscuro}[1]{\colorbox{DarkGray}{\ensuremath{#1}}}
\newcommand{\surf}[2]{\mathbb{T}_{#1}(#2)}
\newcommand{\preetamd}[2]{\mathbb{SJ}_{#1}(#2)}

\newcommand{\inn}{\modulo{{\tt in}}{\CS}}
\newcommand{\innn}{{\tt in}}
\newcommand{\comp}{{\tt comp}}
\newcommand{\outm}{\modulo{{\tt out}}{\CS}}
\newcommand{\lsub}{\lam {\tt sub}}

\newcommand{\boite}{{\tt box}}
\newcommand{\sigt}{\boite_1}
\newcommand{\sigq}{\boite_2}
\newcommand{\rsigt}{\widehat{\boite}}
\newcommand{\void}{\_}

\newcommand{\gm}{\sqsupset}
\newcommand{\geqm}{\sqsupseteq}
\newcommand{\subt}{\triangleleft}
\newcommand{\surt}{\triangleright}

\newcommand{\psymb}{{\tt P}}
\newcommand{\eqp}{\eqw{\psymb}}
\newcommand{\Psymb}{\Pi}
\newcommand{\eqP}{\eqw{\Psymb}}

\newcommand{\hole}{\Box}

\title[PSN  modulo permutations for 
the structural lambda calculus]{Preservation of Strong Normalisation modulo permutations for 
the structural $\lambda$-calculus}

\author[B.~Accattoli]{Beniamino Accattoli\rsuper a}
\address{{\lsuper a}INRIA Saclay and LIX (\'Ecole Polytechnique)}
\email{beniamino.accattoli@gmail.com}
\author[D.~Kesner]{Delia Kesner\rsuper b}
\address{{\lsuper b}Univ. Paris Diderot, Sorbonne Paris Cit\'e, PPS, CNRS}
\email{delia.kesner@pps.jussieu.fr}


\begin{document}

\begin{abstract}
Inspired by a recent  graphical formalism for $\lam$-calculus based on
linear   logic  technology,   we  introduce   an   untyped  structural
$\lam$-calculus, called  $\ldis$, which  combines actions at  a distance
with  exponential  rules  decomposing  the substitution  by  means  of
weakening,  contraction  and   derelicition.   First,  we  prove  some
fundamental properties of $\ldis$  such as confluence and preservation
of   $\beta$-strong  normalisation.    Second,  we   add   a  strong
bisimulation  to  $\ldis$  by  means  of an  equational  theory  which
captures in particular Regnier's $\sig$-equivalence.  We then complete
this  bisimulation   with  two  more   equations  for
(de)composition  of  substitutions and  we  prove  that the  resulting
calculus  still preserves  $\beta$-strong normalization.   Finally, we
discuss some  consequences of our results.
\end{abstract}

\maketitle

\section*{Introduction}


Linear Logic~\cite{LL} has been very influential in computer science,
especially because it provides a tool to explicitly control the use of
resources by limiting the use of the \textit{structural rules} of
weakening and contraction.  Erasure (weakening) and duplication
(contraction) are restricted to formulas marked with an
\textit{exponential} modality, and can
only interact with non-linear proofs marked with a
\textit{bang} modality.  Intuitionistic
and Classical Logic can thus be encoded by a fragment containing such
modalities as, for example, the Multiplicative Exponential Linear
Logic (MELL).


MELL proofs can be represented by sequent trees, but MELL
Proof-Nets~\cite{LL} provide a better geometrical representation of
proofs, eliminating irrelevant syntactical details.  They have been
used extensively to develop different encodings of intuitionistic
logic/lambda-calculus, giving rise to the geometry of
interaction~\cite{GOI}.

Normalisation of proofs (\ie\ \textit{cut elimination}) in MELL
Proof-Nets is performed in particular by \textit{exponential} and
\textit{commutative} rules.  Non-linear proofs are distinguished by
surrounding \textit{boxes}; the exponential rules handle all the
possible operations on them: erasure, duplication and linear replacement,
corresponding respectively to a cut elimination step involving a box
and either a \textit{weakening}, a \textit{contraction} or a
\textit{dereliction}. The commutative rule instead \textit{composes}
non-linear resources.


Different cut elimination systems~\cite{DCKP03,KL07,Kes07}, defined as
\textit{explicit substitution} (ES) calculi, were explained in terms
of, or were inspired from, the fine notion of reduction of MELL
Proof-Nets. They all use the idea that the content of a
substitution/cut is a non-linear resource, \ie\ a box that can be
composed with another one by means of some commutative rules.
They also share common operational semantics defined in terms of
a \textit{propagation system} in which a substitution traverses a term
until the variables are reached.\medskip

\deft{The  structural $\lam$-calculus}.  A graphical  representation for
$\lam$-terms,  $\ljdag$s, has  been  recently proposed~\cite{AG09}.   It
denies  boxes  by  representing  them  with  additional  edges  called
\textit{{\tt j}umps}, and does not need any commutative reduction rule
to  compose non-linear  proofs.  This  paper studies  the term
formalism,  called  $\ldis$-calculus,   resulting  from  reading  back
$\ljdag$s  (and  their correspondent  reductions)  by  means of  their
sequentialisation  theorem~\cite{AG09}.  The  deep connection  between
$\ljdag$s   and   Danos    and   Regnier's   Pure   (\textit{untyped})
Proof-Nets~\cite{Danos99opt}     has      been     already     studied
in~\cite{AccattoliTh}.

Beyond this  graphical and logical interpretation,  the peculiarity of
$\ldis$-calculus  is  that  it  uses  two 
features  which  were  never  combined before:  \textit{action  at  a
  distance} and \textit{multiplicities}.

\textit{Action at a distance} means
  that rewriting rules are specified by means of
  some constructors which are arbitrarily far away from each other. This
approach could be understood as inconvenient
but this is only apparent because rewriting rules can be
locally implemented by means of $\ljdag$s. The distance rules of $\ldis$ do not propagate
substitutions through the term except
for the linear ones which are evaluated exactly as meta-level substitutions,
regardless the distance between the  involved constructors (variable and jump).

\textit{Multiplicities} are intended to count the number of
occurrences of a given variable affected by a jump,
  \ie\ the  rewriting rule to be applied for reducing a term of the
form $t[x/u]$ depends on $|t|_x$, the number
  of free occurrences of the variable $x$ in the term
  $t$. Indeed, we distinguish three cases,
$|t|_x=0$, $|t|_x=1$ and $|t|_x >1$, which
correspond, respectively,  to 
 weakening-box, dereliction-box and contraction-box
cut-elimination rules in Proof Nets.
It is because of the weakening and
contraction rules that we call our language the \textit{structural}
$\lam$-calculus.\medskip

\deft{Content of the paper}.
We start by showing that $\ldis$ admits a simple and elegant
theory \ie\  it enjoys  confluence, full composition (FC), and preservation of
$\beta$-strong normalisation (PSN). The proof of PSN is particularly
concise because of the distance approach.

The main result of the paper is that the theory of $\lj$ admits a
modular extension with respect to propagations of jumps:
an equational theory is added on top
of $\ldis$ and the obtained extension is  shown to
preserve all the good properties we mentioned
  before. Actually, we focus on PSN, since FC and confluence for the
extended $\ldis$-calculus result as straightforward.

In the literature there is a huge number of calculi with expicit
substitutions, ${\tt let}$ constructs or environments, most of them
use some rule to specify commutation (also called
propagation or permutation). In order to encompass these formalisms we
do not approach propagations as \textit{rewriting rules}, but as
equations (which can be used from left to right or vice-versa)
  defining an \textit{equivalence relation}  on terms.

This is only possible because propagations are
not needed in $\ldis$ to compute normal forms, a fact which is a by-product
of the
  \textit{distance} notion.  Moreover, any particular orientation
of the equations (from left to
  right \textit{or} from right to left) results in a terminating
rewriting relation, which implies that the system containing \textit{any} orientation of the 
equations still enjoys PSN.

Equations are introduced in two steps.  We first consider commutations
between independent jumps
and between jumps and  abstractions or left sides of applications. 
This equivalence, written $\eqo$, turns out to be a strong
bisimulation, \ie\ a 
reduction  relation
which is length preserving; thus PSN
for the reduction system $\ldis$ modulo $\eqo$ --- noted $\ldiso$ --- immediately follows. We also
show that $\eqo$ can be seen as a projection of Regnier's
$\sigma$-equivalence~\cite{regnier94} on a syntax with
jumps. Actually, $\eqo$ can be understood as the quotient induced by
the translation~\cite{AccattoliTh} of $\ldis$-terms to Pure
Proof-Nets, which is why it is so well-behaved, and why we call it the
\textit{graphical equivalence}.

The second step is to extend $\eqo$ with general commutations between
jumps and right sides of applications and
contents of jumps.  
 The resulting \textit{substitution equivalence $\eqf$} does not
  only subsume \textit{composition} of jumps, but also
  \textit{decomposition}.  The equations of $\eqf$ correspond
  exactly to the commutative box-box case of Proof-Nets, but they are
  here considered as an \textit{equivalence} --- which is a novelty ---
  and not as a rewriting rule. The reduction relation of $\ldisf$ is a
  rich rewriting system with subtle behaviour, particularly because
  $\eqf$ affects reduction lengths, and thus is not a strong
  bisimulation. Nonetheless, we show that $\ldisf$ enjoys PSN.

This result is non-trivial, and constitutes the main contribution of the paper.
The technique used to obtain PSN for $\ldisf$ consists in
\begin{enumerate}
\item Projecting $\ldisf$ reductions into a calculus that we call $\lauxm$,
\item Proving PSN for $\lauxm$,
\item Infering PSN for $\ldisf$ from (1) and (2).
\end{enumerate}
Actually, $\lauxm$ can be understood as a \textit{memory}
calculus specified by means of \textit{void} jumps --- \ie\ jumps $t[x/u]$ where $x\notin\fv{t}$ --- 
which generalises Klop's $\Lambda_I$-calculus~\cite{Klo}. 
Despite 
the fact that it  appears only
as a technical tool we claim that 
it is a calculus interesting
on its own and can be used for proving termination
results beyond those of this paper.

The last part of the paper presents some interesting consequences of
our main result concerning different variations on
$\ldisf$.  \medskip

\deft{Road Map}.

\begin{enumerate}[$\bullet$]
\item Section~\ref{sec:general-notion} recalls some general 
notions about abstract rewriting.

\item Section~\ref{s:structural-lj} presents the $\ldis$-calculus and 
  shows  that  it   enjoys  basic properties such as full composition,
  simulation   of   one-step
  $\beta$-reduction, and confluence.

\item Section~\ref{s:lj-psn} studies    preservation of 
$\beta$-strong normalisation (PSN). The     PSN    property is proved   using     a   modular  
  technique developed in~\cite{Kes09}, which  results in a
  very short  formal argument in our  case. 
 
\item Section~\ref{s:eq-th} first considers $\ldis$ 
  enriched with the 
  equivalence $\eqo$, which is related to Regnier's
  $\sigma$-equivalence~\cite{regnier94},  and 
then with the equivalence $\eqf$, which also contains composition of
  jumps.

\item Section~\ref{sec:psn} is devoted to the proof of PSN for $\ldis$ modulo $\eqf$,
which  is the main  contribution of the  paper.

\item Section~\ref{s:cons} discusses some consequences of the PSN result of
  Section~\ref{sec:psn}.
\end{enumerate}

This paper
covers some basic results in~\cite{AK10} by extending them
considerably. Indeed, the propagation systems considered
in~\cite{AK10} are just particular cases of the general equational
theory $\eqf$ studied in this paper. The proof technique used here to
show PSN for $\ldis$ modulo $\eqf$
puts in evidence another calculus $\lauxm$ that has interest in itself.   
Moreover, interesting consequences of the main result
are included in Section~\ref{s:cons}.  \medskip

\deft{Related Work}.  Action at a distance has already been used
in~\cite{Milner07,deBruijn87,Ned92}, but none of the previous
approaches takes advantage of distance plus control of resources by
means of multiplicities.  Other works use multiplicities~\cite{KR09}
but not distance so that the resulting formalism contains a lot of
rules, which is really less manageable.  We think that our combined
approach is more primitive than ES, and the resulting theory is much
simpler.  Using distance \textit{and} multiplicities also provides
modularity: the substitution rules become independent from the set of
constructors of the calculus, and thus any change in the language does
not cause any changes in the associated rewriting rules. Our combined
approach does not only capture the well-known notions of
developments~\cite{Hindley78} and superdevelopments~\cite{KvOvR93}, but also allows us to
introduce \supersuperdevelopment s, a more powerful
notion of development defined in~\cite{AK10}.

In the literature there are many calculi which dealt with permutations
of constructors in intuitionistic calculi, but all use \textit{reduction} rules
rather than \textit{equations}, which is less powerful. 
Some  that can be captured by our graphical equivalence appear
in~\cite{Kamareddine00, regnier94,KfouryW95}
and those captured by our substitution equivalence
are~\cite{espiritoSanto2011,HZ09,Yoshida93}.
Intuitionistic calculi inspired from Linear Logic Proof Nets
appear for example in~\cite{KL05,Kes09,KR09}.

\section{Preliminary notions}
\label{sec:general-notion}

As several reduction notions are used along the paper, we first
introduce general definitions of rewriting. \medskip 

A \deft{reduction system} is a pair $(R, \Rew{\R})$ consisting of a set $R$
and a binary relation $\Rew{\R}$ on $R$ called a \deft{reduction relation}. When $(a,b) \in \Rew{\R}$ we write
$a \Rew{\R} b$ and we say that $a$ \deft{$\R$-reduces} to $b$. 
The inverse of $\Rew{\R}$ is written
$\LRew{\R}$, \ie\ $b\ \LRew{\R} a$ iff
$a \Rew{\R} b$. The reflexive and transitive (resp. transitive) closure
of $\Rew{\R}$ is written $\Rewn{\R}$ (resp. $\Rewplus{\R}$). Composition
of relations is denoted by juxtaposition. 
 Given $k \geq 0$, we write $a \Rewnumber{k}{\R} b$ iff 
$a$ is $\R$-related to $b$ in $k$ steps, \ie\ $a \Rewnumber{0}{\R}
  b$ if $a=b$ and $a \Rewnumber{n+1}{\R} b$ if $\exists\ c$ s.t.  $a
  \Rew{\R} c$ and $c \Rewnumber{n}{\R} b$.

Given a \deft{reduction system} $(R, \Rew{\R})$,  we use the following reduction notions: 
\begin{enumerate}[$\bullet$]
  \item $\R$ is \deft{locally confluent} if 
$\LRew{\R} \Rew{\R}\subseteq \Rewn{\R} \LRewn{\R}$,  \ie\ 
if $a \Rew{\R} b$ and $a \Rew{\R} c$, then $\exists d$ s.t.
$b \Rewn{\R} d$ and $c \Rewn{\R} d$.
 \item $\R$ is \deft{confluent} if $\LRewn{\R}\Rewn{\R}\subseteq \Rewn{\R} \LRewn{\R}$,  \ie\ 
if $a \Rewn{\R} b$ and $a \Rewn{\R} c$, then $\exists d$ s.t.
$b \Rewn{\R} d$ and $c \Rewn{\R} d$.
\item $s \in R$ is in \deft{$\R$-normal form}, written $s \in$ \deft{$\R$-nf}, if there is no $s'$ such that $s\Rew{\R} s'$.  
\item $s\in R$ has an \deft{$\R$-normal form} iff there exists $u \in \R$-nf
such that $s \Rewn{\R} u$. 
When  $s$ has a \textit{unique} $\R$-normal form, this one is denoted by $\R(s)$.
\item $s\in R$ is  \deft{$\R$-weakly normalizing}, written $s \in \WN{\R}$,
iff $s$ has an  $\R$-normal form.
\item  $s\in R$   is  \deft{$\R$-strongly  normalizing} or
\deft{$\R$-terminating},   written  $s  \in
  \SN{\R}$, if  there is no infinite  $\R$-reduction sequence starting
  at  $s$.
 \item $s\in R$   is  \deft{$\R$-finitely branching}
       if the set $\set{ s' \mid s \Rew{\R} s'}$ is finite.
\item If $s \in \R$ is  $\R$-strongly  normalizing
and $\R$-finitely branching
then $\eta_{\R}(s)$  
denotes  the
  \deft{maximal length of an $\R$-reduction sequence starting at $s$}.
  This notion is  extended to  lists of terms
   by $\eta_{\R}(s_1 \ldots s_m)= \sum_{i=1}^{m} \eta_{\R}(s_i)$.
\item $\R$ is \deft{weakly normalizing} (resp. \deft{strongly  normalizing} or \deft{terminating})
if every $s \in \R$ is.
\end{enumerate}

A \deft{strong bisimulation between} two
reduction
systems 
$(S, \Rew{S})$ and $(Q, \Rew{\Q})$ is a
relation $\ttE\ \subseteq S \times Q$ s.t. for any pair
$s\ \ttE\ t$:
\begin{enumerate}[$\bullet$]
   \item If $s \Rew{\S} s'$ then 
        there is  $t'$ s.t. 
        $t \Rew{\Q} t'$ and $s'\ \ttE\ t'$, and conversely:
   \item If $t \Rew{\Q} t'$ then there is $s'$ s.t. $s\Rew{\S} s'$ and $s'\ \ttE\ t'$.
  \end{enumerate}

A  \deft{strong bisimulation for $(S, \Rew{S})$} is a 
strong bisimulation between $(S, \Rew{S})$ and itself.
In particular we shall make use of the following property 
 whose proof is straightforward:

\begin{lem}
\label{l:lamj-bis-psn-dif-sys}
Let $\ttE$ be a strong bisimulation between two reduction systems 
$(S, \Rew{\S})$ and $(Q,\Rew{\Q})$. 
\begin{enumerate}
  \item The relation $\ttE$ preserves reduction lengths, \ie\ for any $s\ \ttE\ t$ 
\begin{enumerate}[$\bullet$]
\item If $s \Rewnumber{k}{\S} s'$ then 
      $\exists\ t'$ s.t. $t \Rewnumber{k}{\Q} t'$ and $s'\ \ttE\ t'$.
\item If $t \Rewnumber{k}{\Q} t'$ then  
      $\exists\ s'$ s.t. $s \Rewnumber{k}{\S} s'$ and $s'\ \ttE\ t'$.
\end{enumerate}
\item \label{l:sbisim-pres-psn}The relation $\ttE$ preserves strong normalization, \ie for any $s\ \ttE\ t, 
s\in  \SN{\S}$ if and only if $t\in\SN{\Q}$.
\end{enumerate}
\end{lem}

Given a reduction relation $\Rew{\S}$  and
an equivalence relation $\ttE$ both on $S$, the reduction relation $\Rew{\modulo{\S}{\ttE}}$,
called \deft{reduction $\S$ modulo $\ttE$}, is defined by
$t \Rew{\modulo{\S}{\ttE}} u$ iff $t\ \ttE\ t' \Rew{\S} u'\ \ttE\ u$.

\begin{lem}
\label{l:bisim-conf}
Let $\ttE$ be a strong bisimulation for 
$(S, \Rew{\S})$. Then, 
\begin{enumerate}
  \item \label{l:bisim-conf-i} The relation  $\ttE$ can be postponed
w.r.t $\Rew{\S}$, \ie\ 
$\Rewn{\modulo{\S}{\ttE}}=\Rewn{\S}\ttE$.

  \item \label{l:bisim-conf-ii} If $\Rew{\S}$ is confluent then $\Rew{\modulo{\S}{\ttE}}$ 
   is
    confluent.
\item \label{l:bisim-conf-iii}  If $t \in \SN{\S}$, then $t \in \SN{\modulo{\S}{\ttE}}$.
\end{enumerate}
\end{lem}

\begin{proof}
Point~\ref{l:bisim-conf-i} is straightforward by induction on the
length of $\Rewn{\modulo{\S}{\ttE}}$ using the definition of strong
bisimulation. Points~\ref{l:bisim-conf-ii} and~\ref{l:bisim-conf-iii}
follow from Point~\ref{l:bisim-conf-i}.
\end{proof}

We conclude this section by giving an abstract theorem that we will
use to prove strong normalisation for different notions of reduction
modulo.

\begin{thm}[Termination for reduction modulo by interpretation]
\label{t:equational-abstract}
 Let consider three reduction systems $(A, \Rew{\calA_1})$,    $(A, \Rew{\calA_2})$
and $(B, \Rew{\calB})$. Let 
$\ttE$ (resp. $\ttF$) be an equivalence on $A$ (resp. $B$). 
Consider a relation $\ttR\ \subseteq A \times B$. Suppose
that for all $u,v,U$

\begin{enumerate}[\bf(P1)]
\item[{\bf (P0)}] $u\ \ttR\ U\ \&\ u\ \ttE\ v$ imply  $\exists V$
      s.t. $v\ \ttR\ V\ \&\ U\ \ttF\  V$.
\item[{\bf (P1)}] $u\ \ttR\ U\ \&\ u \Rew{\calA_1} v$ imply   $\exists V$
      s.t. $v\ \ttR\ V\ \&\ U \Rewn{\calB} V$.
\item[{\bf (P2)}] $u\ \ttR\ U\ \&\ u \Rew{\calA_2} v$ imply   $\exists V$
      s.t. $v\ \ttR\ V\ \&\ U \Rewp{\calB} V$.
\item[{\bf (P3)}] The reduction relation $\Rew{\modulo{\calA_1}{\ttE}}$ is  
terminating.
\end{enumerate}

Then, $t\ \ttR\ T\ \&\ T \in \SN{\modulo{\calB}\ttF}$ imply  $t \in
\SN{(\calA_1 \cup \calA_2)/\ttE}$.
\end{thm}

\begin{proof}
Suppose $t\notin \SN{(\calA_1\cup \calA_2)/\ttE}$. Then, there is an
infinite $(\calA_1\cup \calA_2)/\ttE$-reduction sequence starting at $t$,
and since $\Rew{\modulo{\calA_1}{\ttE}}$ is a terminating reduction  relation by
$(P3)$, this reduction has necessarily the form:
\[\begin{array}{ccccccccccccccccc}
   t&\Rewn{\calA_1/\ttE}&t_1&\Rewp{\calA_2/\ttE}&t_2&\Rewn{\calA_1/\ttE}&t_3&\Rewp{\calA_2/\ttE}&t_4&\Rewn{\calA_1/\ttE}\ldots
  \end{array}\]
And can be projected by $(P0)$, $(P1)$ and $(P2)$ into an infinite $\calB$ reduction sequence as follows:
\[\begin{array}{ccccccccccccccccc}
   t&\Rewn{\calA_1/\ttE}&t_1&\Rewp{\calA_2/\ttE}&t_2&\Rewn{\calA_1/\ttE}&t_3&\Rewp{\calA_2/\ttE}&t_4&\Rewn{\calA_1/\ttE}\ldots\\
      \ttR&&\ttR&&\ttR&&\ttR&&\ttR\\
      T&\Rewn{\calB/\ttF}&T_1&\Rewp{\calB/\ttF}&T_2&\Rewn{\calB/\ttF}&T_3&\Rewp{\calB/\ttF}&T_4&\Rewn{\calB/\ttF}\ldots
  \end{array}\]

Since $T \in \SN{\calB/\ttF}$, then we get a contradiction. 
  \end{proof}

\section{The structural $\ldis$-calculus}
\label{s:structural-lj}

We introduce in this section the structural $\ldis$-calculus,
which can simply be understood as a refinement of  $\lam$-calculus.  To be self-contained, 
we start this section by recalling the syntax and semantics of $\lam$-calculus. 
The set of $\lam$-terms,  written  $\termslambda$,
is generated by the following grammar:
\[ (\termslambda) \sep t, u :: = x \mid \lam x. t \mid t u  \]
Dynamics of $\lam$-terms is   given by $\Beta$-reduction (noted $\Rew{\Beta}$) 
which is  defined as the closure by contexts of
the following reduction rule:
$$(\lam x . t) u \mapsto_{\Beta} t\isubs{x/u}$$  
where the meta-operation $t\isubs{x/u}$ on $\lam$-terms is just a particular case
of the meta-operation on $\ldis$-terms given below.

The \deft{structural $\ldis$-calculus} is given by a set of
terms and a set of reduction rules. The set of $\ldis$-terms,  written  $\terms$,
is generated by the following grammar:
\[ (\terms) \sep t, u :: = x \mid \lam x. t \mid t u \mid t[x/u] \]

The term $x$ is \deft{variable}, $\lam x. t$ an \deft{abstraction}, $tu$
an \deft{application} and $t[x/u]$ a \deft{substituted term}. The object
$[x/u]$, which is not a term, is called a \deft{jump}.  The terms $\lam 
x.  t$ and $t[x/u]$ bind $x$ in $t$, \ie\ the sets of \deft{free/bound
  variables} of a term are given by the following definitions:
\[ \begin{array}{lll@{\sep}lll}
\fv{x} & := & \set{x} & \bv{x} & := & \ems \\
\fv{tu} &:=& \fv{t} \cup \fv{u} & \bv{tu} & := & \bv{t} \cup \bv{u}\\
\fv{\lam  x.t}&:=& \fv{t}\setminus\set{x} & \bv{\lam  x.t}& :=& \bv{t} \cup \set{x}\\
\fv{t[x/u]}&:=&(\fv{t}\setminus\set{x}) \cup\fv{u} 
& \bv{t[x/u]}&:=& \bv{t}\cup \set{x}\cup\bv{u} \\  
\end{array} \] 
A jump $[x/u]$ in a
term $t[x/u]$ is called \deft{void} if $x \notin \fv{t}$.  The
equivalence relation generated by the renaming of bound variables is
called \deft{$\alpha$-conversion}.  Thus for example $(\lam y.  x)[x/y]
\equiv_{\alpha} (\lam y'. x')[x'/y]$.  
The notation $\ovl{t}{1}{n}$ is used for the empty sequence of terms if $n=0$ and
for the sequence $[t_1; \ldots; t_n]$ otherwise; $\ovl{t}{1}{n} \subseteq  S$ 
means that all the elements of the sequence belong to the set $S$.
If $i,n\in\nat$ we use $v\ovl{t}{i}{n}$ for the term $v$ if $n<i$ and 
$(v t_i)\ovl{t}{i+1}{n}$ otherwise; similarly, $t\esp{x}{u}{i}{n}$ 
denotes the term $t$ if $n<i$ and 
$t[x_i/u_i]\esp{x}{u}{i+1}{n}$ otherwise; 
$t_1 t_2
\ldots t_n\ (n \geq 1)$ denotes the application 
$(\ldots (t_1 t_2) \ldots \ldots ) t_n$; 

The meta-level \deft{substitution} operation is defined by
induction on terms by using the following equations on $\alpha$-equivalence classes:
\[ \begin{array}{llll}
x \isubs{x/u} & := u \\
y \isubs{x/u} & := y \\
(\lam y. t) \isubs{x/u} & := \lam y. t \isubs{x/u} & \mbox{ if } y \notin \fv{u}   \\
(tv) \isubs{x/u} & := t \isubs{x/u}v \isubs{x/u}   \\
t[y/v] \isubs{x/u} & := t \isubs{x/u}[y/v \isubs{x/u}] & \mbox{ if } y \notin \fv{u}  \\
\end{array} \] 

We write $t \surt u$ or $u \subt t$ when $u$ is a (strict) subterm of $t$.
\deft{Positions} of terms are defined as expected (see~\cite{Terese03}, p. 643, for details);  $t|_p$ denotes
the \deft{subterm of $t$ at position $p$}
and $\pos{x}{t}$ denotes the \deft{set of all the positions} $p$ of
$t$ s.t. $t|_{p}= x$.

We use $|t|$ to denote the size of $t$. We write $|t|_x$ for the
number of free occurrences of the variable $x$ in the term $t$,
called the \deft{multiplicy of $x$ in $t$}. We extend this notion to sets of variables by
  $|t|_{\Gamma}:= \Sigma_{x \in \Gamma} |t|_{x}$. 
A key notion used to define the semantics 
of the $\ldis$-calculus is that of renaming:
given a term $t$ and a subset
$S \subseteq \pos{x}{t} \cap \fv{t} $, we write $\ren{t}{S}{x}{y}$ for
the term $t'$ verifying $t'|_p = t|_p$ if $p \notin \S$
and $t'|_p = y$ if $(t|_p =x\ \&\ p \in S)$. Thus for example,
$\ren{x z x x}{\set{111,2}}{x}{y} = y z x y$. 

When $|t|_{x}= n
\geq 2$, we write $t_{[y]_x}$ for any
\deft{non-deterministic replacement} of $i\ ( 1 \leq i \leq n-1)$
occurrences of $x$ in $t$ by a {\it fresh} variable $y$, \ie\
$t_{[y]_x}$ denotes  any term $\ren{t}{S}{x}{y}$ s.t. 
$|S| \geq 2$ and $S \subset \pos{x}{t}$. 
 Thus for
example, $(x x x x)_{[y]_x}$ may denote $(y x y x)$ or
$(x y y y)$ but not $(y y y y)$.

\deft{Contexts} are generated by the following grammar:
\[ C :: = \hole \mid C v \mid v C \mid v[y/C] \mid C[y/v] \mid \lam y. C\]
We write $\ctx{C}{t}$ to denote the term obtained by replacing the
hole $\hole$ in $C$ by the term $t$. Thus for example  
$\ctx{\lam x. z[y/w\hole]}{x} = \lam x. z[y/wx]$ (remark that 
capture of variables is possible).

The  \deft{binding set} of a context 
is defined as follows: 
\[ \begin{array}{lll@{\sep\sep}lll}
\bs{\hole}&:=&\ems & 
\bs{t{[}x/C{]}}&:=&\bs{C}\\
\bs{t C}&:=&\bs{C} & 
\bs{C[x/v]}&:=& \bs{C}\cup\set{x} \\
\bs{C v}&:=&\bs{C} & 
\bs{\lam x. C}&:=&\bs{C}\cup\set{x}\\
\end{array}\]

We now consider the
rewriting rules of the structural $\lam$-calculus
(Figure~\ref{f:lambdaj}), which decompose the $\beta$-rule into a
finer set of rules. The letter $\List$ in the rule $\B$
  denotes a list $[x_1/u_1]\ldots[x_k/u_k]$ of jumps with
  $k\in\nat$ (so potentially $k=0$) such that $\set{x_1, \ldots, x_k}
  \cap \fv{u} = \ems$.  The $\B$ rule extends the usual ${\tt B}$ rule
  $(\lam x.t) u \Rew{\tt B} t[x/u]$ by allowing to introduce some
  distance between the abstraction $\lam x. t$ and the argument $u$
  which is specified by means of a list of substitutions
  $\slist$. This natural extension comes from reading back a
  multiplicative cut in $\lj$-dags or Pure
  Proof-Nets~\cite{AG09,AccattoliTh}.

The substitution
  rules also deserve some explanation. The side conditions $|t|_x=0$,
  $|t|_x=1$ and $|t|_x>1$ are global on terms but local on graphs,
  simply because in the graph all the occurrences of the same variable are grouped
  together.  Also, the (global) meta-substitution
  operation $t\isubs{x/u}$ used in the right-hand side of the rule
  $\Var$ is completely local on graphs. Similarly, the meta-operation
  $t_{[y]_x}$ used in the right-hand side of the $\DSubs$-rule is an
  algebraic notation for the local operation on graphs which splits  the co-located
  occurrences of $x$ into two disjoint and
  non-empty sets, one of which corresponds  to $x$, while the other is associated to
  the fresh variable $y$. Thus, the structural
$\lam$-calculus can be seen as an algebraic language useful to study
$\lj$-dags and Pure Proof-Nets. 

\begin{figure}[ht]
\[\begin{array}{l@{\sep}lll@{\sep}l}
(\mbox{{\tt B}eta at a {\tt d}istance}) & (\lam x.t)\List\ u & \rRew{\B}&  t[x/u]\List    \\
(\mbox{{\tt w}eakening}) & t[x/u] & \rRew{\Gc} & t &\mbox{if $|t|_x=0$}\\
(\mbox{{\tt d}ereliction}) & t[x/u] & \rRew{\Var} & t\set{x/u} &\mbox{if $|t|_x=1$}\\
(\mbox{{\tt c}ontraction}) & t[x/u] & \rRew{\DSubs} & t_{[y]_x}[x/u][y/u] &\mbox{if $|t|_x>1$}\\
  \end{array}\]
\caption{The $\ldis$-reduction system}
\label{f:lambdaj}
\end{figure}

We close these rules by contexts, as usual:
$\Rew{\R}$ denotes the contextual closure of $\rRew{\R}$, for $\R
\subseteq \set{\B,\Gc,\Var,\DSubs}$. We write $\Rew{\nGc}$ for the
reduction relation $\Rew{\B,\Var,\DSubs}$.  The \deft{reduction
  relation} $\Rew{\ldis}$ (resp.  $\Rew{\dis}$) is generated by all
(resp.  all expect $\B$) the previous rewriting rules modulo
$\alpha$-conversion. \medskip

An expected property of $\lj$ is that the reduction relation $\ldis$ is  stable by substitution. 
\begin{lem} Let $t, u \in terms$.
\begin{enumerate}[$\bullet$]
\item If $t \Rew{\ldis} t'$, then $t\isubs{x/u} \Rew{\ldis}t'\isubs{x/u}$. 
\item If $u \Rew{\ldis} u'$, then $t\isubs{x/u} \Rewn{\ldis}t\isubs{x/u'}$. 
\end{enumerate}
\end{lem}

In the rest of this section we shall prove the following properties of $\ldis$:
full composition (Lemma~\ref{l:fc}), 
simulation of one step $\beta$-reduction (Lemma~\ref{l:l-sim}),
termination and uniqueness of normal forms of the  substitution calculus $\Rew{\dis}$
(Lemmas~\ref{l:dis-terminates} and~\ref{l:uniqueness-j}),
postponement of erasing reductions (Lemma~\ref{l:w-postponement})
and confluence of $\ldis$ (Theorem~\ref{t:confluence}).

\subsection{Jumps and Multiplicities}
\label{ss:struct-subs-and-mul}
The first property we show in this section 
is full composition, stating  that any 
jump $[x/u]$ in a substituted term  $t[x/u]$ can be
reduced to its implicit form $t\isubs{x/u}$.  There are two
  interesting points. The first is that in contrast with most calculi
  of explicit substitutions, full composition holds with no need of
  equivalences. The second is that the proof is by
  induction on $|t|_x$ and not on the structure of $t$.

\begin{lem}[Full Composition (FC)]
\label{l:fc}
Let  $t,  u  \in \terms$.  Then  $t[x/u]
\Rewplus{\dis}  t\isubs{x/u}$. Moreover, $|t|_{x}  \geq  1$ implies
$t[x/u] \Rewplus{\Var,\DSubs} t\isubs{x/u}$.
\end{lem}

\proof
By induction on $|t|_{x}$.
\begin{enumerate}[$\bullet$]
\item If $|t|_{x} = 0$, then $t[x/u] \Rew{\Gc} t = t\isubs{x/u}$.
\item If $|t|_{x} = 1$, then 
      $t[x/u] \Rew{\Var} t\isubs{x/u}$.
\item If $|t|_{x} \geq 2$, then 
\[ \begin{array}{llllll}
t[x/u] &\Rew{\DSubs}& t_{[y]_x}[y/u][x/u]&  \Rewplus{\dis}\ (\ih)\ \\
  & &t_{[y]_x}\isubs{y/u}[x/u]&\Rewplus{\dis}\ (\ih)\ \\
&  &t_{[y]_x}\isubs{y/u}\isubs{x/u} &=&   t\isubs{x/u}
  \rlap{\hbox to 93 pt{\hfill\qEd}}
  \end{array}   \]\bigskip
\end{enumerate}

Due to the very general form of the duplication rule
of $\ldis$, we get the following corollary which together with full composition can be seen as a generalised composition property: 

\begin{cor}
Given  $S \subset \pos{x}{t}$ s.t.  $|S| \geq 2$, then $t[x/u]
\Rewplus{\dis} \ren{t}{S}{x}{y} \set{y/u}[x/u]$, where $y$ is a
fresh variable. 
\end{cor}

\begin{proof}
The term $t[x/u]$  $\DSubs$-reduces to 
$\ren{t}{S}{x}{y}[y/u][x/u]$. We conclude
by full composition.
\end{proof}

 Thus for example $(x (x x))[x/u] \Rewplus{\ldis}
  (x (u x))[x/u]$.  Note that this property is not enjoyed by
  traditional explicit substitution calculi: for instance, in
  $\lx$~\cite{BlooRoselx}, the term $(x (x x))[x/u]$ cannot be reduced to
  $(x (u x))[x/u]$.  However, it holds in
  calculi with partial substitutions, as Milner's calculus $\lsub$~\cite{Milner07}. It is not
  difficult (see \eg~\cite{OConchuirKesner}) to define a translation
  {\tt T} on terms such that $t \Rew{\lsub} t'$ implies ${\tt T}(t)
  \Rewplus{\ldis} {\tt T}(t')$.  This property allows in particular to
  deduce normalisation properties for  $\lsub$ from those  of  $\ldis$.  \\

The one-step simulation of $\lam$-calculus follows directly from full composition:

\begin{lem}[Simulation of $\lam$-calculus]
\label{l:l-sim}
Let $t \in \termslambda$.  If $t\Rew{\beta} t'$ then  $t
\Rewplus{\ldis}  t'$.
\end{lem}

\begin{proof}
By induction on $t \Rew{\beta} t'$.
Let  $t=(\lam  x.  u)  v \Rew{\beta}
u\isubs{x/v}$, then  $t\Rew{\B}  u[x/v] \Rewplus{\dis}\ (Lem.~\ref{l:fc})\ 
u\isubs{x/v}$. All the other cases are straightforward.
\end{proof}

We now introduce a notion that will be useful in various proofs. 
It counts the maximal number of free occurrences of a variable $x$ that may appear during
a $\dis$-reduction sequence from a term $t$. 

The \deft{potential multiplicity} of the variable $x$ in the term $t$, written $\mul{t}{x}$, 
is defined on $\alpha$-equivalence classes as follows:  
if  $x \notin \fv{t}$, then 
$\mul{t}{x}  :=  0$; otherwise:
\[ \begin{array}{llll}
   \mul{x}{x} & := & 1 & \\
   \mul{\lam y. u}{x} & := & \mul{u}{x} & \\
   \mul{u v}{x} & := & \mul{u}{x} + \mul{v}{x} & \\
   \mul{u[y/v]}{x} & := & \mul{u}{x} + \maxi{1}{\mul{u}{y}} \cdot \mul{v}{x} \\
 \end{array} \]
We can formalise the intuition behind $\mul{t}{x}$ as follows.

\begin{lem}
\label{l:mul-occ} 
Let $t \in \terms$.   Then 
\begin{enumerate}[\rm(1)]
\item \label{l:mo-one}$|t|_x\leq \mul{t}{x}$.
\item If $t$ is a $\DSubs$-nf then $|t|_x=\mul{t}{x}$.
\end{enumerate}
\end{lem}

\begin{proof}
Both points  are by induction  on the definition of  $\mul{t}{x}$. The
only  interesting  case  is  when  $t=u[y/v]$: the  \ih\  gives
$|u|_x\leq\mul{u}{x}$,            $|u|_y\leq\mul{u}{y}$            and
$|v|_x\leq\mul{v}{x}$, from which we conclude with the first point. 
For the second one, if $t$ is a $\DSubs$-nf
every   relation   given   by    the   \ih\   is   an   equality   and
$|u|_y=\mul{u}{y}\leq    1$,    otherwise    there    would    be    a
$\DSubs$-redex.                Then               we               get
$\mul{t}{x}=\mul{u}{x}+\maxi{1}{\mul{u}{y}}\cdot\mul{v}{x}=|u|_x+|v|_x=|t|_x$.
\end{proof}

Potential multiplicities enjoy the following properties.

\begin{lem}
\label{l:properties-mul-for-terms}
Let $t \in \terms$. Let $x,y,z$ be pairwise distinct variables.
\begin{enumerate}[\rm(1)]
\item \label{l:pmt-1} If $u \in \terms$ and $y \notin \fv{u}$,
then $\mul{t}{y} =  \mul{t\isubs{x/u}}{y}$.
\item \label{l:pmt-2}  If $|t|_x \geq 2$,
then $\mul{t}{z} =  \mul{t_{[y]_x}}{z}$ and  
$\mul{t}{x} =  \mul{t_{[y]_x}}{x} + \mul{t_{[y]_x}}{y}$, where the two
occurrences of the term $t_{[y]_x}$ denote exactly the same term. 
\item \label{l:pmt-3} 
If  $t \Rew{\dis} t'$, then $\mul{t}{y} \geq   \mul{t'}{y}$.
\end{enumerate}
\end{lem}

\begin{proof}
By induction on $t$.
\end{proof}

By exploiting potential multiplicities we can define a measure of the
global degree of sharing of a given term, and use this
  measure to prove that the $\jop$-reduction subsystem terminates.

We consider multisets of integers. We use $\ems$ to denote the empty
multiset, $\sqcup$ to denote multiset union, and $\geqm$
  (resp. $\gm$) for the standard order (resp. strict order) on
  multisets~\cite{Nipkow-Baader}.  Given an integer
$n$ and a multiset $M$, $n \cdot M$ denotes $\ems$ if $M =\ems$
and the multiset 
$\multiset{n\cdot a_1, \ldots, n\cdot a_n}$ if $M=  \multiset{a_1, \ldots, a_n}$. The
\textbf{$\dis$-measure} of $t \in \terms$, written $\dm{t}$, is given
by:
\[ \begin{array}{llll}
   \dm{x} & := & \ems \\
   \dm{\lam x. u} & := & \dm{u} \\
   \dm{uv} & := & \dm{u} \sqcup  \dm{v} \\
   \dm{u[x/v]} & := & \multiset{\mul{u}{x}} \sqcup \dm{u} \sqcup \maxi{1}{\mul{u}{x}} \cdot \dm{v} \\
   \end{array} \] 

Note that $\dm{u} = \ems$ for $u \in \termslambda$. Potential
multiplicities are 
decreasing by $\dis$-reduction, and we are going to show that the
$\jop$-measure is strictly decreasing; however  both can be incremented by
$\B$-steps. For example, consider $t=(\lam 
  x. x x) y\Rew{\B}(x x)[x/y]=t'$.  We get $\mul{t}{y}=1$,
  $\mul{t'}{y}=2$, $\dm{t}=\emptyset$ and $\dm{t'}=[2]$.

The fact that the $\jop$-measure decreases by $\dis$-reduction is proved as follows:

\begin{lem}
\label{l:properties-dm-for-terms}
Let $t \in \terms$. Then, 
\begin{enumerate}[\rm(1)]
\item \label{l:pdmt-1} $\dm{t} = \dm{t_{[y]_x}}$.
\item \label{l:pdmt-2} If $|t|_x =1$, then
$ \dm{t[x/u]} \gm  \dm{t\isubs{x/u}}$.
\end{enumerate}
\end{lem}

\proof By induction on $t$. The proof of the  first property is 
straightforward.
For the second one we show $\multiset{\mul{t}{x}} \sqcup \dm{t} \sqcup
\maxi{1}{\mul{t}{x}} \cdot \dm{u} \gm \dm{t\isubs{x/u}}$, which proves
the desired property.
\begin{enumerate}[$\bullet$]
\item $t=x$. Then 
$\multiset{1} \sqcup  \dm{u} \gm  \dm{u} = \dm{x\isubs{x/u}}$. 
\item $t = t_1[y/t_2]$. W.l.g we assume $y \notin \fv{u}$. 

If $x \in \fv{t_1}$, we reason as follows:
\[ \begin{array}{llll}
    \multiset{\mul{t}{x}} \sqcup \dm{t} \sqcup  
    \maxi{1}{\mul{t}{x}} \cdot \dm{u} & = \\
    \multiset{\mul{t_1}{x}} \sqcup \multiset{\mul{t_1}{y}}\sqcup \dm{t_1} 
    \sqcup \maxi{1}{\mul{t_1}{y}} \cdot \dm{t_2}  \sqcup 
    \maxi{1}{\mul{t_1}{x}} \cdot \dm{u} & \gm_{\ih} \\
    \multiset{\mul{t_1}{y}}\sqcup  \maxi{1}{\mul{t_1}{y}} \cdot \dm{t_2}   \sqcup
    \dm{t_1\isubs{x/u}} & =_{Lem.~\ref{l:properties-mul-for-terms}:\ref{l:pmt-1}}  \\
    \multiset{\mul{t_1\isubs{x/u}}{y}}\sqcup  
    \maxi{1}{\mul{t_1\isubs{x/u}}{y}} \cdot \dm{t_2}   \sqcup
    \dm{t_1\isubs{x/u}} & = \\
    \dm{t_1\isubs{x/u}[y/t_2]}&= \\
 \dm{t\isubs{x/u}}
    \end{array} \]

If $x \in \fv{t_2}$, then  $1 \leq \mul{t_2}{x}$ by Lemma \ref{l:mul-occ}:\ref{l:mo-one} and so 
$ \maxi{1}{\maxi{1}{\mul{t_1}{y}}
    \cdot \mul{t_2}{x}}=\maxi{1}{\mul{t_1}{y}} \cdot
  \mul{t_2}{x}=\maxi{1}{\mul{t_1}{y}} \cdot
  \maxi{1}{\mul{t_2}{x}}$. Therefore:
\[ \begin{array}{ll}
    \multiset{\mul{t}{x}} \sqcup \dm{t} \sqcup  \maxi{1}{\mul{t}{x}} \cdot \dm{u} & = \\
    \multiset{\maxi{1}{\mul{t_1}{y}} \cdot \mul{t_2}{x}} \sqcup
    \multiset{\mul{t_1}{y}} \sqcup \dm{t_1} \sqcup
    \maxi{1}{\mul{t_1}{y}} \cdot \dm{t_2}  & \\
    \sqcup\     \maxi{1}{\maxi{1}{\mul{t_1}{y}} \cdot \mul{t_2}{x}} \cdot \dm{u} & = \\
 \multiset{\maxi{1}{\mul{t_1}{y}} \cdot \mul{t_2}{x}} \sqcup \multiset{\mul{t_1}{y}} \sqcup \dm{t_1} \sqcup   \maxi{1}{\mul{t_1}{y}} \cdot \dm{t_2}  & \\
    \sqcup\    \maxi{1}{\mul{t_1}{y}} \cdot \maxi{1}{\mul{t_2}{x}}  \cdot \dm{u} & = \\
\multiset{\mul{t_1}{y}}  \sqcup \dm{t_1}\ \sqcup    
    \maxi{1}{\mul{t_1}{y}} \cdot (\multiset{\mul{t_2}{x}} \sqcup \dm{t_2}  \sqcup      \maxi{1}{\mul{t_2}{x}} \cdot \dm{u} ) &  \gm_{\ih} \\
    \multiset{\mul{t_1}{y}}  \sqcup \dm{t_1} \sqcup  \dm{t_2\isubs{x/u}} & = \\
    \dm{t\isubs{x/u}}
    \end{array} \]

\item All the other cases are straightforward.
\qed 
\end{enumerate}

\begin{lem}
\label{l:dm-decreases}
Let $t_0 \in \terms$. Then, 
\begin{enumerate}[\rm(1)]
\item $t_0 \equiv_{\alpha} t_1$ implies $\dm{t_0} = \dm{t_1}$.
\item $t_0 \Rew{\dis} t_1$ implies $\dm{t_0}\gm  \dm{t_1}$.
\end{enumerate}
\end{lem}

\proof
By induction on the relations. The first point is straightforward, hence
we only show the second one. We reason by cases. 
\begin{enumerate}[$\bullet$]
\item $t_0  =  t[x/u]  \Rew{\Gc}  t= t_1$, with $|t|_{x}= 0$. 
Then $\dm{t_0} =\dm{t} \sqcup 1 \cdot \dm{u}  \sqcup  \multiset{0} \gm  \dm{t} = \dm{t_1}$. 
 
\item $t_0  = t[x/u]   \Rew{\Var}  t\isubs{x/u}= t_1$, with $|t|_{x}=  1$. 
Then $\dm{t[x/u]} \gm_{Lem.~\ref{l:properties-dm-for-terms}:\ref{l:pdmt-2}} 
      \dm{t\isubs{x/u}}$. 

\item $t_0  = t[x/u]  \Rew{\DSubs}  t_{[y]_x}[x/u][y/u]= t_1$, with $|t|_{x} \geq 2$
and $y$ fresh. Then, Lemma~\ref{l:properties-mul-for-terms}:\ref{l:pmt-2}
gives $ \multiset{\mul{t}{x}} \gm  \multiset{\mul{t_{[y]_x}}{x}}  \sqcup 
\multiset{\mul{t_{[y]_x}}{y}}$ and  thus:
 \[ \begin{array}{lll}
      \dm{t_0} & = \\
      \multiset{\mul{t}{x}}\sqcup\dm{t} \sqcup \mul{t}{x} \cdot \dm{u}   & =\\ 
      \multiset{\mul{t}{x}}\sqcup\dm{t} \sqcup ( \mul{t_{[y]_x}}{x} +  \mul{t_{[y]_x}}{y}) \cdot \dm{u} & =_{Lem.\ref{l:properties-dm-for-terms}:\ref{l:pdmt-1}} \\
      \multiset{\mul{t}{x}}\sqcup\dm{t_{[y]_x}} \sqcup ( \mul{t_{[y]_x}}{x} +  \mul{t_{[y]_x}}{y}) \cdot \dm{u}   & \gm_{Lem.~\ref{l:properties-mul-for-terms}:\ref{l:pmt-2}} \\ 
      \multiset{\mul{t_{[y]_x}}{x}} \sqcup \multiset{\mul{t_{[y]_x}}{y}}\sqcup\dm{t_{[y]_x}} \sqcup \mul{t_{[y]_x}}{x} \cdot \dm{u} \sqcup \mul{t_{[y]_x}}{y} \cdot \dm{u}  & = \\
    \multiset{\mul{t_{[y]_x}}{x}} \sqcup \multiset{\mul{t_{[y]_x}[x/u]}{y}}\sqcup\dm{t_{[y]_x}} \sqcup \mul{t_{[y]_x}}{x} \cdot \dm{u} \sqcup \mul{t_{[y]_x}[x/u]}{y} \cdot \dm{u}  & = \\
        \multiset{\mul{t_{[y]_x}[x/u]}{y}}\sqcup\dm{t_{[y]_x}[x/u]} \sqcup \mul{t_{[y]_x}[x/u]}{y} \cdot \dm{u}  & =  \dm{t_1}
       \end{array} \]

\item $t_0  = t[x/u] \Rew{} t'[x/u]=t_1$, where $t \Rew{} t'$. Then:
      \[ \begin{array}{llllll}
      \dm{t_0} & = &
     \multiset{\mul{t}{x}}\sqcup \dm{t} \sqcup \maxi{1}{\mul{t}{x}} \cdot \dm{u}   & \gm_{\ih}\\ 
  &&    \multiset{\mul{t}{x}}\sqcup\dm{t'} \sqcup \maxi{1}{\mul{t}{x}} \cdot \dm{u}     & 
\geqm_{Lem.~\ref{l:properties-mul-for-terms}:\ref{l:pmt-3}}\\ 
 && \multiset{\mul{t'}{x}} \sqcup\dm{t'} \sqcup \maxi{1}{\mul{t'}{x}} \cdot \dm{u}   & =   \dm{t_1}
       \end{array} \]

\ignore{
\item $t_0  = t[x/u] \Rew{} t[x/u']=t_1$, where $u \Rew{} u'$. Then
      \[ \begin{array}{lll}
      \dm{t_0} & = \\
      \dm{t} \sqcup \maxi{1}{\mul{t}{x}} \cdot \dm{u}  \sqcup  \multiset{\mul{t}{x}} & \gm_{\ih}\\ 
      \dm{t} \sqcup \maxi{1}{\mul{t}{x}} \cdot \dm{u'}  \sqcup  \multiset{\mul{t}{x}} & =\\ 
      \dm{t_1}
       \end{array} \]
}
       
\item All the other cases are straightforward. 
\qed
\end{enumerate}

The last lemma obviously implies:

\begin{lem}
\label{l:dis-terminates}
The $\dis$-calculus terminates.
\end{lem}

Furthermore:

\begin{lem}
\label{l:uniqueness-j}
The $\dis$-reduction relation is confluent and terminating. Moreover,
if $\dis(t)$ denotes the (unique) $\dis$-normal form of $t$, then the following
properties hold:
\[ \begin{array}{lll@{\sep\sep\sep\sep}lll}
   \dis(x) & = & x & \dis(u v) & = & \dis(u) \dis(v)\\
   \dis(\lam x. u) & = & \lam x. \dis(u) & \dis(u[x/v]) & = & \dis(u)\isubs{x/\dis(v)}\\
   \end{array} \]
\end{lem}

\begin{proof}
One easily shows that $\Rew{\dis}$ is locally confluent, then
Lemma~\ref{l:dis-terminates} allows to apply Newman's Lemma~\cite{Terese03} to conclude with the first
part of the statament.  The second part can be shown by induction on
the structure of terms. Particularly, when $t = u[x/v]$ one has
$u[x/v] \Rewn{\dis} \dis(u)[x/\dis(v)] \Rewplus{\dis}
(Lem.~\ref{l:fc})\ \dis(u)\isubs{x/\dis(v)}$. It is then sufficient to
note that $\dis$-normal forms are stable by substitutions of
$\dis$-normal forms.
\end{proof}

We conclude this section by showing
another important property of $\ldis$ concerning
the  postponement  of  erasing  steps. We first need the following lemma:

\begin{lem}
\label{l:struct-unary-wop-post}
Let $t \in \terms$. Then:
\begin{enumerate}[\rm(1)]
\item \label{l:struct-unary-wop-post-i} $t \Rew{\Gc}\Rew{\nGc} t'$ implies $t \Rew{\nGc}\Rewplus{\Gc} t'$. 
\item \label{l:struct-unary-wop-post-iv} $t \Rewp{\Gc}\Rew{\nGc}t'$ implies $t \Rew{\nGc}\Rewp{\Gc}t'$
\end{enumerate}
\end{lem}

\begin{proof}
Point~\ref{l:struct-unary-wop-post-i}
is  by induction on the relations and case analysis.
Point~\ref{l:struct-unary-wop-post-iv} is by induction on the length
of $\Rewp{\Gc}$ using 
Point~\ref{l:struct-unary-wop-post-i}.
\end{proof}

Let us use $\tau:t\Rew{}^* t'$ as a notation for a reduction sequence, the symbol ';' for the concatenation of reduction sequences and $\weakmes{\tau}$ for the number of $\Rew{\nGc}$ steps in $\tau$. Then we obtain:

\begin{lem}[$\Gc$-postponement]
\label{l:w-postponement}
Let $t \in \terms$. If $\tau:t\Rewn{\ldis} t'$ then $\exists\ \tau':t\Rewn{\nGc}\Rewn{\Gc} t'$ s.t. $\weakmes{\tau}=\weakmes{\tau'}$.
\end{lem}

\begin{proof}
By   induction  on  $k=\weakmes{\tau}$.   The  case   $k=0$  is
  straightforward.  Let $k>0$.  If $\tau:t\Rew{\nGc}u\Rewn{\ldis} t'$
then   simply   conclude  using   the   \ih\   on  the   sub-reduction
$\rho:u\Rewn{\ldis}t'$.    Otherwise  the sequence $\tau$   starts   with  a
  $\Gc$-step. If all the steps  in $\tau$ are $\Gc$, then we trivially
  conclude.      Otherwise   $\tau=\tau_{\Gc};\Rew{\nGc};\rho$  where
$\tau_{\Gc}$ is  the maximal  prefix of $\tau$  made out  of weakening
steps  only.   By  Lemma~\ref{l:struct-unary-wop-post}:\ref{l:struct-unary-wop-post-iv}
we  get  that
$t\Rew{\nGc}\Rewp{\Gc};\rho\  t'$  and  we  conclude  by  applying  the
\ih\ to $\Rewp{\Gc};\rho$.
\end{proof}

\subsection{Confluence}
\label{ss:struct-confluence}
Confluence of calculi with  ES can be easily proved
by using Tait  and Martin L\"of's technique (see for  example the
case of $\les$~\cite{Kes07}). This  technique is based
on the definition of a simultaneous reduction relation
 which enjoys the  diamond property. It is completely standard so we give the statements of the lemmas and omit the proofs. \medskip

The \deft{simultaneous reduction relation $\paralp{\ldis}$} is defined on terms in $\dis$-normal form as follows:
\begin{enumerate}[$\bullet$]
\item $x \paralp{\ldis} x$
\item If $t  \paralp{\ldis} t'$, then $\lam x. t  \paralp{\ldis} \lam x. t'$
\item If $t  \paralp{\ldis} t'$ and $u \paralp{\ldis} u'$, then $t u  \paralp{\ldis} t' u'$
\item If $t  \paralp{\ldis} t'$ and
         $u \paralp{\ldis} u'$, then $(\lam x. t) u  \paralp{\ldis} \fc(t'[x/u'])$
\end{enumerate}
Note that the third and fourth cases overlap, thus for example,
$(\lam x. II)II \paralp{\ldis} (\lam x. I)I$
and $(\lam x. II)II \paralp{\ldis} I$, where $I$ denotes the identity
function $\lam y. y$.

 A first lemma ensures that $\paralp{\ldis}$ can be simulated by $\Rew{\ldis}$.

\begin{lem}
\label{l:paral-Rewn}
If $t \paralp{\ldis} t'$, then $t \Rewn{\ldis} t'$. 
\end{lem}

\begin{proof}
By induction on $t \paralp{\ldis} t'$. 
\end{proof}

A second lemma ensures that $\Rew{\ldis}$ can be projected through $\fc(\cdot)$ on $\paralp{\ldis}$.

\begin{lem}
\label{l:Rew-xc-paral}
If $t \Rew{\ldis} t'$, then $\fc(t) \paralp{\ldis} \fc(t')$. 
\end{lem}

\begin{proof}
By induction on $t \Rew{\ldis} t'$.
\end{proof}

The two lemmas combined essentially say that $\paralp{\ldis}$ is
confluent if and only if $\Rewn{\ldis}$ is confluent. Then we show the
diamond property for $\paralp{\ldis}$, which implies that
$\Rew{\ldis}$ is confluent:

\begin{lem}
\label{l:paral-diamond}
The relation $\paralp{\ldis}$ enjoys the diamond property.
\end{lem}

\begin{proof}
By induction on $\paralp{\ldis}$ and case analysis.
\end{proof}

Then we conclude:

\begin{thm}[Confluence]
\label{t:confluence}
For all $i\in \set{1,2}$, for all $t, u_i \in \terms$ s.t. $t \Rewn{\ldis} u_i$,
$\exists v$ s.t. $u_i \Rewn{\ldis} v$. 
\end{thm}

\begin{proof}
Let $t \Rewn{\ldis} u_i$ for $i=1,2$. Lemma~\ref{l:Rew-xc-paral}
gives $\fc(t) \paralpn{\ldis} \fc(u_i)$ for $i=1,2$. Lemma~\ref{l:paral-diamond}
implies $\paralp{\ldis}$ is confluent so that $\exists v$ such that
$\fc(u_i) \paralpn{\ldis} v$ for $i=1,2$. We can then close the diagram
with $u_i\Rewn{\dis} \fc(u_i) \Rewn{\ldis} v$ by Lemma~\ref{l:paral-Rewn}.
\end{proof}

While confluence holds for all calculi with explicit substitutions,
\deft{metaconfluence} does not. 
The idea is to
switch to  an enriched language  with a new  kind of (meta)variable  of the
form $X^\Delta$, to  be intended as a named  context hole expected to
be  replaced by  terms whose  free  variables form a subset of $\Delta$. 
This form of metaterm is for example used in  the
framework of higher-order unification~\cite{HuetThEtat}. In
presence  of   meta-variables  not   all  the  substitutions   can  be
computed.  For instance in the metaterm $X^{y}[y/z]$ the jump $[y/z]$  is
blocked.  Consider:
\[ (X^{\set{z_1}} Y^{\set{z_2}})[z_1/x][z_2/x]\  \LRew{\DSubs}
    (X^{\set{z}} Y^{\set{z}})[z/x] \Rew{\DSubs} (X^{\set{z_1}} Y^{\set{z_2}})[z_2/x][z_1/x] \] 

These metaterms are different normal forms. 
However, it is  enough to  add the following equation  to recover
confluence:
\[  t[x/u][y/v] \sim_{\CS} t[y/v][x/u] \mbox{ if } y \notin \fv{u}\ \&\ x \notin \fv{v} \]  

A proof of confluence of $\ldis$ modulo $\CS$ for metaterms can be found in~\cite{Renaudth}.

\section{Preservation of $\beta$-Strong Normalization for $\ldis$}
\label{s:lj-psn}

A reduction system $\R$ for a language containing the set
$\termslambda$ of all $\lam$-terms is said to enjoy the \deft{PSN
  property} iff every $\lam$-term which is
$\beta$-strongly normalizing is also
$\R$-strongly normalizing. Formally, for all
$t \in \termslambda$, if $t \in \SN{\beta}$, then $t \in \SN{\R}$. 

The PSN property, when it holds, is usually non-trivial to prove. We are
going to show that $\lj$ enjoys PSN by giving a particularly compact proof. The proof
  technique has been developed by D.~Kesner~\cite{Kes09}; it reduces PSN to a
  property called $\iep$, which relates termination of
\deft{I}mplicit substitution to termination of \deft{E}xplicit
substitution. It is an abstract technique not
depending on the particular rules of the calculus with
explicit substitutions.

A reduction system $\R$ for a language $\terms_{\R}$ containing the
set $\termslambda$ is said to enjoy the \deft{$\iep$ property} iff for
$n\geq 0$ and for all $t, u \in \termslambda$, $\ovl{v}{1}{n}
  \subseteq \termslambda$:
\[\begin{array}{ccc}
u \in    \SN{\R}\ \&\ t\isubs{x/u} \ovl{v}{1}{n}  \in \SN{\R}\ \&\  t[x/u] \ovl{v}{1}{n} \in \terms_{\R} &\mbox{ imply } 
&   t[x/u] \ovl{v}{1}{n}   \in \SN{\R}
   \end{array}\]

Of course one generally considers a system $\R$ which can simulate the
$\lam$-calculus,  so that the following properties seem to be natural requirements to get PSN.

\begin{thm}[Natural Requirements for PSN]
\label{t:ie-implies-psn}
Let $\R$ be a calculus verifying the following facts:
\begin{enumerate}[\bf(F1)]
\item[{\bf (F0)}] \label{f:uno} If $\ovl{t}{1}{n} \subseteq\termslambda\cap\SN{\R}$, then $x \ovl{t}{1}{n}
  \in \SN{\R}$.
\item[{\bf (F1)}] \label{f:dos} If $u \in \termslambda\cap\SN{\R}$, then $\lam x. u \in    \SN{\R}$.
\item[{\bf (F2)}] \label{f:tres} If $v \in \termslambda\cap\SN{\R}\ \&\ u\isubs{x/v} \ovl{t}{1}{n} \in \termslambda\cap\SN{\R}$, then 
$(\lam x.  u) v \ovl{t}{1}{n} \in  \SN{\R}$.

\end{enumerate}
Then, $\R$ enjoys PSN. 
\end{thm}

\proof
We show that $t\in \SN{\beta}$ implies $t \in \SN{\R}$ by induction on
the pair $\pair{\eta_{\beta}(t)}{|t|}$, using the lexicographic
ordering. We reason by cases.
\begin{enumerate}[$\bullet$]
\item If $t = x\ovl{t}{1}{n}$, then  $t_i \in \SN{\beta}$
and $\pair{\eta_{\beta}(t_i)}{|t_i|} <_{lex} \pair{\eta_{\beta}(t)}{|t|}$. We have 
$t_i \in \SN{\R}$ by the  \ih\ and thus  $x\ovl{t}{1}{n} \in \SN{\R}$ by fact {\bf F0}.
\item If $t =  \lam x.u$, then  $u  \in \SN{\beta}$
and $\pair{\eta_{\beta}(u)}{|u|} <_{lex} \pair{\eta_{\beta}(t)}{|t|}$. 
We have 
$u  \in \SN{\R}$ by the  \ih\ and thus  $\lam x. u \in \SN{\R}$ by fact {\bf F1}.
\item If $t = (\lam x.  u) v\ovl{t}{1}{n}\in \SN{\beta}$, 
then $u\isubs{x/v} \ovl{t}{1}{n} \in \SN{\beta}$ and $v \in \SN{\beta}$. 
Indeed, $\eta_{\beta}(u\isubs{x/v}\ovl{t}{1}{n})  < \eta_{\beta}(t)$
and  $\eta_{\beta}(v) < \eta_{\beta}(t)$. 
We have that 
both terms are in $\SN{\R}$ by  the  \ih\
Then {\bf F2} guarantees that $t \in \SN{\R}$.
\qed
\end{enumerate}
Now we show that 
$\lj$  satisfies the three natural requirements of the last
theorem, and thus it satisfies PSN.

\begin{lem}[Adequacy of $\iep$] 
\label{l:adequacy}
If $\ldis$ verifies $\iep$, then $\ldis$ satisfies PSN.
\end{lem}

\begin{proof} By Theorem~\ref{t:ie-implies-psn} it is sufficient to
show {\bf F0}, {\bf F1} and {\bf F2}.  The first two properties are
  straightforward.  For the third one,  assume $v \in \termslambda \cap \SN{\ldis}$
  and $u\isubs{x/v}\ovl{t}{1}{n} \in \termslambda \cap \SN{\ldis}$.
  Then in particular $u, v, \ovl{t}{1}{n}\in \termslambda \cap
  \SN{\ldis}$. We show that $t = (\lam x.  u) v \ovl{t}{1}{n} \in
  \SN{\ldis}$ by induction on $\eta_{\ldis}(u) + \eta_{\ldis}(v) +
  \Sigma_{i}\ \eta_{\ldis}(t_i)$.  For that, it is sufficient to show
  that every $\ldis$-reduct of $t$ is in $\SN{\ldis}$.  If the
  $\ldis$-reduct of $t$ is internal we conclude by the \ih\ Otherwise
  $t = u[x/v] \ovl{t}{1}{n}$ which is in $\SN{\ldis}$ by the
  \iep\ property.
\end{proof}

As a consequence, in order to get PSN for $\lj$ we only need to prove 
the \iep\ property. For that, we first generalise
the \iep\ property in order to deal with  possibly many
substitutions.

A reduction system $\R$ for a language $\terms_{\R}$ containing
  the set $\termslambda$ is said to enjoy the \deft{Generalised $\iep$ property}, written \deft{$\giep$, } iff for
  all $t, \ovl{u}{1}{m}\ (m \geq 1), \ovl{v}{1}{n}\ (n \geq 0)$ in
  $\terms_{\R}$, if $\ovl{u}{1}{m} \subseteq
  \SN{\R}\ \&\ t\sp{x}{u}{1}{m} \ovl{v}{1}{n}
 \in \SN{\R}$, then
  $t\esp{x}{u}{1}{m} \ovl{v}{1}{n} \in \SN{\ldis}$, where $x_i \neq
  x_j$ for $i,j=1\ldots m$ and $x_i \notin \fv{u_j}$ for $i,j=1\ldots  m$.

\begin{thm}[$\giep$ for $\ldis$]
\label{t:ieg}
The $\ldis$-calculus enjoys the \giep\ property.
\end{thm}

\textbf{Notation}: To improve readability of the proof we shall
abbreviate the notation $\esp{x}{u}{1}{m}$ by $\espp{x}{u}{1}{m}$. Similarly for implicit substitutions.

\proof 
Suppose $\ovl{u}{1}{m} \in    \SN{\ldis}\ \&\ 
             t\sp{x}{u}{1}{m} \ovl{v}{1}{n}  \in    \SN{\ldis}$.
We show $t_0= t \esp{x}{u}{1}{m} \ovl{v}{1}{n}\in    \SN{\ldis}$ by
induction on:
$$\langle \eta_{\ldis}(t\sp{x}{u}{1}{m} \ovl{v}{1}{n}),\ \ 
\fo{t}{\ovl{x}{1}{m}},\ \   \eta_{\ldis}(\ovl{u}{1}{m})  \rangle$$
where
$\fo{t}{x_i} = 3^{|t|_{x_i}}$ and 
 $\fo{t}{\ovl{x}{1}{m}} = \Sigma_{i \in m} \fo{t}{x_i}$.

To show $t_0  \in \SN{\ldis}$ it is sufficient to show that
every $\ldis$-reduct of $t_0$ is in $\SN{\ldis}$.

\begin{enumerate}[$\bullet$]
\item $t_0 \Rew{\ldis} t\espp{x}{u}{1}{j-1} [x_j/u'_j] \espp{x}{u}{j+1}{m} \ovl{v}{1}{n}= t'_0$ with $u_j \Rew{\ldis} u'_j$.  Then we get:
  \begin{enumerate}[$-$]
\item $\eta_{\ldis}(t\spp{x}{u}{1}{j-1}\isubs{x_j/u'_j}\spp{x}{u}{j+1}{m}\
 \ovl{v}{1}{n}) \leq \eta_{\ldis}(t\spp{x}{u}{1}{m} \ovl{v}{1}{n})$,
 \item $\fop{\ovl{x}{1}{m}}$ does not change, and
 \item $\eta_{\ldis}(\ovl{u}{1}{j-1} u'_j \ovl{u}{j+1}{m}) < \eta_{\ldis}(\ovl{u}{1}{m})$.
 \end{enumerate}
We conclude by the \ih\  since $\ovl{u}{1}{j-1} u'_j \ovl{u}{j+1}{m} \in \SN{\ldis}$ and  our hypothesis $t\sp{x}{u}{1}{m}\ovl{v}{1}{n}\in \SN{\ldis}$ is equal or reduces to $t\spp{x}{u}{1}{j-1}\isubs{x_j/u'_j}\spp{x}{u}{j+1}{m}\ovl{v}{1}{n} \in \SN{\ldis}$ (depending on $|t|_{x_j}$).

\item $t_0  \Rew{\ldis} t' \espp{x}{u}{1}{m} \ovl{v}{1}{n}= t'_0$ with $t \Rew{\ldis}
  t'$. Then we have that:
$$\eta_{\ldis}(t'\spp{x}{u}{1}{m} \ovl{v}{1}{n}) < 
\eta_{\ldis}(t\spp{x}{u}{1}{m} \ovl{v}{1}{n})$$
We conclude by the \ih\ since 
  $t'\spp{x}{u}{1}{m} \ovl{v}{1}{n} \in \SN{\ldis}$.

\item $t_0  \Rew{\ldis} t \espp{x}{u}{1}{m} v_1 \ldots  v'_i \ldots  v_n= t'_0$ with $v_i \Rew{\ldis}
  v'_i$.
 Then we have that:
$$\eta_{\ldis}(t\spp{x}{u}{1}{m} v_1 \ldots  v'_i \ldots  v_n) < 
\eta_{\ldis}(t\spp{x}{u}{1}{m} \ovl{v}{1}{n})$$
We conclude by the \ih\ since 
  $t\spp{x}{u}{1}{m} v_1 \ldots  v'_i \ldots  v_n \in \SN{\ldis}$.

\item $t_0 \Rew{\Gc} t\espp{x}{u}{1}{j-1} \espp{x}{u}{j+1}{m} \ovl{v}{1}{n}= t'_0$, with $|t|_{x_j}= 0$. Then we have that:
$$\eta_{\ldis}(t\spp{x}{u}{1}{j-1} \spp{x}{u}{j+1}{m} \ovl{v}{1}{n} )=
\eta_{\ldis}(t\spp{x}{u}{1}{m} \ovl{v}{1}{n})$$
But $\fo{t}{\ovl{x}{1}{j-1}\ovl{x}{j+1}{m}} <  \fo{t}{\ovl{x}{1}{m}}$.
We conclude by the \ih\ since 
$t\spp{x}{u}{1}{j-1}\spp{x}{u}{j+1}{m} \ovl{v}{1}{n} =  t\spp{x}{u}{1}{m} \ovl{v}{1}{n} \in \SN{\ldis}$
by hypothesis.

\item $t_0 \Rew{\Var} t  \espp{x}{u}{1}{j-1} \isubs{x_j/u_j} \espp{x}{u}{j+1}{m} \ovl{v}{1}{n} = t'_0$
with $|t|_{x_j} = 1$. Note that $t'_0= t \isubs{x_j/u_j}  \espp{x}{u}{1}{j-1} \espp{x}{u}{j+1}{m} \ovl{v}{1}{n}$.  Then we get:
$$\eta_{\ldis}(t\isubs{x_j/u_j}\spp{x}{u}{1}{j-1}\spp{x}{u}{j+1}{m} \ovl{v}{1}{n}) = 
\eta_{\ldis}(t\spp{x}{u}{1}{m} \ovl{v}{1}{n})$$
Since the \jump s
are independent, then $(\ovl{x}{1}{j-1} \cup  \ovl{x}{j+1}{m}) \cap \fv{u_j} = \ems$ implies
$\fo{t\isubs{x_j/u_j}}{\ovl{x}{1}{j-1}\ovl{x}{j+1}{m}} < \fo{t}{\ovl{x}{1}{m}}$.
   We conclude since $t\spp{x}{u}{1}{j-1}\isubs{x_j/u_j}\spp{x}{u}{j+1}{m} \ovl{v}{1}{n}=
t\spp{x}{u}{1}{m} \ovl{v}{1}{n} \in \SN{\ldis}$ by hypothesis.

\item $t_0\Rew{\DSubs} t_{[y]_{x_j}}\espp{x}{u}{1}{j-1} [x_j/u_j][y/u_j] \espp{x}{u}{j+1}{m}  \ovl{v}{1}{n} = t'_0$
with $|t|_{x_j} \geq 2$ and $y$ fresh. Then, \\ 
 $\eta_{\ldis}(t_{[y]_{x_j}}\spp{x}{u}{1}{j-1} \isubs{x_j/u_j}\isubs{y/u_j}\spp{x}{u}{j+1}{m} \ovl{v}{1}{n}) = 
\eta_{\ldis}(t\spp{x}{u}{1}{m} \ovl{v}{1}{n}))$ and\\
$\fo{t_{[y]_{x_j}}}{\ovl{x}{1}{j-1} x_j y\ovl{x}{j+1}{m}} < \fo{t}{\ovl{x}{1}{m}}$. 
In order to apply the \ih\ to $t_{[y]_{x_j}}$  we need:  
\begin{enumerate}[$-$]
\item $\ovl{u}{1}{j-1}, u_j, u_j,  \ovl{u}{j+1}{m} \in \SN{\ldis}$. This holds by hypothesis.
\item $t_{[y]_{x_1}}\spp{x}{u}{1}{j-1}  \isubs{x_j/u_j}\isubs{y/u_j}\spp{x}{u}{j+1}{m} \ovl{v}{1}{n} \in \SN{\ldis}$. 
This holds since the term is equal to $t\spp{x}{u}{1}{m} \ovl{v}{1}{n}$ which is $\SN{\ldis}$ by hypothesis.
\end{enumerate}

Note that this is the case that forces the
use of a  generalised sequence of substitutions: if we were proving the
statement for $t [x/u] \ovl{v}{1}{n}$ using as hypothesis $u \in
\SN{\ldis}\ \&\ t\isubs{x/u} \ovl{v}{1}{n} \in \SN{\ldis}$ then there
would be no way to use the \ih\ to get $t_{[y]_{x}}
[x/u][y/u] \ovl{v}{1}{n}\in \SN{\ldis}$.

\item $t_0 =  (\lam x. t')\espp{x}{u}{1}{m} v_1\ovl{v}{2}{n} \Rew{\B}
     t'[x/v_1]\espp{x}{u}{1}{m} \ovl{v}{2}{n} =t'_0$. We
     have that:
     $$u_0= (\lam  x. t')\spp{x}{u}{1}{m} v_1\ovl{v}{2}{n} \in \SN{\ldis}$$
     holds by hypothesis. Then:
\[ 
u_0 \Rew{\B}  t'\spp{x}{u}{1}{m}[x/v_1] \ovl{v}{2}{n} = t'[x/v_1]\spp{x}{u}{1}{m} \ovl{v}{2}{n}=u'_0 \]
Thus $\eta_{\ldis}(u'_0) < \eta_{\ldis}(u_0)$ and $u'_0\in
\SN{\ldis}$. Since $ \ovl{u}{1}{m} \in \SN{\ldis}$ by hypothesis we
can apply the \ih\ and get $t'_0 \in \SN{\ldis}$.
\qed\medskip
\end{enumerate}

\noindent The following is a consequence of Thereom~\ref{t:ieg}: just take
the number of substitutions $m$ to be $1$ and consider only
the $\giep$ property for $\termslambda  \subset \terms$.

\begin{cor}[\iep\ for $\ldis$]
\label{t:ieldis}
The $\ldis$-calculus enjoys the \iep\ property.
\end{cor}

Corollary~\ref{t:ieldis}, then Lemma~\ref{l:adequacy}
and finally  Theorem~\ref{t:ie-implies-psn} imply:

\begin{cor}[PSN for $\ldis$]
\label{coro:struct-psn-lj}
The $\lj$-calculus enjoys PSN, \ie\ if $t \in \termslambda \cap \SN{\beta}$, then $t \in \SN{\ldis}$.
\end{cor}

Note that Lemma \ref{l:adequacy} and Theorem \ref{t:ieg}, which
contains the arguments for PSN, do not use full composition, nor
termination of $\Rew{\dis}$, confluence or postponement of erasures:
none of the properties of $\lj$ plays a role in this compact proof of
PSN, which is quite surprising. The crucial point is the formulation
at a distance of the rewriting rules. Indeed, we will later show that
such a simple proof does not longer work when rules propagating jumps
are added to the system.

\section{An equational theory for $\ldis$}
\label{s:eq-th}
Sections~\ref{s:structural-lj} and~\ref{s:lj-psn} show  that the basic
theory of $\ldis$ enjoys good properties such as full composition,
confluence and PSN. In most calculi with explicit
substitutions, where substitutions are propagated through constructors
and do not act at a distance, full composition can only be obtained by
adding an equivalence relation $\eqcs$ defined as the contextual and
reflexive-transitive closure of the following equation:
\[ \begin{array}{lll@{\hspace{.5cm}}l}
   t[x/s][y/v] & \sim_{\CS} & t[y/v][x/s] & \mbox{ if } x\notin\fv{v}\ \&\ y\notin\fv{s}  \\
   \end{array} \] 
Otherwise  a term like $x[y/z][x/w]$ cannot reduce to its implicit form 
$w[y/z] = x[y/z]\isubs{x/w}$
(and so full
composition does not hold). Interestingly, $\ldis$ enjoys full
composition without using equation $\CS$, which is remarkable since 
plain
rewriting is much easier than rewriting
modulo an equivalence relation.\medskip

However, as mentioned at the end of
Section~\ref{ss:struct-confluence}, the equation $\CS$ is necessary to
recover confluence on metaterms.  It is then natural to wonder what
happens when $\eqcs$ is added to $\lj$. The answer is extremely
positive since $\eqcs$ preserves all the good properties of $\ldis$,
and this holds in a very strong sense. In fact, $\eqcs$ is a strong
bisimulation for $(\terms,\Rew{\lj})$ (\cf\ Lemma~\ref{l:eqo-bisim}),
so that $\eqcs$ can be postponed w.r.t. $\Rew{\lj}$
(\cf\ Lemma~\ref{l:bisim-conf})  and $\ldis$ modulo $\eqcs$ enjoys
  PSN
  (\cf\ Lemma~\ref{l:lamj-bis-psn-dif-sys}:\ref{l:sbisim-pres-psn}).

As already mentioned in the introduction, 
 $\lj$-terms and
$\lj$-dags~\cite{AccattoliTh} are strongly bisimilar, 
but the translation of $\lj$-terms to $\lj$-dags is not
injective, \ie\ there are different $\lj$-terms 
which are mapped to the \textit{same} $\lj$-dag. 
It is then interesting to characterise  the quotient induced by the
translation~\cite{AccattoliTh}, which turns out to be $\eqcs$: indeed $t\eqcs u$ if and only if $t$ and
$u$ are mapped to the same $\lj$-dag $G$, and since they both behave
like $G$ (\ie\  are strongly bisimilar to $G$), then  they behave the same
(\ie\  they are strongly bisimilar).\medskip

The $\lj$-calculus is also 
interesting since it can be mapped
 to another graphical language, Danos' and Regnier's \deft{Pure
  Proof-Nets}, being able to capture
\textit{untyped} $\lam$-calculus. It is possible to endow Pure Proof-Nets
with an operational semantics\footnote{Danos' and Regnier's original
  operational semantics does not match exactly $\lj$ because they use
  a big-steps rule for eliminating exponential cuts, which corresponds
  to use just one substitution rule
  $t[x/u]\Rew{}t\isubs{x/u}$. However, the refinement of Pure
  Proof-Nets where duplications are done small-steps is very natural
  from an explicit substitution point of view, altough --- to our
  knowledge --- it has never been considered before.} which makes them
strongly bisimilar to $\lj$. The quotient
induced by the translation  from $\lj$-terms into Pure Proof-Nets is
given by the \deft{graphical equivalence} $\eqo$ 
which is the contextual
and reflexive-transitive closure of the 
equations in Figure~\ref{f:eqo}.

\begin{figure}[ht]
\[ \begin{array}{lll@{\hspace{.5cm}}l}
   t[x/s][y/v] & \sim_{\CS} & t[y/v][x/s] & \mbox{ if } x\notin\fv{v}\ \&\ y\notin\fv{s}  \\
   \lam y. t [x/s] & \preeqsigu & (\lam y. t) [x/s]  & \mbox{ if } y\notin \fv{s} \\
   t[x/s] v & \preeqsigt & (t v)[x/s]& \mbox{ if } x\notin\fv{v} \\
   \end{array} \]
\caption{The graphical  equivalence $\eqo$} 
\label{f:eqo}
\end{figure}

This means that Pure Proof-Nets quotient more than
$\lj$-dags\footnote{$\lj$-dags can be mapped on Pure Proof-Nets, and
  once again the map is a strong bisimulation.}. As
for $\equiv_{\CS}$, $\eqo$ is a strong bisimulation
(\cf\ Lemma~\ref{l:eqo-bisim}), and thus confluence and PSN
of $\ldis$ 
automatically lift to $\Rew{\modulo{\ldis}{\osym}}$
(\cf\ Theorem~\ref{thm:eqo-conf-psn}), which is the reduction relation
$\lj$ modulo $\eqo$.\medskip

Another way to explain the  $\osym$-equivalence 
is by  means of \textit{linear} constructors. Indeed, 
the body of an abstraction cannot be duplicated nor erased by the
abstraction itself---in this sense an abstraction is \textit{linear}
in its body. Similarly, explicit substitutions are linear with
respect to their left subterm, while they are \textit{non-linear} with
respect to their right subterm, \ie\ the content of the jump, which may be
duplicated or discarded. Applications are linear in their left
subterm but they are non-linear in their argument, 
because they can wrap
it in a jump. This linear/non-linear classification reflects the fact
that jumps and arguments (and only them) are associated
to $!$-boxes in Proof-Nets, the non-linear construction of Linear Logic.  The
equations defining $\eqo$ can be understood as a permutation between a jump 
and a linear subterm of the
adjacent constructor.

It is then natural to wonder if $\eqo$ can be extended with
equations permuting jumps with non-linear subterms (see Figure
\ref{f:boite}, page \pageref{f:boite}), without breaking confluence
and PSN. The answer is yes; the obtained equational
theory is called the \textbf{substitution equivalence} $\eqf$, and the
fact that $\ldis$ modulo $\eqf$ enjoys PSN is the main result of this
paper.\medskip

Extending $\eqo$ to non-linear permutations
is delicate from a termination point of view, since the use of
non-linear equations affects reduction lengths. Indeed, the natural
but na\"{\i}ve extension of $\eqo$ breaks PSN. By analyzing a
counter-example to PSN we define $\eqf$ so that PSN turns out to be true. The
proof of this fact, however, is more involved than that for $\eqcs$ and
$\eqo$, mainly because $\eqf$ is not a strong bisimulation.
Therefore, we shall develop a new technique for proving
PSN modulo $\eqf$.\medskip

Section~\ref{s:regnier} starts over by explaining
the equivalence $\eqo$ in terms of Regnier's $\sigma$-equivalence~\cite{regnier94}, 
providing a different point of view with respect to what was already
mentioned. Section \ref{s:propp-intro} discusses 
how to extend $\eqo$ to $\eqf$ by
    showing the difficulties to prove PSN for  the obtained extension.  Section~\ref{sec:psn} develops the proof of PSN for $\ldis$ modulo $\eqf$.

\subsection{The graphical equivalence}
\label{s:regnier}

\deft{Regnier's equivalence} $\equiv_{\rsig}$ is the smallest
equivalence on $\lam$-terms closed by contexts and containing the 
equations in Figure~\ref{f:regnier}. 
\begin{figure}[ht]
\[ \begin{array}{lll@{\hspace{.5cm}}l}
(\lam x. \lam y.t) u & \preeqw{\rsig_1} & \lam y. ((\lam x.t) u) & \mbox{if } y\notin\fv{u}\\
(\lam x.t v) u     & \preeqw{\rsig_2} & (\lam x. t) u v    & \mbox{if } x\notin \fv{v}\\
\end{array}\]
\caption{The  equivalence $\equiv_{\rsig}$ }
\label{f:regnier}
\end{figure}

\noindent Regnier proved that two $\rsig$-equivalent terms have
essentially the same {\it operational} behavior: $\eqw{\rsig}$ is contained
in the equational theory generated by $\beta$-reduction, \ie\ 
$\eqw{\rsig}\subset\eqw{\beta}$, and if $t\eqw{\rsig} t'$ then 
the maximal $\beta$-reduction
sequences from $t$ and $t'$ have the same length (the
so-called \textit{Barendregt's norm}). That is why Regnier calls $\eqw{\rsig}$ an \textit{operational
  equivalence}. \medskip

It is then natural to expect that 
the previous property can be \textit{locally} reformulated in terms of a
strong
bisimulation, namely, 
 \[ \begin{array}{cccc@{\hspace{1cm}}c@{\hspace{1cm}}ccccc}
  t				& \Rew{\beta}	& u	&&				&&  t			& \Rew{\beta}	& u	\\
  \eqw{\rsig}	&		&	&& \mbox{ implies }	&&\eqw{\rsig}	&		&\eqw{\rsig}	\\
  t'				&		&	&&				&&t'				&\Rew{\beta}	& u' \\
 \end{array}
\]

Unfortunately, this is not
the case. Consider the following example, where 
grey boxes are used to help the identification of redexes and their reductions: 
\[ \begin{array}{cccccccccc}
t&=& \grisar{\lam y. (\grisarOscuro{(\lam x.y) z_1}) z_2} & \grisar{\Rew{\Beta}} &
     \grisarOscuro{(\lam x.z_2) z_1} & =  & u \\
\\
&&\eqw{\rsig_1}&&\not \eqw{\rsig}
\\\\
t'&=&(\grisarOscuro{(\lam x. (\lam y. y)) z_1}) z_2 & \grisarOscuro{\Rew{\Beta}} & (\lam y. y) z_2 & = & u'
\end{array}\]

The
term $t'$ has only one redex whose reduction gives $u'$
which is not $\eqw{\rsig}$-equivalent to $u$,  the reduct of $t$. The
diagram can be completed only by unfolding the whole reduction:
\[ \begin{array}{cccccccc}
t&=&\grisar{\lam y. (\grisarOscuro{(\lam x.y) z_1}) z_2}  & \grisar{\Rew{\Beta}} &
\grisarOscuro{(\lam x.z_2) z_1} &\grisarOscuro{\Rew{\beta}}& z_2\\
\\
&&\eqw{\rsig_1}&&&&= (\subseteq \eqw{\rsig})
\\\\
t'&=&(\grisarOscuro{(\lam x. (\lam y. y)) z_1}) z_2   & \grisarOscuro{\Rew{\Beta}} & (\lam y. y) z_2&\Rew{\beta}& z_2
\end{array}\]

Note that the second step from $t'$ reduces a {\it created} redex. \medskip

We are now going to analyze $\eqw{\rsig}$ 
in the framework of $\ldis$.  For that, Regnier's equivalence
can be understood on $\ldis$-terms by first removing the $\B$-redexes. Indeed,
let us take the clauses defining $\eqw{\rsig}$ and let us make a 
$\B$-reduction step on both sides, thus eliminating the multiplicative
redexes as in Regnier's definition:
\[ \begin{array}{ccc@{\sep\sep\sep\sep}ccc}
 \grisar{(\lam x. \lam y. t) u} & \preeqw{\rsig_1}&  \lam y. (\grisar{(\lam x.t) u})
& (\grisar{(\lam x. t) u}) v& \preeqw{\rsig_2} & \grisar{(\lam x. (t v)) u}\\ \\
\downarrow_\B & &  	\downarrow_\B &
\downarrow_\B & &  	\downarrow_\B \\ \\
 \grisar{(\lam y. t)[x/u]}	&  &\lam y. (\grisar{t [x/u]}) &
\grisar{t[x/u]} v & & \grisar{(t v)[x/u]}
\end{array}\] 

Now,  $\eqw{\rsig}$ can be seen
as a change of the positions of jumps in a given term and particularly
as a permutation equivalence of jumps concerning the linear constructors of the
calculus.

This is not so surprising since such permutations turn into simple
equalities when one extends the standard translation of $\lam$-calculus
into Linear Logic Proof-Nets to $\ldis$-terms (see for
example~\cite{KL07}). Another interesting observation is the relationship between $\eqw{\rsig}$ and
the equivalence $\equiv_{\CS}$ introduced in
Section~\ref{ss:struct-confluence}. To understand this point we
proceed the other way around by expanding jumps into $\beta$-redexes:
\[ \begin{array}{cccccccc}
\grisar{\grisarOscuro{t[y/v]}[x/u]} & \equiv_{\CS} & \grisarOscuro{\grisar{t[x/u]}[y/v]}\\ \\
 	\grisarOscuro{\uparrow_\B} & &  	\grisarOscuro{\uparrow_\B}\\ \\
	(\grisar{\grisarOscuro{(\lam y. t) v}) [x/u]} & & 
        \grisarOscuro{(\lam y.\grisar{t [x/u]}) v}\\ \\
 	\grisar{\uparrow_\B} & &  	\grisar{\uparrow_\B}\\ \\
	\grisar{(\lam x. (\grisarOscuro{(\lam y. t) v})) u} & & \grisarOscuro{(\lam y. (\grisar{(\lam x. t) u})) v}\\ \\
\end{array}\] 
Note that 
the relation between the
resulting terms is
contained in $\eqw{\rsig}$, that is why it was not visible
in $\lam$-calculus:
\[ \begin{array}{cccccccc}
(\lam x. ((\lam y. t) v)) u & \preeqw{\rsig_2} & (\lam x.\lam y. t) u v& \preeqw{\rsig_1} & (\lam y. ((\lam x. t) u)) v
\end{array}\] 

In~\cite{AccattoliTh} it has been proved that two $\lj$-terms $t$
  and $t'$ are translated to the same Pure Proof-Net if and only if
  $t \equiv_{\sig, \CS}\  t'$. More precisely, this relation
can be given by the \deft{graphical equivalence}
$\eqo$ already defined  in Figure~\ref{f:eqo}. 
\ignore{
\begin{figure}[ht]
\[ \begin{array}{lll@{\hspace{.5cm}}l}
   t[x/s][y/v] & \sim_{\CS} & t[y/v][x/s] & \mbox{ if } x\notin\fv{v}\ \&\ y\notin\fv{s}  \\
   \lam y. (t [x/s]) & \preeqsigu & (\lam y. t) [x/s]  & \mbox{ if } y\notin \fv{s} \\
   t[x/s] v & \preeqsigt & (t v)[x/s]& \mbox{ if } x\notin\fv{v} \\
   \end{array} \] 
\caption{The $\eqo$-equivalence relation}
\label{f:eqo}
\end{figure}
}

The equations defining $\eqo$ are specified by 
means of \textit{local}
permutations, but it is 
also possible to define
$\eqo$ in terms of global permutations. First, define 
a \deft{spine context} $S$ as:
\[ S  ::= \hole  \mid \lam x. S \mid S t \mid S[x/t]\]
and then define $\eqo$ as the context closure of the following equation $\sim_\osym$:
\[\ctx{S}{t[x/u]}\sim_\osym \ctx{S}{t}[x/u] \ \ \ \ \  \mbox{if $\bs{S}\cap \fv{u}=\emptyset$ and $|t|_x=|\ctx{S}{t}|_x$}\]

The two definitions are easily seen to be equivalent.  We shall now
prove that $\eqo$ is a strong bisimulation, which will immediately
imply (Lemma~\ref{l:lamj-bis-psn-dif-sys}) that $\eqo$ preserves
reduction lengths. This property is stronger than the one proved by
Regnier for $\eqw{\rsig}$, since it holds for any reduction sequence,
not only for the maximal ones.  \medskip

\begin{lem}
\label{l:eqo-stability}
Let $\ttE$ be the equivalence relation $\CS$ or $\osymb$, and $t,t'\in\terms$ s.t. $t \eqw{\ttE} t'$. Let $u \in \terms$. Then:
\begin{enumerate}[\rm(1)]
  \item \label{l:eqo-stability-minus} $|t|_x  = |t'|_x$.
  \item \label{l:eqo-stability-zero} For all $S \subseteq \pos{x}{t}$ 
  there is 
  $S' \subseteq \pos{x}{t'}$ s.t.
  $|S| = |S'|$ and  
  $\ren{t}{S}{x}{y}  \eqw{\ttE} \ren{t'}{S'}{x}{y}$. 
  \item \label{l:eqo-stability-two}  $t\isubs{x/u} \eqw{\ttE} t'\isubs{x/u}$.
  \item \label{l:eqo-stability-one}  $u\isubs{x/t} \eqw{\ttE} u\isubs{x/t'}$.
  \end{enumerate}      
 \end{lem}

\begin{proof}
 Straightforward inductions.
\end{proof}

\begin{lem}
\label{l:eqo-bisim}
The relations $\eqcs$ and $\eqo$ are   strong bisimulations for $\ldis$. 
\end{lem}

\proof
We prove the statement for $\eqo$. The proof for $\eqcs$ is obtained by 
simply forgetting the cases $\set{ \sim_{\sig_1}, \sim_{\sig_2}}$.
Assume $t_0 \eqo t_1$ holds in $n$-steps, which is written as
$t_0\eqo^n t_1$, and let $t_1\Rew{\ldis} s_1$. We show $\exists\ s_0$
s.t.  $t_0\Rew{\ldis} s_0 \eqo s_1$  by induction on
$n$.  

The inductive step for $n> 1$ is straightforward. For 
 $n=1$ we reason by induction on the definition of
$t_0\eqo^1 t_1$, given by the closure under contexts of the equations
$\set{\sim_{\CS}, \sim_{\sig_1}, \sim_{\sig_2}}$.

We only show here the cases where $t_0 \eqo t_1$ is contextual, all the other ones being straightforward. 
\begin{enumerate}[$\bullet$]
\item If $t_0=t [x/u]\eqo t' [x/u]=t_1\Rew{\ldis} t' [x/u']=s_1 $,
where $t \eqo t'$ and $u \Rew{\ldis} u'$,  then we close the diagram by $t_0  \Rew{\ldis} t [x/u'] \eqo s_1$. 
\item The case $t_0=t [x/u]\eqo t [x/u']=t_1\Rew{\ldis} t' [x/u']=s_1 $,
where $u \eqo u'$ and $t \Rew{\ldis} t'$ is analogous to the previous  one.
\item If $t_0=t [x/u]\eqo t' [x/u]=t_1\Rew{\ldis} t'' [x/u]=s_1 $, 
where $t \eqo t' \Rew{\ldis} t''$, then by the \ih\ $\exists\ t'''$ s.t. $t\Rew{\ldis}t''' \eqo t''$. 
We close the diagram by  $t_0\Rew{\ldis}t''' [x/u]  \eqo^* s_1$.
\item The case $t_0=t [x/u]\eqo t [x/u']=t_1\Rew{\ldis} t [x/u'']=s_1 $,
where $u \eqo u' \Rew{\ldis} u''$ is analogous to the previous one.
\item If $t_0=t [x/u]\eqo t [x/u']=t_1\Rew{\Gc} t=s_1$, where
$u \eqo u'$ and $|t|_x=0$,  then $t_0\Rew{\Gc} t=s_1$.
\item If $t_0=t [x/u]\eqo t' [x/u]=t_1\Rew{\Gc} t=s_1$, where 
      $t \eqo t'$ and $|t'|_x=0$,  then the previous remark implies
      $|t|_x =0$ and we close the diagram by  $t_0\Rew{\Gc} t\eqo t'=s_1$.
\item If $t_0=t [x/u]\eqo t [x/u']=t_1\Rew{\DSubs}
  t_{[y]_x}[x/u'][y/u']=s_1$, where $u \eqo u'$ and $|t|_x >1$, then we
  close the diagram by $t_0\Rew{\DSubs} t=t_{[y]_x}[x/u][y/u] \eqo^2
  t_{[y]_x}[x/u'][y/u']$.
\item  If $t_0=t [x/u]\eqo t' [x/u]=t_1\Rew{\DSubs}
  t'_{[y]_x}[x/u][y/u]=s_1$, where $t \eqo t'$ and $|t'|_x >1$, then
  we first write $t'_{[y]_x}$ as $\ren{t'}{S'}{x}{y}$, where
  $S' \subset \pos{t'}{x}$ and $|S'| \geq 2$.
  Lemma~\ref{l:eqo-stability}:\ref{l:eqo-stability-minus}
  gives
  $|t|_x >1$ and Lemma~\ref{l:eqo-stability}:\ref{l:eqo-stability-zero}
  gives $S \subset \pos{t}{x}$ verifying $|S| = |S'|$ and 
  $\ren{t'}{S'}{x}{y} \eqo \ren{t}{S}{x}{y}$. 
  Then, we close the diagram with
  $t_0\Rew{\DSubs} \ren{t}{S}{x}{y}[x/u][y/u]\eqo
  t'_{[y]_x}[x/u][y/u]$.
\item If $t_0=t [x/u]\eqo t [x/u']=t_1\Rew{\Var}
  t\isubs{x/u'}=s_1$, where $u \eqo u'$ and $|t|_x=1$, then  $t [x/u]\Rew{\Var}
  t\isubs{x/u}\eqo t\isubs{x/u'}$,  where the last equivalence 
  holds by Lemma~\ref{l:eqo-stability}:\ref{l:eqo-stability-one}.
\item If $t_0=t[x/u]\eqo t'[x/u]=t_1\Rew{\Var} t'\isubs{x/u}=s_1$,
  where $t \eqo t'$ and $|t'|_x =1$. Then, $t [x/u]\Rew{\Var}
  t\isubs{x/u}\eqo t'\isubs{x/u}$ where the last equivalence holds by
  Lemma~\ref{l:eqo-stability}:\ref{l:eqo-stability-minus}-\ref{l:eqo-stability-two}.
\qed\medskip
\end{enumerate}

\noindent A consequence (\cf\ Lemma~\ref{l:bisim-conf}) of the previous lemma is that
both $\eqcs$ and  $\eqo$ can be postponed, which implies in particular the following.

\begin{thm}
\label{thm:eqo-conf-psn}
The reduction systems $(\terms, \Rew{\ldis/\CS})$ 
and $(\terms, \Rew{\ldiso})$ are both confluent and enjoy PSN.
\end{thm}

\begin{proof}
Confluence follows from Lemma~\ref{l:eqo-bisim} and
Theorem~\ref{t:confluence} by application of
Lemma~\ref{l:bisim-conf}:\ref{l:bisim-conf-ii}, while PSN follows from
Lemma~\ref{l:eqo-bisim} and Corollary~\ref{coro:struct-psn-lj} by
application of Lemma \ref{l:lamj-bis-psn-dif-sys}.
\end{proof}

Actually, $\Rew{\ldiso}$ is equal to $\eqo\Rew{\ldis}\eqo$. In
  the framework of rewriting modulo an equivalence relation there are
  various, non-equivalent, forms of confluence. The one given by Theorem \ref{thm:eqo-conf-psn} is
  the weakest one, but the 
  \textit{Church-Rosser modulo} property also holds in our framework. 

\begin{thm}[Church-Rosser modulo $\CS$ and $\osymb$]
\label{cr-modulo}
Let $\ttE$ be the equivalence relation $\CS$ or $\osymb$. 
If $t_0 \eqttE t_1$, $t_0 \Rewn{\ldisttE} u_0$ and
$t_1 \Rewn{\ldisttE} u_1$, then $\exists v_0, v_1$ s.t.
$u_0 \Rewn{\ldisttE} v_0$, $u_1 \Rewn{\ldisttE} v_1$
and $ v_0 \eqttE v_1$.
\end{thm}

\begin{proof}
By  Lemma~\ref{l:eqo-bisim} and Lemma~\ref{l:bisim-conf}. 
\end{proof}

We finish this section with the following interesting property.

\begin{lem}
\label{l:dis-o}
The reduction relation $\modulo{\dis}{\osym}$ is strongly normalizing.
\end{lem}

\begin{proof} The proof uses the measure $\dm{}$ used to prove
Lemma~\ref{l:dis-terminates} and the fact that 
$t \equiv_{\osym} t'$ implies $\dm{t} = \dm{t'}$.
\end{proof}

\subsection{The  substitution equivalence}
\label{s:propp-intro}
Composition of explicit substitutions is a sensible topic in the
literature, it is interesting to know if
$\lj$ can be extended with a safe notion of (structural) composition.

The structural $\lam$-calculus is peculiar 
since composition of substitution is 
provided natively, but only \textit{implicitly} and at a distance.  Indeed, a term
$t[x/u][y/v]$ s.t.  $y\in \fv{u}\ \&\ y\in \fv{t}$  reduces in various
steps to:
$$t[x/u\isubs{y/v}][y/v]$$ but not to the \textit{explicit
  composition} $t[x/u[y/v]][y/v]$. One of the aims of this paper is to
prove that adding explicit composition to $\ldis$ preserves PSN and confluence.

The second aim concerns \textit{explicit
  decomposition}. Indeed, some calculi~\cite{OH06,MaraistOTW99,Schw99,HZ09,Hasegawa} explicitly
\textit{decompose} substitutions, \ie\ reduce $t[x/u[y/v]]$ to
$t[x/u][y/v]$. We show that PSN and confluence hold even when extending
$\lj$ with such a rule.\medskip

More generally, having a core system, $\lj$, whose operational
semantics does not depend on propagations, we study how to modularly add
propagations  by  keeping the good properties. We have
already shown that $\lj$ is stable with respect to 
the graphical equivalence, which can be seen as handling propagations
of jumps with respect to \textit{linear} constructors. We proved that $\ldiso$ is confluent and
enjoys PSN (Theorem~\ref{thm:eqo-conf-psn}). What we investigate here is if we
can extend it to propagations with respect to \textit{non-linear}
constructors.\medskip

The idea is to extend $\eqo$ to $\eqfz$, where $\eqfz$
is the the contextual and reflexive-transitive closure
of the relation
generated by $\set{\CS, \sig_1, \sig_2}$ plus:
\[ \begin{array}{llll@{\hspace{.5cm}}l}
(t v)[x/u]     & \preeqw{\sigt^0} & t v[x/u]    & \mbox{if } x \notin \fv{t}\\
t[y/v][x/u]    & \preeqw{\sigq^0} & t[y/v[x/u]] & \mbox{if } x\notin \fv{t}\\
\end{array}\]
In terms of global permutations $\eqfz$ can be defined as the
  context closure of $\ctx{C}{t[x/u]}\sim_\fz\ctx{C}{t}[x/u]$ where
  $|t|_x=|\ctx{C}{t}|_x$, and $C$ is \textit{any}
  context (not just a spine context) which does not capture the
  variables of $u$.  These equations are constructor preserving (same
kind and number of constructors), in contrast to more traditional
explicit substitution calculi containing for instance the following 
rule:
\[ \begin{array}{cccc}
(t u)[y/v]& \Rew{@} & t[y/v] u[y/v]
\end{array}\]
which achieves two actions at the same time: 
duplication and propagation of a jump. In $\ldisfz$ there is a
neat separation between propagations and duplications, so that no
propagation affects the number of constructors.  The rule $\Rew{@}$
can be simulated in $\ldisfz$ only in the very special case where $t$
and $u$ both have occurrences of $y$. In our opinion this is not a
limitation: the rule $\Rew{@}$ is particularly inefficient since it
duplicates even if there is no occurrence of $y$ at all, thus it is
rather a good sign that $\ldisfz$ cannot simulate $\Rew{@}$.\medskip

The reduction relation $\ldisfz$ does not enjoy PSN, since it is a bit na\"ive
on the way it handles void substitutions. The following counter-example has been found by Stefano
Guerrini. Let  $u=(z z) [z/y]$, then:
\[ \begin{array}{lllllllll}
   t&=&u [x/u] = (z z) [z/y][x/u] & \equiv_{\sigq^0}  & 
   (z z) [z/y[x/u]] & \Rew{\DSubs} \\\\
   &&(z_1 z_2)  [z_1/y[x/u]]  [z_2/y[x/u]] & \Rewplus{\Var} & 
   y[x/u] (y [x/u]) & \equiv_{\sigma_2,\sigt^0, \alpha}  \\\\
   &&(y y) [x_1/u] [x/u] & \equiv_{\sigq^0} & (y y) [x_1/u[x/u]]
   \end{array} \]
The term $t$ reduces to a term containing $t$ and so there is a loop
of the form $t \Rewplus{} C_0[t] \Rewplus{} C_0[C_1[t]] \Rewplus{}
\ldots$.  Now, take $t_0=(\lam x.((\lam z. z z) y)) ((\lam z. z z) y)$,
which is strongly normalizing in the $\lam$-calculus.  Since $t_0$
$\ldisfz$-reduces to $t$, $t_0$ is not $\ldisfz$-strongly normalizing
and thus $\ldisfz$ does not enjoy PSN. It is worth to note that, in
contrast to \mellies\ counterexample for $\lsigma$~\cite{Mellies1995a}, the $\B$-rule has
no role in building the diverging reduction: the fault comes only from the
jump subsystem $\dis$ modulo $\eqfz$.\medskip

The key point of the previous counter-example is that the jump $[x/u]$
is free to float everywhere in the term since $x$ has no occurrence in
$t$. Such behavior can be avoided by imposing the constraint "$x\in
\fv{v}$" to $\sigt^0$ and $\sigq^0$. This has also a natural graphical
justification in terms of Pure Proof-Nets (\cite{AccattoliTh}, Chapter 6, page 149), since such constraint 
turns $\sigt^0$ and $\sigq^0$ into the exact analogous \ignore{form} of the
commutative box-box rule of Linear Logic Proof-Nets, but used here as
an equivalence relation.  We then modify
$\preeqw{\sigt^0}$ and $\preeqw{\sigq^0}$ by
  introducing the equivalence $\equiv_{\boite}$ as the contextual and
  reflexive-transitive closure of the  equations in Figure~\ref{f:boite}.
\begin{figure}[ht]
\[ \begin{array}{llll@{\hspace{.5cm}}l}
(t v)[x/u]     & \preeqw{\sigt} & t v[x/u]    & 
  \mbox{if } x \notin \fv{t}\ \&\  x\in \fv{v}\\
t[y/v][x/u]    & \preeqw{\sigq} & t[y/v[x/u]] & 
  \mbox{if } x\notin \fv{t}\ \&\ x\in \fv{v}\\
\end{array}\]
\caption{The equivalence $\equiv_{\boite}$}
\label{f:boite}
\end{figure}

\noindent Now, we redefine $\eqfz$ in the following way. The \deft{substitution
  equivalence} $\eqf$ is the smallest equivalence closed by contexts
containing all the equations in Figure~\ref{f:eqf}. 

\begin{figure}[ht]
\[ \begin{array}{llll}
t[x/s][y/v] & \sim_{\CS} & t[y/v][x/s] &  \mbox{if } x\notin\fv{v}\ \&\ y\notin\fv{s}  \\
   \lam y. (t [x/s]) & \preeqsigu & (\lam y. t) [x/s]  &  \mbox{if } y\notin \fv{s} \\
   t[x/s] v & \preeqsigt & (t v)[x/s]&  \mbox{if } x\notin\fv{v} \\
(t v)[x/u]     & \preeqw{\sigt} & t v[x/u]    & 
  \mbox{if } x \notin \fv{t}\ \&\ x\in \fv{v}\\
t[y/v][x/u]    & \preeqw{\sigq} & t[y/v[x/u]] & 
  \mbox{if } x\notin \fv{t}\ \&\  x\in \fv{v}\\
\end{array} \]
\caption{The substitution equivalence $\eqf$}
\label{f:eqf}
\end{figure}

\noindent Alternatively,
  $\eqf$ can be defined by the context closure of the following
  global permutating equations:
\[
\begin{array}{lllll}
\ctx{C}{t[x/u]}&\sim_{\fsymb}& \ctx{C}{t}[x/u] &\mbox{if $\bs{C}\cap \fv{u}=\emptyset$ and $|t|_x=|\ctx{H}{t}|_x>0$}\\
\ctx{S}{t[x/u]}&\sim_{\fsymb}& \ctx{S}{t}[x/u] &\mbox{if $\bs{S}\cap \fv{u}=\emptyset$ and $|t|_x=|\ctx{S}{t}|_x=0$}
\end{array}\]
where $C$ is any context and $S$ is a spine context.\medskip

It is now natural to study $\ldis$-reduction
modulo $\eqf$.  It is easy to prove that the
jump calculus terminates with respect to the
new equivalence $\eqf$ so that the previous counterexample to PSN is
ruled out. We need an auxiliary lemma about potential multiplicities and the $\jop$-measure. 

\begin{lem}
\label{l:mul-for-eq}
Let $t_0,t_1\in \terms$. If  $t_0 \eqf  t_1$ then:
\begin{enumerate}[\rm(1)]
\item $\mul{t_0}{z} = \mul{t_1}{z}$ for every variable $z$.
\item \label{p:mul-for-eq-two} $\dm{t_0} = \dm{t_1}$.
\end{enumerate}
\end{lem}

\begin{proof}
By induction on $\eqf$. The base cases are easy calculations, the
inductive cases use the \ih\ 
\end{proof}

Lemma~\ref{l:dm-decreases} and
Lemma~\ref{l:mul-for-eq}:\ref{p:mul-for-eq-two} together proves the
following corollary.

\begin{cor}
\label{l:dis-f}
The reduction relation $\modulo{\dis}{\fsymb}$ is terminating.
\end{cor}

\ignore{
For the na\"ive equations it is possible to show that $t \equiv_{\sigt^0} u$ implies
$\dm{t} = \dm{u}$, while  $t \equiv_{\sigq^0} t'$ does not  imply $\dm{t} = \dm{t'}$, as Lemma \ref{xxx} will show.
}

\section{Preservation of $\beta$-Strong Normalization for $\ldisf$}
\label{sec:psn}

The structural $\lam$-calculus modulo $\eqf$ is an incredibly subtle and complex
rewriting system, and proving PSN is 
is not an easy task. Some of the difficulties are:
  
\begin{enumerate}[$\bullet$]
\item \textit{The relation $\eqf$ is not a strong bisimulation}.  It
  is not difficult to see that $\ldis$ is confluent modulo $\eqf$
  (essentially the same proof than for $\ldis$). However, $\eqf$ does
  not preserve reduction lengths to normal form, \ie\  it is not a
  strong bisimulation. Two  examples can be given by analysing the
  interaction  between $\eqf$ with erasure and duplication. 
  Here is an example for erasure:  
\[ \begin{array}{cccccccccc}
z[x/y][y/u]& \Rew{\Gc}&z[y/u]\\
\eqw{\sigq}&&\downarrow_{\Gc}\\
z[x/y[y/u]]&\Rew{\Gc}& z
   \end{array} \]   
and here another  one for duplication:
\[ \begin{array}{cccccccccc}
(x x)[x/y][y/z]& \Rew{\DSubs}&(x x_1)[x/y][x_1/y][y/z]&\Rew{\DSubs}&(x x_1)[x/y_1][x_1/y_2][y_1/z][y_2/z]\\
\eqw{\sigq}&&&&\eqf\\
(x x)[x/y[y/z]]&\Rew{\DSubs}& (x x_1)[x/y[y/z]][x_1/y[y/z]]&\eqw{\alpha}&(x x_1)[x/y_1[y_1/z]][x_1/y_2[y_2/z]]
   \end{array} \]      
Indeed, if $\eqf$ would have been a strong bisimulation, then in both diagrams
the two terms of the second column would be $\eqf$-equivalent, while
they are not (remark that $\eqf$ preserves the number of constructors
so that those terms cannot be $\eqf$-equivalent).

\item \textit{The relation $\eqf$ cannot be postponed}. The last
  example shows also that $\eqf$ cannot be postponed. 
This is illustrated by the upper left corner of
the previous figure:
\[ \begin{array}{cccccccccc}
(x x)[x/y][y/z]& \Rew{\DSubs}&(x x_1)[x/y][x_1/y][y/z]\\
\eqw{\sigq}\\
(x x)[x/y[y/z]]
   \end{array} \]
Observe that this phenomenon is caused by
the equation $\sim_{\sigq}$. Remark that both
composition (\ie\ $\Rew{\sigq}$) and decomposition 
($\LRew{\sigq}$) are used in Guerrini's
counterexemple. 
\ignore{Surprisingly, from our study we learned that dealing
with decomposition is more complicate than dealing with composition. 
\item \textit{Erasures cannot be postponed}.  Consider 
\[ \begin{array}{lllllllll}
\lam x. (y [y/y'[z/x]])&\Rew{\Gc}& \lam x. (y [y/y'])&\preeqsigu&(\lam x. y )[y/y'] 
   \end{array} \]
the two steps cannot be permuted. This is a phenomenon
concerning $\ldiso$ too, except that $\eqo$ can be postponed
with respect to $\ldis$ (Lemma~\ref{l:eqo-bisim}),
and then $\Rew{\Gc}$ can be postponed
with respect to $\Rew{\nGc}$. Unfortunately, $\eqf$ cannot be delayed in $\ldisf$ and so 
neither $\Rew{\Gc}$ can.
}
\item \label{it:in-out-instability}\textit{There is no canonical representant of equivalence
  classes which is stable by reduction}. Indeed, there are two natural
  canonical representants in $\ldisf$. Given $t$ we can define ${\tt
    in}(t)$ as the term obtained by moving all substitution towards
  the variables as much as possible and ${\tt
    out}(t)$ the term obtained moving substitutions far from variables
  as much as possible. Consider 
$t=x[x/(\lam y. z[z/y]) x'] \Rew{\B} x[x/ z[z/y][y/ x']]=t'$,
then 
${\tt out}(t) = x[x/(\lam y. z[z/y]) x']$
does not reduce to ${\tt out}(t')= x[x/ z][z/y][y/ x']$.
Similarly, for the other representative
since ${\tt in}(t)=(x[y/z] z)[z/z']$
does not reduce to ${\tt in}(t')=x z[z/z']$.\medskip
\end{enumerate}

\noindent In~\cite{AK10} we proved that $\ldis$ enjoys PSN in the cases where
the equations $\set{\preeqw{\sigt}, \preeqw{\sigq}}$ are both oriented from left to
right or from right to left. Here we prove PSN considering them as
equivalences. Surprisingly, the proof of this more
  general result is relatively more compact
  and concise than the one(s) in~\cite{AK10}. Indeed, even if we need
  to pass through an auxiliary calculus, such a calculus can be proved
  to enjoy PSN without using labels, in contrast to our previous
  result and proof. \medskip

Let us explain our technique. Even if there is no canonical representative form of an
$\fsymb$-equivalence class which is stable by reduction
(\cf\ Section~\ref{s:propp-intro}), there is an even more natural way
to reason about PSN in the presence of the non-trivial 
equations
$\set{\sigt,\sigq}$ which consists in projecting
$\ldisf$ over a simpler equational calculus. Since both the calculus
and the projection are quite peculiar we introduce them
gradually.\medskip

A usual na\"{\i}ve idea consists in projecting $\ldisf$ 
into the $\lam$-calculus by means of a function computing the complete
unfolding of jumps. This gives the following diagram:
\begin{equation}
\begin{array}{ccccc}
t&\Rew{\ldis}&u\\
\downarrow^*_{\jop}&&\downarrow^*_{\jop}\\
\fc(t)&\Rewn{\beta}&\fc(u)
\end{array}
\label{eq:naive-proj}
\end{equation}
This principle could be easily exploited in order to prove some
properties of $\ldisf$ (such as confluence), however, this projection
erases divergent sub-terms, therefore it cannot be used to prove
PSN. For instance, consider $t=x[y/\Omega]$
(where $\Omega$ is a non-terminating term), which is only
$\ldis$-weakly normalizing, whereas $\fc(t)=x$ is in normal form. It
is easy to show that projection of terms without void jumps preserves
divergence and thus PSN.  Unfortunately, erasures cannot be postponed
in $\ldisf$.\medskip

Roughly speaking, the projection gives $\fc(t)\Rewn{\beta}\fc(u)$ so
that there are some steps $t\Rew{\ldis} u$ s.t. $\fc(t)=\fc(u)$.  It
is not really a problem if such (erased) steps are finite, but here
there may be infinite sequences of such (erased) steps. It is then quite natural to
change the complete unfolding $\fc$ into a \textit{non-erasing}
unfolding $\wfc$, which does not project void jumps:

\begin{equation}
 \begin{array}{lll}
   \wfc(x) & := & x\\
   \wfc(\lam x. u) & := & \lam x. \wfc(u)\\
   \wfc(u v) & := & \wfc(u) \wfc(v)\\
   \wfc(t[x/u]) & := & \left \{ \begin{array}{ll}
                        \wfc(t) \isubs{x/\wfc(u)} & \mbox{ if } x \in \fv{t} \\
                        \wfc(t) [x/\wfc(u)] & \mbox{ if } x \notin \fv{t} \\
                        \end{array} \right . 
   \end{array}  
\end{equation} 
Note that there are still some erased steps, as for instance
$t=x[x/y]\Rew{\Var} y=u$, where $\wfc(t) = y = \wfc(u)$, but intuition
tells that $\wfc$ preserves divergence, because diverging terms
  are no longer erased by the projection. Note also that the image of the projection of
the previous reduction step $t \Rew{\Var} u$ is no longer a reduction
step in the $\lam$-calculus, so that we need to specify which are the
rewriting rules and the equations of the image of the
translation.  \medskip

For didactive purpose let us assume that we are able to turn the image
of the projection into a calculus --- let us say $\lauxm$ --- such that
$\wfc$ projects $\ldisf$ into $\lauxm$ and preserves divergence. Two
important remarks are: since $\wfc(\cdot)$ preserves divergence, then
PSN for $\lauxm$ implies PSN for $\ldisf$; also, the $\lauxm$-calculus
does not contain the equations $\set{\sigt,\sigq}$ because
they were turned into equalities thanks to their side conditions.

It is then reasonable  to expect that proving PSN for $\lauxm$
 is easier
than PSN for $\ldisf$. Our proof technique can then be summarised as follows:
\begin{enumerate}[(1)]
  \item Introduce $\lauxm$;
  \item Prove PSN for $\lauxm$; 
  \item Show that $\wfc(\cdot)$ preserves divergence from $\ldisf$ to $\lauxm$;
  \item Conclude PSN for $\ldisf$.
\end{enumerate}

Section~\ref{s:lauxm} presents the rewriting rules of
$\lauxm$, thus completing point 1. Section~\ref{s:iep-for-lauxm} deals with point
2 and Section \ref{s:projection} with points 3 and 4.\medskip

We believe that the isolation of $\lauxm$ is an important contribution
of this paper. Indeed, it is easy to
see that $\lauxm$ should contain at least the three following rewriting rules:

\[ \begin{array}{lll@{\hspace{.5cm}}lll}
(\lam x.t) \slist\ u & \rRew{\beta} &  t\isubs{x/u}\slist & \mbox{  if }  x \in \fv{t}\\
(\lam x.t) \slist\ u & \rRew{\B} &  t[x/u]\slist & \mbox{  if }  x \notin \fv{t}\\
t[x/u] & \rRew{\Gc} & t & \mbox{  if }  x \notin \fv{t} \\
\end{array}\] 

More precisely, 

\begin{enumerate}[$\bullet$]
  \item The reduction step $t=(\lam x. x) y\Rew{\B} x[x/y]=u$ projects into $\wfc(t) = (\lam x. x)\ y \Rew{\beta} y= \wfc(u)$. 
\item The reduction step $t=(\lam 
 x. z) y\Rew{\B} z[x/y]=u$ should map to itself, \ie\ $\wfc(t) = (\lam x. z) y \Rew{\B} z[x/y]= \wfc(u)$. 
\item The reduction step $t=z[x/y]\Rew{\Gc}z=u$ 
 should map to itself, \ie\
$\wfc(t)= z[x/y]\Rew{\Gc} z = \wfc(t)$.
\end{enumerate}
However,  projecting on
such a simple calculus still does not work. There are 
three  phenomena  we should
take care of:

\begin{enumerate}[(1)]
\item 
\deft{Equations}. 
As we already mentioned $\eqw{\sigt}$ and $\eqw{\sigq}$
vanish, that is, 
$t\eqw{\sigt,  \sigq} u$ implies
$\wfc(t)=\wfc(u)$. The graphical equivalence, instead,
do not vanish, and must be added to the intermediate calculus, thus 
getting  the reduction  relation to be considered modulo $\eqo$.

  \item \deft{Generalised erasure}. Consider:
$$t=z[x/y_1 y_2][y_1/v_1] [y_2/v_2]\Rew{\Gc} z[y_1/v_1] [y_2/v_2]=u$$
where $\wfc(t)=z[x/v_1 v_2]$ and $\wfc(u)=z[y_1/v_1] [y_2/v_2]$. 
Hence the $\Gc$-rule $t[x/s]\Rew{} t$ if $|t|_x=0$
must be generalised in order to replace
the jump $[x/s]$ by many (eventually none) jumps containing subterms of $s$.  We shall
then use the following (Hydra like) rule :
\[ \begin{array}{lll@{\hspace{.5cm}}lll}
t[x/u] & \mapsto_{\New} & t[x_{1}/u_{1}]\ldots [x_n/u_n] & \forall i\
            (x_i \mbox{ fresh }\ \&\ u_i \subt
            u\ \&\ \fv{u_i} \subseteq \fv{u}\ \&\ n \geq 0)\\
\end{array}\] 
Where $u_i\subt u$ means that $u_i$ is a subterm of $u$. The condition upon
free variables is necessary in order to avoid unwanted captures
inducing degenerated behaviors. Note that the particular case $n=0$ gives
the $\Gc$-rule.
\item \deft{Unboxing}: An erasing step
$t\Rew{\Gc} u$ can cause jumps
to \textit{move} towards the \textit{root} of the term. Consider:
$$t=(z z[x/y])[y/v]\Rew{\Gc} (z z)[y/v]=u$$
where $\wfc(t)=z z[x/v]$ and $\wfc(u)=(z z)[y/v]=_\alpha (z z)[x/v]$.  
Hence, to project this step over $\lauxm$ we need a rule moving
jumps towards the \textit{root} of the term, which could have in
principle the general form:
$$\ctx{C}{t[x/u]}\Rew{} \ctx{C}{t}[x/u]$$ 
This rule is the one that shall demand a more involved --- but
still reasonable --- technical development. Indeed, 
reduction that moves  \textit{any} jump towards the root  modulo $\eqo$  may cause non-termination:
$$\lam x. x[y/z]\Rew{}(\lam x. x)[y/z]\eqo\lam x. x[y/z]\Rew{}\ldots$$ 
In order to avoid this problem we
restrict the general form of the rule to a certain kind of contexts, those
whose hole is contained in at least one \textit{box}, \ie\  
the argument of an application or the argument of a
jump. \medskip
\end{enumerate}

\noindent We now develop a PSN proof for $\ldisf$. 
Section~\ref{s:lauxm}  formally defines the intermediate calculus
$\lauxm$, 
while Section~\ref{s:iep-for-lauxm} proves PSN
for the intermediate calculus  $\lauxm$ and
Section~\ref{s:projection} proves the properties of the projection which allows us to conclude PSN for $\ldisf$. 

Let us conclude this section by observing that the 
  generalised erasure and the unboxing rules are 
introduced to project the $\Gc$-rule and not the 
equations 
$\set{\sigt, \sigq}$. Said in other words, 
to prove PSN of the simpler calculus $\ldis$ (resp. $\ldiso$) through the $\wfc$
  projection into $\laux$ (resp. $\lauxm$), one still needs the generalised erasure and the unboxing rules. That is why we
  believe that the technique  developed here is really interesting by itself.

\subsection{The $\lauxm$-calculus}
\label{s:lauxm}

The $\lauxm$-calculus can be understood as a memory calculus
based on \textit{void} jumps. It is given by a set of
  terms, written $\termsv$,  generated by the following grammar, where only
void jumps are allowed:
\[  (\termsv) \sep t, u :: = x \mid \lam x. t \mid t u \mid t[\void/u] \]
The notation $t[\void/u]$ just means that the constant $\void$
has no (free) occurrence in the term $t$ and $\vec{[\void/s]}$ 
denotes a list of void jumps $[\void/s_1]\ldots [\void/s_n]$. 

To define the operational semantics we need to define a particular
kind of context. More precisely, if $C$ denotes a context then a
\deft{boxed context} $B$ is given by the following grammar:
\[ B:: = t C \mid  t[\void/C] \mid B t \mid B[\void/t] \mid \lam y. B \]
We now consider the  reduction rules
 and equations in Figure~\ref{f:laux}. 
The notation $\List$ in the rules $\B$ and $\beta$
means a list $[\void/u_1]\ldots[\void/u_k]$ 
of void jumps where  $k\in\nat$ (so
potentially $k=0$).

\begin{figure}[ht]
\[ \begin{array}{lll@{\hspace{.5cm}}lll}
(\lam x.t) \slist\ u & \rRew{\beta} &  t\isubs{x/u}\slist & \mbox{if }  x \in \fv{t}\\
(\lam x.t) \slist\ u & \rRew{\B} &  t[\void/u]\slist & \mbox{if }  x \notin \fv{t}\\
t[\void/u] & \rRew{\New} & t[\void/u_{1}]\ldots [\void/u_n] & 
            \forall i\ (u_i \subt u\ \&\ \fv{u_i} \subseteq \fv{u}\ \&\ n \geq 0)\\
\ctx{B}{t[\void/u]} &  \rRew{\unboxed}   & \ctx{B}{t}[\void/u] & B \mbox{ does not bind } u\\\\
 t[\void/s][\void/v] & \sim_{\CS} & t[\void/v][\void/s] 
  \\
   \lam y. (t [\void/s]) & \preeqsigu & (\lam y. t) [\void/s]  & \mbox{ if } y\notin \fv{s} \\
   t[\void/s] v & \preeqsigt & (t v)[\void/s]  \\
\end{array}\] 
\caption{The $\lauxm$-reduction system\label{f:lauxm}}
\label{f:laux}
\end{figure}

 \noindent Note that the $\Gc$-rule $t[x/u] \Rew{} t$ with $x \notin \fv{t}$
 of $\ldis$ is a particular case of the $\New$-rule.  Remark
 also that the unboxing rule of $\lauxm$ moves \textit{void} jumps
 outside terms, which was forbidden in the equation
   $\sigq$ of $\ldisf$. However, this does not break PSN because 
 there is no \textit{boxing} rule in $\lauxm$. Indeed, Guerrini's counterexample
 uses both boxing and unboxing.

We write $t \Rew{\lauxm} t'$ iff $t \eqo t_1 \Rew{\laux} t'_1 \eqo t'$, where
$\Rew{\laux}$ is the reduction relation generated by the previous
rewriting rules $\set{\beta,\B,\New,\unboxed}$ and $\eqo$ is the 
equivalence relation defined in Figure~\ref{f:eqo}
but restricted here to the $\laux$-syntax. As before, $\Rew{\R}$ denotes the contextual closure
of $\rRew{\R}$, for  $\R \subseteq \set{\beta, \B,\New,\unboxed}$.

We now show some properties of the new memory reduction system which are 
used in Section~\ref{s:iep-for-lauxm} to show PSN.

\begin{lem}
\label{l:sous-terme-substitution}
Let $u,v,s \in \termsv$. If $u \subt s$ and $x \notin \fv{u}$, then
$u \subt s\isubs{x/v}$. 
\end{lem}

\begin{proof} By induction on $s$. 
\end{proof}

\begin{lem}
\label{l:newu-pass-to-sub} 
Let $t_0, t_1, u_0, u_1 \in \termsv$. 
\begin{enumerate}[$\bullet$]
\item If $t_0\Rewnmod{\New, \unboxed}{\osym} t_1$ then $t_0\isubs{x/u_0}
\Rewnmod{\New, \unboxed}{\osym} t_1\isubs{x/u_0}$.
\item   If $u_0\Rewnmod{\New, \unboxed}{\osym} u_1$ then $t\isubs{x/u_0}\Rewnmod{\New, \unboxed}{\osym} t\isubs{x/u_1}$.
\end{enumerate}
\end{lem}

\begin{proof}
Straightforward.
\end{proof}

\begin{lem}
\label{l:gen-new} 
Let $t,v,u,s_i \in \termsv$.
Let $x \in  \fv{v}$.
Then $t[\void/v\isubs{x/u}] \Rewn{\New}
     t\vec{[\void/s]}[\void/u]$, where  
     $s_i \subt v$ and $\fv{s_i} \subseteq \fv{v}$ and $x \notin  \fv{s_i}$.      
\end{lem}

\begin{proof}
Straightforward, the case $t[\void/v\isubs{x/u}] = 
     t\vec{[\void/s]}[\void/u]$ happening in particular
when $v = x$ and $\vec{[\void/s]}$ is empty.
\end{proof}

\begin{lem}
\label{l:out-subs}
Let $t,u,v\in\termsv$. If $y \in \fv{t}$ then
$t\isubs{y/v [\void/u]} \Rewnmod{\New, \unboxed}{\osym}  t\isubs{y/v} [\void/u]$. 
\end{lem}

\begin{proof}
By induction on $t$. 
\end{proof}

\begin{lem}
\label{l:substituion-new-u-o}
Let $t_0, t_1, v \in \termsv$. 
If $t_0 \Rewplus{\New} t_1$,
$x \in \fv{t_0}$ and $x \not \in  \fv{t_1}$,
then $t_0\isubs{x/v} 
\Rewnmod{\New,\unboxed}{\osym}
  t_1[\void/v]$. 
\end{lem}

\proof
By induction on the number of steps from $t_0$ to $t_1$, and in the
base case by induction on the reduction step from $t_0$ to $t_1$. 
\begin{enumerate}[$\bullet$]
\item $t_0 = u_0[\void/u_1] \Rew{\New} u_0[\void/v_1] \ldots [\void/v_m] = t_1$,  
      where 
     $v_i \subt u_1$ and $\fv{v_i} \subseteq \fv{ u_1}$. 
     Then $x \in  u_1$ and $x \notin u_0$ and  $x \notin  v_i$ so that 
     \[ 
           u_0[\void/u_1]\isubs{x/v}  = \\
           u_0[\void/u_1\isubs{x/v}]  \Rewn{\New} \ (Lem.~\ref{l:gen-new})\ 
           u_0[\void/v_1] \ldots [\void/v_m] [\void/v] \]
\item $t_0  = \lam y. u_0 \Rew{}  \lam y. u_1 =t_1$, where $u_0 \Rew{} u_1$.
        Then,
        \[ (\lam y. u_0)\isubs{x/v}     = 
           \lam y. u_0\isubs{x/v}    \Rewnmod{\New,\unboxed}{\osym}\ (\ih)\  
           \lam y. u_1[\void/v]        \equiv_{\sig_1} (\lam y. u_1)[\void/v] \]   
  \item $t_0  = u_0 v_0  \Rew{}  u_1 v_0 =t_1$, where $u_0 \Rew{} u_1$.
        Then,
        \[ (u_0 v_0) \isubs{x/v}  =  
            u_0 \isubs{x/v} v_0    \Rewnmod{\New,\unboxed}{\osym}\ (\ih)\ 
         u_1 [\void/v] v_0   \equiv_{\sig_2} (u_1 v_0)[\void/v] \] 
  \item $t_0  = u_0[\void/v_0] \Rew{} u_1[\void/v_0] =t_1$, where $u_0 \Rew{} u_1$.
 Then, 
        \[ u_0[\void/v_0]\isubs{x/v}  = 
        u_0 \isubs{x/v}[\void/v_0]  \Rewnmod{\New, \unboxed}{\osym}\ (\ih)\
        u_1[\void/v][\void/v_0]  \equiv_{\CS} 
        u_1[\void/v_0][\void/v] \]  
\item All the remaining cases are straightforward.       
\qed 
\end{enumerate}

\begin{cor}
\label{c:stability-substitution}
Let $t_0, t_1, v \in \termsv$. 
If $t_0 \Rewplusmod{\New,\unboxed}{\osym} t_1$, $x \in \fv{t_0}$ and $x \notin  \fv{t_1}$,
then $t_0\isubs{x/v} \Rewnmod{\New,\unboxed}{\osym}
  t_1[\void/v]$. 
\end{cor}

\proof
By induction on the number of $\New$-steps in the reduction 
$t_0 \Rewplusmod{\New,\unboxed}{\osym} t_1$.
Note first that $\Rew{\unboxed}$ and $\eqo$ do not loose free variables.

\begin{enumerate}[$\bullet$] 
\item If there is only one $\New$-step, then the reduction is of the form $t
  \Rewnmod{\unboxed}{\osym} u_0 \Rew{\New} u_1
  \Rewnmod{\unboxed}{\osym} t'$.  We have $t\isubs{x/v}
  \Rewnmod{\unboxed}{\osym} u_0\isubs{x/v}
  \Rewnmod{\New, \unboxed}{\osym}\ (Lem.~\ref{l:substituion-new-u-o})\
  u_1[\void/v] \Rewnmod{\unboxed}{\osym} t'[\void/v]$.
\item If there are $n> 1$ $\New$-steps, then the reduction is of the
  form $t_0 \Rewnmod{\unboxed}{\osym} u_0 \Rew{\New} u_1
  \Rewplusmod{\New,\unboxed}{\osym} t_1$, with $n-1<n$ $\New$-steps from $u_1$ to
  $t_1$ we consider two cases.

If $x \in \fv{u_0}\cap\fv{u_1}$, then $x$ is lost in the subsequence 
$u_1
  \Rewplusmod{\New,\unboxed}{\osym} t_1$. We thus have
$t_0\isubs{x/v} \Rewnmod{\unboxed}{\osym} u_0\isubs{x/v} \Rew{\New} u_1\isubs{x/v}
  \Rewnmod{\New,\unboxed}{\osym}\ (\ih)\ t_1[\void/v]$.

If $x \in \fv{u_0}\setminus\fv{u_1}$, then
$t_0\isubs{x/v} \Rewnmod{\unboxed}{\osym} u_0\isubs{x/v}  
\Rewnmod{\New,\unboxed}{\osym} u_1[\void/v]\ (Lem.~\ref{l:substituion-new-u-o})\
\Rewnmod{\New,\unboxed}{\osym} t_1[\void/v]$.
\qed
\end{enumerate}

\subsection{Preservation of $\beta$-Strong Normalization for $\lauxm$ }
\label{s:iep-for-lauxm}

The proof of PSN for $\lauxm$ we are going to develop in this
section is based on the \iep\ property (\cf\ Section~\ref{s:lj-psn}) and follows the main lines of
that of Theorem~\ref{t:ieg}.  
Indeed, given $u \in \SN{\lauxm}$ and
$t\isubs{x/u} \ovl{v}{1}{n} \in \SN{\laux}$ we show that
$s=t[\void/u] \ovl{v}{1}{n} \in \SN{\laux}$ by using a measure on terms
which decreases for every  one-step $\lauxm$-reduct of $s$.
However, 
PSN for $\lauxm$ is much more involved: first because of the nature of the reduction rules $\set{\New, \unboxed}$, 
second because of the equivalence $\eqo$. 

A first remark is that jumps
in $\lauxm$ are all void  so in particular 
they
cannot be duplicated. As a consequence, 
there is no need at first sight to generalise the \iep\ property to terms of
the form $t\espv{x}{u}{1}{m} \ovl{v}{1}{n}$ as we did before 
(Theorem~\ref{t:ieg}). However,
there are now new ways of getting jumps 
\textit{on the surface} of
the term. Indeed, if $t=\lam y. t'[\void/v]$ and $y\notin\fv{v}$ one 
has $s =t[\void/u] \ovl{v}{1}{n} \eqo (\lam y. t')[\void/v][\void/u] \ovl{v}{1}{n}$
Things are even more complicated since jumps
can also be moved between the arguments $v_1,\ldots,v_n$
as in: 
$$s\eqo ((\lam y. t'[\void/v]) v_1)[\void/u] \ovl{v}{2}{n}$$ 
The opposite phenomenon can happen too, \ie\
the jump $[\void/u]$ can enter inside $t$, for instance:
$$s\eqo \lam y.(t'[\void/v][\void/u]) \ovl{v}{1}{n}$$ 
The main point is that the
measure we shall use to develop the proof of the \iep\ property needs to be stable by 
the equivalence $\eqo$, \ie\ 
if $s \eqo s'$, then $s$ and $s'$
must have this same measure. 

In order to handle this phenomenon we are going to split $s$ in two
parts: the multiset $\preetamd{}{s}$ of \deft{jumps} of $s$
which \textit{are} or \textit{can} get to the \deft{surface}, and the 
\deft{trunk}
$\surf{}{s}$, \ie\  the term obtained from $s$ by removing all the jumps in
$\preetamd{}{s}$. This \textit{splitting} of the term is then used to
generalise the statement of the \iep\ as follows:

\begin{center}
If $\surf{}{s}\in\SN{\lauxm}$ and $u\in\SN{\lauxm}$ for every 
$[\void/u]\in \preetamd{}{s}$ then $s\in\SN{\lauxm}$.
\end{center}

%

An intuition behind the scheme of this proof is that the term
$\surf{}{s}$ and the jumps in $\preetamd{}{s}$ are dynamically
independent, in the sense that any reduction of $s$ can be seen as an
interleaving of a reduction (eventually empty) of $\surf{}{s}$ and
reductions (eventually empty) of elements of $\preetamd{}{s}$. Indeed,
the void jumps in $\preetamd{}{s}$ \textit{cannot be affected} by a
reduction of $\surf{}{s}$, since none of their free variables is bound
in $s$, and \textit{cannot affect} a reduction of $\surf{}{s}$ since
they are void. The unboxing rule slightly complicates things, but
morally that is why the new generalised form of the
\iep\ property holds.\medskip

The attentive reader
may wonder why we cannot handle the equivalence $\eqo$ by using a strong
bisimulation argument, as in the case of $\ldiso$ (\cf\ Theorem~\ref{thm:eqo-conf-psn}). 
Unfortunately, $\eqo$ is not a strong bisimulation 
for $\laux$ as the following example shows:
 \[ \begin{array}{cccc@{\hspace{1cm}}c@{\hspace{1cm}}ccccc}
  x[\void/t[\void/x] v]				& \Rew{\New}	& x[\void/t[\void/x]]	\\
  \eqo	&		&	\eqo\\
  x[\void/(t v)[\void/x]]				&	\Rew{\New}& ?	\\
 \end{array}
\]

Before starting with the technical details of the proof let us add two
more important remarks. First, we have just used $\surf{}{s}$ and
$\preetamd{}{s}$ for didactic purposes, the actual definitions are
parametrised with respect to a set of variables (those which can be
captured in the context containing $s$).  Moreover, in order to
simplify the proofs we will not work  directly with $\preetamd{}{s}$:
we are going to define a parametrised predicate $\snsudd{\Gam}{s}$,
which is true when all the jumps in $\preetamd{}{s}$ are in $\SN{\lauxm}$, and
a parametrised measure $\etamd{\Gam}{s}$,  built out from the elements
of $\preetamd{}{s}$.  Second, the unboxing rule makes some inductive
reasonings non-trivial, so we isolate them in an intermediate lemma
(Lemma.~\ref{l:lauxmes-red-unb2}).\medskip

Given $s \in \termsv$ and 
a set of variables $\Gam$, 
the trunk $\surf{\Gam}{s}$ is given by the following inductive definition:
\[ \begin{array}{llll}
   \surf{\Gam}{x}  &:=& x \\
   \surf{\Gam}{t u} &:=& \surf{\Gam}{t} u  \\
   \surf{\Gam}{\lam x. t} &:=& \lam x. \surf{\Gam \cup \set{x}}{t}\\
   \surf{\Gam}{t[\void/u]} &:=& \surf{\Gam}{t} 
        & \mbox{ if } \fv{u} \cap \Gam = \ems\\
   \surf{\Gam}{t[\void/u]} &:=& \surf{\Gam}{t}[\void/u] & \mbox{ otherwise } \\
   
   \end{array} \] 
Note that $x \in \fv{s}$ and $x \in \Gam$ implies
$x \in \surf{\Gam}{s}$. Next, we define a predicate on $\termsv$ which is true when all surface jumps contain terminating terms: 
\[ \begin{array}{llll}
   \snsudd{\Gam}{x}  &:=& true \\
   \snsudd{\Gam}{t u} &:=&  \snsudd{\Gam}{t} \\
   \snsudd{\Gam}{\lam x. t} &:=&  \snsudd{\Gam \cup \set{x}}{t}\\
    \snsudd{\Gam}{t[\void/u]} &:=&  \snsudd{\Gam}{t}\ \&\    u\in \SN{\lauxm}  
        & \mbox{ if } \fv{u} \cap \Gam = \ems\\
    \snsudd{\Gam}{t[\void/u]} &:=&  \snsudd{\Gam}{t} & \mbox{ otherwise } \\
    \end{array} \] 
Observe that $s \in \SN{\lauxm}$ implies in particular
$\snsudd{\Gam}{s}$ for any set $\Gam$.

For any term $s\in \termsv$ s.t. $\snsudd{\Gam}{s}$ we define the following multiset measure:
\[ \begin{array}{llll}
   \etamd{\Gam}{x}  & := & \ems \\
   \etamd{\Gam}{t u} &:=& \etamd{\Gam}{t} \sqcup \etamd{\Gam}{u}\\
   \etamd{\Gam}{\lam x. t} &:=&  \etamd{\Gam \cup \set{x}}{t}\\
   \etamd{\Gam}{t[\void/u]} &:=& \etamd{\Gam}{t} \sqcup
             \multiset{\pair{\eta_{\lauxm}(u)}{|u|}}  & \mbox{ if } \fv{u} \cap \Gam = \ems\\
   \etamd{\Gam}{t[\void/u]} &:=& \etamd{\Gam}{t} \sqcup
     \etamd{\Gam}{u} & \mbox{ otherwise } \\
   
   \end{array} \] 
Now, we can reformulate a generalised statement for the \iep\ property on {\bf V}oid jumps as follows:

\begin{center}
(\viep)  For all  $t \in \termsv$, if   $\surf{\ems}{t} \in  \SN{\lauxm}$ and 
$\snsudd{\ems}{t}$, then $t \in \SN{\lauxm}$. 
\end{center}

Some lemmas about basic properties of $\surf{\Gam}{t}$, $\snsudd{\Gam}{t}$ and $\etamd{\Gam}{t}$ follow.

\begin{lem}
\label{l:surf-snsudd-no-var}
Let $t \in \termsv$ and $x \notin \fv{t}$. Then 
$\surf{\Gam\cup\set{x}}{t} =\surf{\Gam}{t}$ and 
$\snsudd{\Gam\cup\set{x}}{t}$ iff $\snsudd{\Gam}{t}$.
\end{lem}

\begin{proof}
Straightforward.
\end{proof}

\begin{lem}
\label{l:surf-sub}
Let $t, u \in \termsv$ s.t. $\fv{t} \subseteq \Gam$. Then,
\begin{enumerate}[\rm(1)]
  \item \label{p:surf-sub-one}$t\Rewn{\New}\surf{\Gam}{t}$.
  \item \label{p:surf-sub-two}If $x\notin \Gam$ then 
        $\surf{\Gam\cup\set{x}}{t}\isubs{x/u}\Rewn{\New}\surf{\Gam}{t\isubs{x/u}}$.
\end{enumerate}
\end{lem}

\proof \hfill
\begin{enumerate}[(1)]
\item Straightforward induction on $t$. 
\item By induction on $t$.
\begin{enumerate}[$\bullet$]
  \item $t=x$.  Then $\surf{\Gam\cup\set{x}}{t}\isubs{x/u}=u \Rewn{\New}\ 
  (Point~\ref{p:surf-sub-one})\ \surf{\Gam}{u}= \surf{\Gam}{t\isubs{x/u}}$. 
  \item $t=y$.  Then $\surf{\Gam\cup\set{x}}{t}\isubs{x/u}=y=\surf{\Gam}{t\isubs{x/u}}$.
\item The cases $t=\lam y.v$ and $t=uv$ are straightforward using the \ih
 \item $t=v[\void/w]$. 
  Let us analyse one particular case in detail, the other ones being similar
    can be proved by application of the definitions and the \ih\ Let us suppose 
  $\Gam\cap\fv{w}=\ems$, $x\in\fv{w}$ and
  $\Gam\cap\fv{u}=\ems$. Then 

$\surf{\Gam\cup\set{x}}{t}\isubs{x/u}=
  \surf{\Gam\cup\set{x}}{v}[\void/w]\isubs{x/u}=\surf{\Gam\cup\set{x}}{v}\isubs{x/u}[\void/w\isubs{x/u}] \mbox{ and}$

$\surf{\Gam}{t\isubs{x/u}}=\surf{\Gam}{v\isubs{x/u}[\void/w\isubs{x/u}]}=\surf{\Gam}{v\isubs{x/u}}$. 
The  \ih\ gives 

  $\surf{\Gam\cup\set{x}}{v}\isubs{x/u}\Rewn{\New}\surf{\Gam}{v\isubs{x/u}}$
  and so we conclude with
$$\surf{\Gam\cup\set{x}}{v}\isubs{x/u}[\void/w\isubs{x/u}]\Rewn{\New}\surf{\Gam}{v\isubs{x/u}}[\void/w\isubs{x/u}]\Rew{\New}\surf{\Gam}{v\isubs{x/u}}\eqno{\qEd}$$ 
\end{enumerate}
\end{enumerate}

\begin{lem}
\label{l:surf-pred-sub}
Let $t,u \in \termsv$ and $x \notin \Gam$. If $\snsudd{\Gam \cup \set{x}}{t}$,
$\snsudd{\Gam}{u}$ and $\surf{\Gam\cup\set{x}}{t
}\isubs{x/u}\in\SN{\lauxm}$ then $\snsudd{\Gam}{t\isubs{x/u}}$.
\end{lem}

\begin{proof} 
 By induction on $t$ using Lemma~\ref{l:surf-snsudd-no-var}.
\end{proof}

\begin{lem}
\label{l:mes-eqo}
Let $t_0 \in \termsv$ s.t.   $\surf{\Gam}{t_0} \in  \SN{\lauxm}$ and 
$\snsudd{\Gam}{t_0}$. If $t_0\eqo t_1$ then 
$\snsudd{\Gam}{t_1}$ and $\surf{\Gam}{t_0} \eqo \surf{\Gam}{t_1}$
and $\etamd{\Gam}{t_0} = \etamd{\Gam}{t_1}$. Thus in particular
the equality $\eta_{\lauxm}(\surf{\Gam}{t_0}) =\eta_{\lauxm}(\surf{\Gam}{t_1})$ holds.
\end{lem}

\begin{proof}
By induction on $t_0\eqo t_1$.
\end{proof}

The next lemma deals with the unboxing rule, which requires a complex induction.

\begin{lem}
\label{l:lauxmes-red-unb2}
Let $t_0 \in \termsv$ s.t.   $\surf{\Gam}{t_0} \in  \SN{\lauxm}$ and 
$\snsudd{\Gam}{t_0}$. If $t_0=\ctx{B}{s[\void/u]}\Rew{\unboxed} \ctx{B}{s}[\void/u]= t_1$, 
where $B$ does not bind $u$,  then $\snsudd{\Gam}{t_1}$ and: 
\begin{enumerate}[$\star$]
  \item If $\Gam\cap \fv{u}=\ems$ then 
  \begin{enumerate}[$-$]
  \item Either $\surf{\Gam}{t_0}=\surf{\Gam}{t_1}$ and $\etamd{\Gam}{t_0}
\gm \etamd{\Gam}{t_1}$,
  \item Or $\surf{\Gam}{t_0} \Rew{\New} \surf{\Gam}{t_1}$;
  \end{enumerate}
  
  \item If $\Gam\cap \fv{u}\neq \ems$ then
  $\surf{\Gam}{t_0} \Rewplus{\modulo{\unboxed, \New}{\osym}} \surf{\Gam}{t_1}$.
\end{enumerate}
\end{lem}

\proof
By induction on $B$.
\begin{enumerate}[$\bullet$]
\item Base cases:
\begin{enumerate}[$-$]
\item $B =v C$: then 
  $t_0=v \ctx{C}{s[\void/u]} \Rew{\unboxed} (v \ctx{C}{s})[\void/u] = t_1$. Hence
  $\surf{\Gam}{t_0}=\surf{\Gam}{v} \ctx{C}{s[\void/u]}$ and
  $\snsudd{\Gam}{t_0}$ iff $\snsudd{\Gam}{v}$. There are two cases:
\begin{enumerate}[(1)]
  \item \underline{$\Gamma \cap\fv{u}=\ems$}: then 
    $\surf{\Gam}{t_0}\Rew{\New}\surf{\Gam}{v} \ctx{C}{s}=\surf{\Gam}{t_1}$. To show 
    $\snsudd{\Gam}{t_1}$ we need  $\snsudd{\Gam}{v}\ \&\ 
    u\in\SN{\lauxm}$. The first point is equivalent to
    $\snsudd{\Gam}{t_0}$, which holds by hypothesis, the second 
    holds since  $u$ is a
    subterm of $\surf{\Gam}{t_0} \in \SN{\lauxm}$.
\item \underline{$\Gamma \cap\fv{u}\neq\ems$}: then 
  $\surf{\Gam}{t_0}\Rew{\unboxed}
  (\surf{\Gam}{v} \ctx{C}{s})[\void/u]=\surf{\Gam}{t_1}$. To show 
  $\snsudd{\Gam}{t_1}$ we need  $\snsudd{\Gam}{v}$, which holds by hypothesis. 
 \end{enumerate}
\item $B =v [\void/C]$: there are four cases:
\begin{enumerate}[(1)]

\item \emph{$\Gam\cap\fv{u}=\ems\ \&\ \Gam\cap\fv{\ctx{C}{s}}=\ems$}:
    then 
    $\surf{\Gam}{t_0}=\surf{\Gam}{v}=\surf{\Gam}{t_1}$. Also, $\snsudd{\Gam}{t_0}$
    so that  
    $\snsudd{\Gam}{v}\ \&\ \ctx{C}{s[\void/u]}\in\SN{\lauxm}$. To show 
    $\snsudd{\Gam}{t_1}$ we need 
    $\ctx{C}{s}\in\SN{\lauxm}\ \&\ u\in\SN{\lauxm}$, which clearly
    follows from $\ctx{C}{s[\void/u]}\in\SN{\lauxm}$. We still 
    need to  show that
    $\etamd{\Gam}{t_0} \gm\etamd{\Gam}{t_1}$ which holds because
    $\etamd{\Gam}{t_1}$ is just $\etamd{\Gam}{t_0}$ where 
    the multiset $           \multiset{\pair{\eta_{\laux}(\ctx{C}{s[\void/u]})}{|\ctx{C}{s[\void/u]}|}}
    \in \etamd{\Gam}{t_0}$ is replaced by the strictly smaller multiset
    $\multiset{\pair{\eta_{\laux}(\ctx{C}{s})}{|\ctx{C}{s}|},
      \pair{\eta_{\laux}(u)}{|u|}}$. 

\item \emph{$\Gam\cap\fv{u}=\ems\ \&\ \Gam\cap\fv{\ctx{C}{s}}\neq\ems$}: 
   then 
  \[ \surf{\Gam}{t_0}=\surf{\Gam}{v}[\void/\ctx{C}{s[\void/u]}]\Rew{\New}\surf{\Gam}{v}[\void/\ctx{C}{s}]= \surf{\Gam}{t_1}\] 
  Also, 
  $\snsudd{\Gam}{t_0}$ implies $\snsudd{\Gam}{v}$. 
  To show 
  $\snsudd{\Gam}{t_1}$ we need $\snsudd{\Gam}{v}\ \&\ u\in\SN{\lauxm}$,
  which then holds by hypothesis and because $u$ is a subterm of
  $\surf{\Gam}{t_0} \in \SN{\lauxm}$.

 \item \emph{$\Gam\cap\fv{u}\neq\ems\ \&\  \Gam\cap\fv{\ctx{C}{s}}=\ems$}: then
   \[ \surf{\Gam}{t_0}=\surf{\Gam}{v}[\void/\ctx{C}{s[\void/u]}]\Rew{\unboxed}\surf{\Gam}{v}[\void/\ctx{C}{s}][\void/u]\Rew{\New}\surf{\Gam}{v}[\void/u]=\surf{\Gam}{t_1}\]

   Also,  
   $\snsudd{\Gam}{t_0}$ implies  $\snsudd{\Gam}{v}$. To show
   $\snsudd{\Gam}{t_1}$ we need 
   $\snsudd{\Gam}{v}\ \&\ \ctx{C}{s}\in\SN{\lauxm}$, which holds
   by the hypothesis and the fact that $\ctx{C}{s}$ is a subterm of a $\New$-reduct of 
   $\surf{\Gam}{t_0} \in \SN{\lauxm}$.

\item \emph{$\Gam\cap\fv{u}\neq\ems\ \&\  \Gam\cap\fv{\ctx{C}{s}}\neq\ems$}: then
  \[ \surf{\Gam}{t_0}=\surf{\Gam}{v}[\void/\ctx{C}{s[\void/u]}]\Rew{\unboxed}\surf{\Gam}{v}[\void/\ctx{C}{s}][\void/u]=\surf{\Gam}{t_1}\]

  Also 
  $\snsudd{\Gam}{t_0}$ implies  $\snsudd{\Gam}{v}$, which is equivalent to 
  $\snsudd{\Gam}{t_1}$. 

\end{enumerate}
\end{enumerate}
\item Inductive cases:
\begin{enumerate}[$-$]

\item $B =B'[\void/v]$: 
  We have $t_0=\ctx{B'}{s[\void/u]}[\void/v] \Rew{\unboxed} \ctx{B'}{s}[\void/v][\void/u] = t_1$.
  Also 
  $\ctx{B'}{s[\void/u]} \Rew{\unboxed} \ctx{B'}{s}[\void/u]$ and the hypothesis
  $\surf{\Gam}{t_0} \in \SN{\lauxm}$  and $\snsudd{\Gam}{t_0}$ imply  in particular
  $\surf{\Gam}{\ctx{B'}{s[\void/u]}} \in \SN{\lauxm}$ and $\snsudd{\Gam}{\ctx{B'}{s[\void/u]}}$.
  The \ih\ gives $\snsudd{\Gam}{\ctx{B'}{s}[\void/u]}$ and we distinguish 
  several  cases: 
  \begin{enumerate}[(1)]
  \item \emph{$\Gam\cap\fv{u}=\ems\ \&\ \Gam\cap\fv{v}=\ems$}: 
        The hypothesis  $\snsudd{\Gam}{t_0}$ implies in particular
        $v \in \SN{\lauxm}$ and the \ih\ 
        $\snsudd{\Gam}{\ctx{B'}{s}[\void/u]}$ gives
        $\snsudd{\Gam}{\ctx{B'}{s}}\ \&\ u \in \SN{\lauxm}$, so 
        we conclude also $\snsudd{\Gam}{t_1}$. We now consider two cases:
        \begin{enumerate}[(a)]
        \item If $u_0=\surf{\Gam}{\ctx{B'}{s[\void/u]}} = \surf{\Gam}{\ctx{B'}{s}[\void/u]}$
              and $\etamd{\Gam}{\ctx{B'}{s[\void/u]}}  \gm \etamd{\Gam}{\ctx{B'}{s}} \sqcup
                  \multiset{\pair{\eta_{\lauxm}(u)}{|u|}}$, 
        then $\surf{\Gam}{t_0} = u_0 =              \surf{\Gam}{t_1}$ and 
        $\etamd{\Gam}{t_0} = \etamd{\Gam}{\ctx{B'}{s[\void/u]}}  \sqcup
                  \multiset{\pair{\eta_{\lauxm}(v)}{|v|}} \gm \etamd{\Gam}{\ctx{B'}{s}} \sqcup
                  \multiset{\pair{\eta_{\lauxm}(u)}{|u|}} \sqcup
                  \multiset{\pair{\eta_{\lauxm}(v)}{|v|}} = \etamd{\Gam}{t_1}$.
        
        \item If $u_0=\surf{\Gam}{\ctx{B'}{s[\void/u]}} \Rew{\New}
          \surf{\Gam}{\ctx{B'}{s}[\void/u]}=u_1$, then $\surf{\Gam}{t_0} =
          u_0 \Rew{\New}
          u_1  = \surf{\Gam}{t_1}$.
        \end{enumerate}        
  \item \emph{$\Gam\cap\fv{u}=\ems\ \&\ \Gam\cap\fv{v}\neq\ems$}: 
        The  \ih\ 
        $\snsudd{\Gam}{\ctx{B'}{s}[\void/u]}$ gives
        $\snsudd{\Gam}{\ctx{B'}{s}}\ \&\ u \in \SN{\lauxm}$, so 
        we conclude also $\snsudd{\Gam}{t_1}$. We now consider two cases:  
         
        \begin{enumerate}[(a)]
        \item If $u_0= \surf{\Gam}{\ctx{B'}{s[\void/u]}} = \surf{\Gam}{\ctx{B'}{s}[\void/u]}$
              and $\etamd{\Gam}{\ctx{B'}{s[\void/u]}}  \gm \etamd{\Gam}{\ctx{B'}{s}} \sqcup
                  \multiset{\pair{\eta_{\lauxm}(u)}{|u|}}$, 
        then $\surf{\Gam}{t_0} = u_0[\void/v] = 
              \surf{\Gam}{\ctx{B'}{s}}[\void/v] = 
              \surf{\Gam}{t_1}$ and 
        $\etamd{\Gam}{t_0} = \etamd{\Gam}{\ctx{B'}{s[\void/u]}}  \sqcup
                  \etamd{\Gam}{v} \gm \etamd{\Gam}{\ctx{B'}{s}} \sqcup
                  \multiset{\pair{\eta_{\lauxm}(u)}{|u|}} \sqcup
                  \etamd{\Gam}{v}  = \etamd{\Gam}{t_1}$.
        \item If $u_0=\surf{\Gam}{\ctx{B'}{s[\void/u]}} \Rew{\New}
          \surf{\Gam}{\ctx{B'}{s}[\void/u]}=u_1$, then 
          $$ \surf{\Gam}{t_0} =
          u_0[\void/v] \Rew{\New}
          u_1[\void/v] = \surf{\Gam}{\ctx{B'}{s}}[\void/v] = 
          \surf{\Gam}{t_1} $$  
        \end{enumerate}   
  \item \emph{$\Gam\cap\fv{u} \neq \ems$}: Then the \ih\ gives
    $u_0=\surf{\Gam}{\ctx{B'}{s[\void/u]}}
    \Rewplus{\modulo{\New,\unboxed}{\osym}}
    \surf{\Gam}{\ctx{B'}{s}[\void/u]}=u_1$. We consider the following cases.

    \begin{enumerate}[(a)]
    \item \emph{$\Gam\cap\fv{v}= \ems$}:  then
\begin{center}$ \surf{\Gam}{t_0} = u_0  \Rewplus{\modulo{\New,\unboxed}{\osym}}
     u_1 = \surf{\Gam}{\ctx{B'}{s}}[\void/u] = \surf{\Gam}{t_1}$\end{center}

Also $\snsudd{\Gam}{t_0}$ 
          implies $v \in \SN{\lauxm}$ and the \ih\ $\snsudd{\Gam}{\ctx{B'}{s}[\void/u]}$
          implies  $\snsudd{\Gam}{\ctx{B'}{s}}$, we thus conclude
          $\snsudd{\Gam}{t_1}$. 
    \item \emph{$\Gam\cap\fv{v}\neq\ems$}: then
\begin{center}$
\begin{array}{llllllllll}
 \surf{\Gam}{t_0} &=& u_0[\void/v]  
          \Rewplus{\modulo{\New,\unboxed}{\osym}}
   u_1[\void/v] &=
   & \surf{\Gam}{\ctx{B'}{s}}[\void/u][\void/v]
         &\eqo\\
         &&&&          \surf{\Gam}{\ctx{B'}{s}}[\void/v][\void/u] &=& \surf{\Gam}{t_1} 
         \end{array}$\end{center}

  Also, the  \ih\ $\snsudd{\Gam}{\ctx{B'}{s}[\void/u]}$
          implies  $\snsudd{\Gam}{\ctx{B'}{s}}$, we therefore conclude
          $\snsudd{\Gam}{t_1}$.

    \end{enumerate}

  \end{enumerate}

\item The cases $B =\lam y.B'$ and
             $B=B' w$ are similar to the previous ones. 

\qed
\end{enumerate} 
\end{enumerate}

The following lemma states that the measure we use for proving
\viep\ for $\lauxm$ decreases with every rewriting step.

\begin{lem}
\label{l:lauxmes-red}
Let $t_0 \in \termsv$ s.t.   $\surf{\Gam}{t_0} \in  \SN{\lauxm}$ and 
$\snsudd{\Gam}{t_0}$. If $t_0\Rew{\laux} t_1$ then $\snsudd{\Gam}{t_1}$ and
\begin{enumerate}[$-$]
  \item Either $\surf{\Gam}{t_0} \Rewplus{\lauxm} \surf{\Gam}{t_1}$ or 
  \item $\surf{\Gam}{t_0}=\surf{\Gam}{t_1}$ and $\etamd{\Gam}{t_0} \gm \etamd{\Gam}{t_1}$.
\end{enumerate}
\end{lem}

\proof
By induction on $t_0\Rew{\laux} t_1$.
\begin{enumerate}[$\bullet$]
\item Base cases:
\begin{enumerate}[$-$]
  \item If $t_0=(\lam x. s)\List\ u\Rew{\B} s [\void/u]\List = t_1$, where $x \notin \fv{s}$. 

Let $\List:=[\void/v_1]\ldots[\void/v_k]$,
$Q:=\set{v_i\ |\ \Gam\cap\fv{v_i}\neq\ems, i\in\set{1,\ldots,k}}$ and
$\overline{Q}:=\set{ v_i\ |\ \Gam\cap\fv{v_i}=\ems,
  i\in\set{1,\ldots,k}}$. Define $\List_Q$  the sublist of $\List$ containing only the elements in $Q$.  We have
\begin{enumerate}[$\star$]
  \item $\snsudd{\Gam}{t_0}$ iff $\snsudd{\Gam\cup\set{x}}{s} =_{Lem.~\ref{l:surf-snsudd-no-var}} \snsudd{\Gam}{s}$ and
    $v_j\in\SN{\lauxm}$ for every $v_j\in \overline{Q}$.
  \item $\surf{\Gam}{t_0}=(\lam x.\surf{\Gam\cup\set{x}}{s})\List_Q\ u=_{Lem.~\ref{l:surf-snsudd-no-var}}(\lam     x.\surf{\Gam}{s})\List_Q\ u$.
\end{enumerate}
There are two cases:
\begin{enumerate}[(1)]
  \item \emph{$\Gam\cap\fv{u}\neq\ems$}. We have
    $\surf{\Gam}{t_1}=\surf{\Gam}{s}[\void/u]\List_Q$. Then
    $\surf{\Gam}{t_0}\Rew{\B} \surf{\Gam}{t_1}$. Moreover,
    $\snsudd{\Gam}{t_1}$ iff $\snsudd{\Gam}{s}$ and $v_j\in\SN{\lauxm}$
    for every $v_j\in \overline{Q}$ , which holds by 
    the hypothesis $\snsudd{\Gam}{t_0}$.
\item \emph{$\Gam\cap\fv{u}=\ems$}. We have
  $\surf{\Gam}{t_1}=\surf{\Gam}{s}\List_Q$.  Then $\surf{\Gam}{t_0}\Rew{\B}
  \surf{\Gam}{s} [\void/u]\List_Q\Rew{\New}
  \surf{\Gam}{s}\List_Q=\surf{\Gam}{t_1}$.  Moreover, $\snsudd{\Gam}{t_1}$
  iff $\snsudd{\Gam}{s}$ and $u\in\SN{\lauxm}$ and $v_j\in\SN{\lauxm}$ for every $v_j\in
  \overline{Q}$. The first and third parts follow
from the hypothesis $\snsudd{\Gam}{t_0}$ while the second one follows from the hypothesis
  $\surf{\Gam}{t_0} \in \SN{\lauxm}$.
\end{enumerate}

  \item $t_0=(\lam x. s)\List\ u\Rew{\beta} s \isubs{x/u}\List= t_1$, where $x \in \fv{s}$.  

Let $\List$, $Q$, $\overline{Q}$ and $\List_{Q}$ be as in the previous case.
We have 
\begin{enumerate}[$\star$]
  \item $\snsudd{\Gam}{t_0}$ iff $\snsudd{\Gam\cup\set{x}}{s}$ and
    $v_j\in\SN{\lauxm}$ for every $v_j\in \overline{Q}$.
  \item $\surf{\Gam}{t_0}=(\lam x.\surf{\Gam\cup\set{x}}{s})\List_Q\ u$ with $x \in \surf{\Gam\cup\set{x}}{s}$.
\end{enumerate}

Then 
    $\surf{\Gam}{t_0}\Rew{\beta}
    \surf{\Gam\cup\set{x}}{s} \isubs{x/u}\List_Q \Rewn{\New\ (Lem.~\ref{l:surf-sub})}\ \surf{\Gam}{s\isubs{x/u}}\List_Q=\surf{\Gam}{t_1}$. Thus in particular $\surf{\Gam\cup\set{x}}{s} \isubs{x/u} \in \SN{\lauxm}$.

 Since $u$ is a
    subterm of $\surf{\Gam}{t_0}$, then  $u\in\SN{\lauxm}$ and so
    $\snsudd{\Gam}{u}$. Then $\snsudd{\Gam}{t_1}$ iff $\snsudd{\Gam}{s
      \isubs{x/u}}$ and $v_j\in\SN{\lauxm}$ for every $v_j\in \overline{Q}$. The first part  holds by  Lemma~\ref{l:surf-pred-sub}, the second one from the
hypothesis $\snsudd{\Gam}{t_0}$. 

\item $t_0 = u[\void/v] \Rew{\New}  u[\void/v_1] \ldots [\void/v_k]= t_1$, where 
  $k \geq 0$, $v_j \subt v$ for all $j$ and $\fv{v_j} \subseteq \fv{v}$. There are two cases:
  \begin{enumerate}[(1)]
  \item \emph{$\Gam\cap\fv{v}=\ems$}: we have that $\snsudd{\Gam}{t_0}$ implies
    $\snsudd{\Gam}{t_1}$. Then $\surf{\Gam}{t_0} = \surf{\Gam}{u} =
    \surf{\Gam}{t_1}$, moreover the multiset
    $\multiset{\pair{\eta_{\laux}(u)}{|u|}}$ of
$\etamd{\Gam}{t_0}$ is replaced by the following multiset of
        $\etamd{\Gam}{t_1}$: 
      $\multiset{\pair{\eta_{\laux}(v_1)}{|v_1|}, \ldots,
        \pair{\eta_{\laux}(v_k)}{|v_k|}}$.  Since $\eta_{\laux}(v) \geq
        \eta_{\laux}(v_i)$ and $|v| > |v_i|$ we thus
        conclude $\etamd{\Gam}{t_0} \gm \etamd{\Gam}{t_1}$.

\item \emph{$\Gam\cap\fv{v}\neq\ems$}: let $Q$ and 
  $\overline{Q}$ as in de $\B$-case. Then  $\snsudd{\Gam}{t_1}$ iff
  the terms in $\overline{Q}$ are $\SN{\lauxm}$ and $\snsudd{\Gam}{u}$
  holds: the former requirement holds because
  $\surf{\Gam}{t_0}=\surf{\Gam}{u}[\void/v]$ and so $v\in\SN{\lauxm}$, the
  latter because  $\snsudd{\Gam}{t_0}$ iff 
  $\snsudd{\Gam}{u}$. Last, $\surf{\Gam}{t_1}=\surf{\Gam}{u}\List_Q$,
  where $\List_Q$ is the list of substitutions associated to the elements
  in $Q$, then 
$$\surf{\Gam}{t_0}=\surf{\Gam}{u}[\void/v]\Rew{\New}\surf{\Gam}{u}\List_Q=\surf{\Gam}{t_1}$$

  \end{enumerate}

\item $t_0= \ctx{B}{s[\void/u]} \Rew{\unboxed} \ctx{B}{s}[\void/u] = t_1$. This case holds by Lemma~\ref{l:lauxmes-red-unb2}.
\end{enumerate}

\item Inductive cases:
\begin{enumerate}[$-$]
\item $t_0= u[\void/v] \Rew{\laux} u[\void/v'] = t_1$
  where $v  \Rew{\laux} v'$.  We consider three cases.
 
  \begin{enumerate}[(1)]
  \item  \emph{$\fv{v} \cap \Gam = \ems\ \&\ \fv{v'} \cap \Gam =\ems$}: 
  We have $\surf{\Gam}{t_0} = \surf{\Gam}{u} = \surf{\Gam}{t_1}$.
  Also  $\snsudd{\Gam}{t_0}$ implies $v \in \SN{\lauxm}$ so that 
   $v' \in \SN{\lauxm}$ and thus $\snsudd{\Gam}{t_1}$.
  Finally, 
  \[ \begin{array}{lll} 
    \etamd{\Gam}{t_0}  =  &  \etamd{\Gam}{u} \sqcup
    \multiset{\pair{\eta_{\lauxm}(v)}{|v|}} & \gm \\ 
    & \etamd{\Gam }{u} \sqcup
             \multiset{\pair{\eta_{\lauxm}(v')}{|v'|}}  & = 
   \etamd{\Gam}{t_1} 
   \end{array} \]
  
\item \emph{$\fv{v} \cap \Gam \neq \ems\ \&\ \fv{v'} \cap \Gam \neq
  \ems$ } : We have
  $\snsudd{\Gam}{t_0}=\snsudd{\Gam}{u}=\snsudd{\Gam}{t_1}$.  Also
  $\surf{\Gam}{t_0}=\surf{\Gam}{u}[\void/v]$ and
  $\surf{\Gam}{t_1}=\surf{\Gam}{u}[\void/v']$, thus
  $\surf{\Gam}{t_0} \Rewplus{\lauxm} \surf{\Gam}{t_1}$.

\item  \emph{$\fv{v} \cap \Gam \neq  \ems\ \&\ \fv{v'} \cap \Gam = \ems$}:   
 We have that $\surf{\Gam}{t_0}\in \SN{\lauxm}$ implies  $v\in\SN{\lauxm}$, 
so that $v'\in\SN{\lauxm}$ and $\snsudd{\Gam}{t_1}$. 

Then $\surf{\Gam}{t_0}=\surf{\Gam }{u}[\void/ v] 
   \Rew{\New} \surf{\Gam }{u}= \surf{\Gam}{t_1}$.  

   \end{enumerate}
\ignore{
\item $t_0 = u v  \Rew{\laux} u' v = t_1$, where $u  \Rew{\laux} u'$.
This case is straightforward  by the \ih\
\item $t_0 = u v \Rew{\laux} u v' = t_1$, where $v  \Rew{\laux} v'$.
This case is also straightforward  by the \ih\
\item $t_0=\lam x. s\Rew{\laux} \lam x. s' = t_1$: just use the \ih\
\item $t_0= u[\void/v] \Rew{\laux} u'[\void/v] = t_1$: just use the \ih }
\item All the other cases are straightforward.
\qed
\end{enumerate}
\end{enumerate}

\begin{thm}[\viep\ for $\lauxm$]
\label{t:ie-laux}
Let $t \in \termsv$ s.t.   $\surf{\ems}{t} \in  \SN{\lauxm}$ and 
$\snsudd{\ems}{t}$, then $t \in \SN{\lauxm}$. 
\end{thm}

\begin{proof}
We proceed by induction on the measure $m(t)=\pair{
  \eta_{\lauxm}(\surf{\ems}{t})}{ \etamd{\ems}{t}}$. 
To show $t \in \SN{\lauxm}$ it is sufficient to show
$t' \in \SN{\lauxm}$ for every $\lauxm$-reduct of $t$.
Take any of such reducts $t'$. Then Lemmas~\ref{l:mes-eqo} and ~\ref{l:lauxmes-red}
guarantee 
$\surf{\ems}{t'} \in  \SN{\lauxm}$ and 
$\snsudd{\ems}{t'}$. Moreover, $\eqo$ preserves $m(t)$ and 
$\Rew{\lauxm}$ strictly decreases $m(t)$. We thus
apply the \ih\ to conclude. 
\end{proof}

The following  is a  consequence of the  previous theorem: let  $t, u,
\ovl{v}{1}{n} \in  \lam$-terms   and $s=t
    [\void/u]\ovl{v}{1}{n}$.  If $\surf{\ems}{s} = t \ovl{v}{1}{n} \in
    \SN{\laux}$ and $\snsudd{\ems}{s}$ holds, \ie\ $u \in \SN{\laux}$,
    then $s=t[\void/u]\ovl{v}{1}{n} \in \SN{\laux}$. Hence:

\begin{cor}[\iep\ for $\lauxm$]
\label{t:ielaux}
The $\lauxm$-calculus enjoys the \iep\ property.
\end{cor}

\begin{cor}[PSN for $\lauxm$]
\label{c:psn-lauxm}
The $\lauxm$-calculus enjoys PSN, \ie\ if $t \in \termslambda \cap \SN{\beta}$, then 
$t \in \SN{\lauxm}$.
\end{cor}

\begin{proof} By Theorem~\ref{t:ie-implies-psn} it is sufficient to
show {\bf F0}, {\bf F1} and {\bf F2}. The first two properties are straightforward. 
To show ${\bf F2}$ assume $v \in \SN{\laux}$ and
$u\isubs{x/v} \ovl{t}{1}{n} \in \SN{\laux}$, both are $\lam$-terms.
Then in particular $u, v, \ovl{t}{1}{n}\in \SN{\laux}$. We show that
$t = (\lam x.  u) v \ovl{t}{1}{n} \in \SN{\laux}$ by induction on
$\eta_{\laux}(u) + \eta_{\laux}(v) + \Sigma_{i}\ \eta_{\laux}(t_i)$.
For that, it is sufficient to show that every $\laux$-reduct of $t$ is
in $\SN{\laux}$.  If the $\laux$-reduct of $t$ is internal we conclude
by the \ih\ If $t \Rew{\beta} u\isubs{x/v} \ovl{t}{1}{n}=t'$ with $x
\in \fv{u}$, then $t' \in \SN{\laux}$ by hypothesis. If $t \Rew{\B} u[\void/v]
\ovl{t}{1}{n}=t'$, then $t' \in \SN{\laux}$ by the
\iep\ property (Corollary~\ref{t:ielaux}).  There is no other possible $\laux$-reduct of $t$
which is a  $\lam$-term and has no jumps.
\end{proof}

\subsection{Projecting  $\ldisf$ into $\lauxm$}
\label{s:projection}

In order to relate the $\ldisf$ and the $\lauxm$ calculi 
we define a projection function from $\ldis$-terms to $\laux$-terms:
\[ \begin{array}{lll}
   \proj(x)      & := & x \\
   \proj(\lam x.t) & := & \lam x.\proj(t) \\
   \proj(tu)     & := & \proj(t)\proj(u)  \\
   \proj(t[x/u]) & := & \left \{ \begin{array}{ll}
                        \proj(t) \isubs{x/\proj(u)} & \mbox{ if } x \in \fv{t} \\
                        \proj(t) [\void/\proj(u)] & \mbox{ if } x \notin \fv{t} \\
                        \end{array} \right . \\
   \end{array} \]
Notice that $\fv{t} = \fv{\proj(t)}$. Also, $\proj(t) = t$ if $t \in \termslambda$.

We now state some basic static properties of $\proj$.

\begin{lem}
\label{l:basic-properties-proj} Let $t,u \in \terms$. Then, 
$\proj(t\isubs{x/u}) = \proj(t)\isubs{x/\proj(u)}$.
\end{lem}

\begin{proof}
By induction on $t$.
\end{proof}

\begin{lem}[Projection]
\label{l:projection}
Let $t_0 \in \terms$. Then, 
\begin{enumerate}[\rm(1)]
\item \label{ppp-i} $t_0 \Rew{\B} t_1$ implies $\proj(t_0) \Rewplus{\beta, \B} \proj(t_1)$.
\item\label{ppp-iii}  $t_0 \Rew{\Gc, \Var, \DSubs} t_1$ implies 
$\proj(t_0) \Rewnmod{\New, \unboxed}{\osym} \proj(t_1)$.
\item \label{ppp-iv} $t_0 \eqo t_1$ implies $\proj(t_0) \equiv_{\osym} \proj(t_1)$.
\item \label{ppp-v} $t_0 \equiv_{\sigt, \sigq} t_1$ implies $\proj(t_0) = \proj(t_1)$.
\end{enumerate}
\end{lem}

\proof \hfill
\begin{enumerate}[$\bullet$]
\item Base cases:
\begin{enumerate}[$-$]
\item $t_0 =  (\lam x.t)\slist\ u  \Rew{\B}  t[x/u]\slist = t_1$.

Let $\mlist = [\void/\proj(v_i)]^1_m$ (resp. $\rho$) be the sequence of jumps
(resp. the meta-level substitution) resulting from the 
projection of $t_0$, \ie\ $\proj(t_0) = (\lam x.\proj(t)) \mlist\ \rho\ \proj(u)$.

If $x \in \fv{t}$, then:
\[\begin{array}{lllll}
\proj(t_0) &=& (\lam x.\proj(t)\rho) [\void/ \proj(v_i) \rho]^1_m  \proj(u)& \Rew{\beta}\\
&& \proj(t)\rho\isubs{x/\proj(u)}   [\void/ \proj(v_i) \rho]^1_m   &=&\\
&& \proj(t)\isubs{x/\proj(u)}\rho   [\void/ \proj(v_i) \rho]^1_m  &= \proj(t_1)
  \end{array} \]

If $x \notin \fv{t}$, then: 
\[\begin{array}{lllll}
\proj(t_0) &= &(\lam x.\proj(t)\rho) [\void/ \proj(v_i) \rho]^1_m \proj(u)& \Rew{\B}\\
&& \proj(t)\rho [\void/\proj(u)] [\void/ \proj(v_i) \rho]^1_m &=&\\ 
            &&\proj(t)[\void/\proj(u)] [\void/ \proj(v_i) \rho]^1_m  \rho &=& 
            \proj(t_1)
\end{array} \] 

\item $t_0 = t[x/u]  \Rew{\Gc}  t= t_1$  where  $|t|_{x}= 0$.  
Then $\proj(t[x/u]) =\proj(t)[\void/\proj(u)] \Rew{\New} \proj(t)$. 
\item  $t_0  =  t[x/u]  \Rew{\Var} t\isubs{x/u}= t_1$  where  $|t|_{x}= 1$.
Then $\proj(t[x/u]) =\proj(t)\isubs{x/\proj(u)} 
=_{Lem.~\ref{l:basic-properties-proj}} \proj(t\isubs{x/u})$.
\item  $t_0 = t[x/u]  \Rew{\DSubs}  t_{[y]_x}[x/u][y/u]= t_1$  where 
$|t|_{x} \geq 2$. Then 
$\proj(t[x/u]) =\proj(t)\isubs{x/\proj(u)} =
                     \proj(t_{[y]_x})\isubs{y/\proj(u)}\isubs{x/\proj(u)} 
=_{Lem.~\ref{l:basic-properties-proj}}
                     \proj(t_{[y]_x}[x/u][y/u])$. 
\item $t_0 = t[x/u][y/v]  \equiv_{\CS}   t[y/v][x/u]= t_1$  where  $y\notin
    \fv{u}\  x\notin    \fv{v}$.
There are two cases:  

If $x \in \fv{t}$ or $y \in \fv{t}$, then we obtain $\proj(t_0) = \proj(t_1)$.

If $x \notin \fv{t}$ and $y \notin \fv{t}$, then 
$$\proj(t_0) = \proj(t)[\void/\proj(u)][\void/\proj(v)] \equiv_{\CS}
                \proj(t)[\void/\proj(v)][\void/\proj(u)] =\proj(t_1)$$

\item $t_0 = (\lam y.t)[x/u]  \equiv_{\sig_1}  \lam y.t[x/u]= t_1$   
where $y \notin  \fv{u}$.
There are two  cases:

If $x \in \fv{\lam y. t}$, then $\proj(t_0) = (\lam y. \proj(t))\isubs{x/\proj(u)} = 
                    \lam y. \proj(t)\isubs{x/\proj(u)} =
                    \proj(t_1)$.

If $x \notin \fv{\lam y. t}$, then $\proj(t_0) = (\lam y. \proj(t))[\void/\proj(u)] \equiv_{\sig} 
                    \lam y. \proj(t)[\void/\proj(u)] =
                    \proj(t_1)$.

\item $t_0 = (tv)[x/u]  \equiv_{\sig_2}  t[x/u]v= t_1$   where 
$x \notin  \fv{v}$.

There are two cases:

If $x \in \fv{t}$, then 
$\proj(t_0) = (\proj(t) \proj(v))\isubs{x/\proj(u)}  = 
                   \proj(t)\isubs{x/\proj(u)} \proj(v)  =
                    \proj(t_1)$. 

If  $x \notin \fv{t}$, then 
$\proj(t_0) = (\proj(t) \proj(v))[\void/\proj(u)]  \equiv_{\sig} 
                   \proj(t)[\void/\proj(u)] \proj(v)  =
                    \proj(t_1)$.

\item $t_0 \equiv_{\sigt,\sigq} t_1$. Then trivially  $\proj(t_0) = \proj(t_1)$.

\end{enumerate}

\item The inductive cases:  

\begin{enumerate}[$-$]
\item $t_0 = u[x/v] \Rew{} (\mbox{resp. }  \equiv)\ u'[x/v] = t_1$,
  where $u \Rew{} (\mbox{resp. } \equiv)\ u'$. If $x \in \fv{u}\ \&\ x
  \in \fv{u'}$ or $x \notin \fv{u}\ \&\ x \notin \fv{u'}$ then the
  property is straightforward by the \ih\ So let us suppose $x \in
  \fv{u}\ \&\ x \notin \fv{u'}$ (so that the reduction step is
  necessarily a $\Gc$-step). We have $\proj(u)
  \Rewnmod{\New,\unboxed}{\osym}\ (\ih)\ \proj(u')$. But $x \in
  \fv{u}\ \&\ x \notin \fv{u'}$ 
  implies $x \in
  \fv{\proj(u)}\ \&\ x \notin \fv{\proj(u')}$
  so that in particular we have  $\proj(u)
  \Rewplusmod{\New,\unboxed}{\osym}\ \proj(u')$.
Then $\proj(t_0) =
  \proj(u)\isubs{x/\proj(v)} 
  \Rewnmod{\New,\unboxed}{\osym}
  \proj(u')[\void/\proj(v)] = \proj(t_1)$ holds by Corollary~\ref{c:stability-substitution}.

\ignore{
\item If $t_0$ is an abstraction or an application the property is straightforward
      by the \ih\
\item $t_0 = u[x/v] \Rew{} (\mbox{resp. } \equiv)\ u[x/v'] = t_1$, 
      where $v \Rew{} (\mbox{resp. } \equiv)\ v'$,
      then the property is also straightforward
      by the \ih\
}
 \item All the other cases are straightforward. 
\qed
\end{enumerate}
\end{enumerate}

Here are some interesting examples:
%
%
%
%
%
%
%
%
\[ \begin{array}{|lll|lll|}
   \hline
   t & \Rew{} & t'          & \proj(t) & \Rewn{} & \proj(t') \\
   \hline
   f[y/x][x/u] & \Rew{\Gc} & f[x/u] & 
   f[\void/u]  & =  &  f[\void/u] \\
   f[y/x z][x/u][z/v] & \Rew{\Gc} & f[x/u][z/v] & 
   f[\void/uv]  & \Rew{\New} & f[\void/u][\void/v] \\
   f[y/xx][x/u] &  \Rew{\Gc} & f[x/u] & 
   f[\void/uu]  &  \Rewplus{\New} & f[\void/u] \\
   (f[w/f[y/x z]] g)[x/u] [z/v] & \Rew{\Gc} & 
      (f[w/f] g) [x/u] [z/v] & 
   (f[\void/f[\void/uv]] g)  & \Rew{\New, \unboxed} & (f[\void/f] g)[\void/u][\void/v] \\
   \hline
   \end{array} \]

The previous property allows us to conclude with one of the main
results of this paper. 

\begin{thm}[PSN for $\ldisf$]
\label{t:psn-ldisf}
The $\ldisf$-calculus enjoys PSN, \ie\ if $t \in \termslambda \cap \SN{\beta}$, then 
$t \in \SN{\ldisf}$.
\end{thm}

\begin{proof} 
We apply Theorem~\ref{t:equational-abstract}, where $\calA = \laux$,
$\calA_1= \set{\Gc, \Var,\DSubs}$, 
$\calA_2=\set{\B}$, $\ttE$ is
$\equiv_{\fsymb}$, $\ttF$ is $\equiv_{\osym}$ and
$t\ \R\ \proj(t)$. Property {\bf (P0)} holds by
Lemma~\ref{l:projection}:\ref{ppp-iv}-\ref{ppp-v}, Property {\bf (P1)}
holds by Lemma~\ref{l:projection}:\ref{ppp-iii}, Property {\bf (P2)}
holds by Lemma~\ref{l:projection}:\ref{ppp-i} and
Property {\bf (P3)} holds by Corollary~\ref{l:dis-f}.  Now, take $t \in
\termslambda \cap \SN{\beta}$ so that Corollary~\ref{c:psn-lauxm}
gives $t \in \SN{\lauxm}$. Since $\proj(t) = t$, then $t \in
\SN{\ldisf}$ by application of the Theorem.
\end{proof}

\section{Consequences of the main result}
\label{s:cons}

In this section we show how the strong result obtained in
  Section~\ref{s:projection} can be used to
  prove PSN for different variants of
the $\ldisf$-calculus.
\subsection{Adding $\set{\unboxed}$ to $\ldisf$}
\label{s:ldisf+n+u}
We show that the reduction relation
$\unboxed$ of
$\lauxm$ can be added to $\ldisf$ without breaking PSN. The main point
of this extension is to show that it is safe to consider unboxing (for void jumps)
 together with the box equations (for non-void jumps). 
For that, we first extend the 
rule $\unboxed$
to act on the whole set $\terms$ and not only on $\termsv$ (but
  they still concern void substitutions only). Boxed contexts are extended to non-void jumps as expected, namely:
\[ B:: = t C \mid  t[x/C] \mid B t \mid B[x/t] \mid \lam y. B \]
Then the rule is given by:
\[ \ctx{B}{t[x/u]}   \rRew{\unboxed}   \ctx{B}{t}[x/u], \sep  \mbox{where } B \mbox{ does not bind } u\ \&\ x\notin\fv{t} \] 
 Indeed, the $\proj$ function maps $\unboxed$-reduction steps of
 $\modulo{\set{\ldis, \unboxed}}{\fsymb}$ into $\set{\New,
   \unboxed}$-reduction steps of $\lauxm$, as the next lemma shows.

\begin{lem}[Extended Projection]
\label{l:projectionext}
Let $t_0 \in \terms$. Then, 
$t_0 \Rew{\unboxed} t_1$ implies  $\proj(t_0)  \Rewnmod{\New,\unboxed}{\osym}  \proj(t_1)$.
\end{lem}

\proof
By induction on the reduction relations. 
\begin{enumerate}[$\bullet$]
\item $t_0 = \ctx{B}{t[x/u]} \Rew{\unboxed}
  \ctx{B}{t}[x/u]= t_1$  where $B$ does not
  bind $u$  and $x \notin \fv{t}$.  We show a stronger property, namely: 

 If $t_0 = \ctx{C}{t[x/u]} \Rew{}
  \ctx{C}{t}[x/u]= t_1$  where $C$ does not
  bind $u$ and $x \notin \fv{t}$, then $\proj(t_0)  \Rewnmod{\New,\unboxed}{\osym}  \proj(t_1)$.   
Then the property we want show is just a particular case of the stronger property. By $\alpha$-conversion
  we assume w.l.g. 
  that
  $x$ is not even free in $\ctx{C}{t}$.

We reason by induction on $C$.
  \begin{enumerate}[$-$]
    \item $t_0 = \ctx{}{t[x/u]} \Rew{\unboxed} \ctx{}{t}[x/u]
        = t_1$. Then $t_0 = t_1$ so that $\proj(t_0) = \proj(t_1)$. 
    \item $t_0 = \ctx{C'}{t[x/u]} v\Rew{\unboxed} (\ctx{C'}{t} v)[x/u]
        = t_1$.  
       \ignore{We have $\proj(t_0) = \proj(\ctx{C'}{t[x/u]}) \proj(v)
        \Rewnmod{\New,\unboxed}{\osym}(\ih)\ \proj(\ctx{C'}{t} [x/u]) \proj(v)$.

        If $x \notin \fv{\ctx{C'}{t}}$, then: 
        \[ \begin{array}{ll}   
         \proj(t_0) = \proj(\ctx{C'}{t} [x/u]) \proj(v) =  \\
         \proj(\ctx{C'}{t}) [x/\proj(u)] \proj(v) \equiv_{\sig_2}\\
         (\proj(\ctx{C'}{t})\proj(v)) [x/\proj(u)]  = \proj(t_1)
         \end{array} \] 
      
        If $x \in \fv{\ctx{C'}{t}}$, then: 
        \[ \begin{array}{ll} 
           \proj(\ctx{C'}{t} [x/u]) \proj(v) =  \\
         \proj(\ctx{C'}{t}) \isubs{x/\proj(u)} \proj(v) = \\
         (\proj(\ctx{C'}{t})\proj(v)) \isubs{x/\proj(u)}  = \proj(t_1) 
         \end{array} \]  
      }
       Then we conclude by using the \ih\ and the
        equivalence $\equiv_{\sig_2}$. 
      
      \item $t_0 = v \ctx{C'}{t[x/u]} \Rew{\unboxed} (v \ctx{C'}{t})[x/u] = t_1$.  
        \ignore{We have 
        $\proj(t_0) = \proj(v)\proj(\ctx{C'}{t[x/u]}) 
        \Rewnmod{\New,\unboxed}{\osym} (\ih)\ 
        \proj(v) \proj(\ctx{C'}{t}[x/u])$. 

        If $x \notin \fv{\ctx{C'}{t}}$, then: 
         \[ \begin{array}{lll} 
          \proj(v)\ \proj(\ctx{C'}{t}[x/u]) = \\
         \proj(v) (\proj(\ctx{C'}{t}) [x/\proj(u)])\Rew{\unboxed}\\
         (\proj(v) \proj(\ctx{C'}{t})) [x/\proj(u)]  = \proj(t_1)
          \end{array} \] 
      
        If $x \in \fv{\ctx{C'}{t}}$, then: 
      \[ \begin{array}{ll} 
         \proj(v) \proj(\ctx{C'}{t}[x/u]) = \\
         \proj(v) (\proj(\ctx{C'}{t}) \isubs{x/\proj(u)})= \\
         (\proj(v) \proj(\ctx{C'}{t})) \isubs{x/\proj(u)}  = \proj(t_1)
         \end{array} \] 
        }
        Then we conclude by using the \ih\ and the
        reduction  $\Rew{\unboxed}$. 
        
  \item $t_0 = \lam y. \ctx{C'}{t[x/u]}\Rew{\unboxed} (\lam 
        y. \ctx{C'}{t})[x/u]= t_1$.  
        \ignore{We have $\proj(t_0) =\l
        y. \proj(\ctx{C'}{t[x/u]}) \Rewnmod{\New,\unboxed}{\osym} (\ih)\ 
        \l
        y. \proj(\ctx{C'}{t}[x/u])$.

        If $x \notin \fv{\ctx{C'}{t}}$, then: 
        $\l
        y. \proj(\ctx{C'}{t}[x/u]) = \lam y. (\proj(\ctx{C'}{t}) [x/\proj(u)]) 
        \equiv_{\sig_1} \proj(t_1)$.
      
        If $x \in \fv{\ctx{C'}{t}}$, then $\l
        y. \proj(\ctx{C'}{t}[x/u]) = \lam y. (\proj(\ctx{C'}{t}) \isubs{x/\proj(u)}) 
	= \proj(t_1)$.
        }
        Then we conclude by using the \ih\ and the
        equivalence $\equiv_{\sig_1}$. 
        
\item  $t_0 = v[y/\ctx{C'}{t[x/u]}]\Rew{\unboxed}
        v[y/\ctx{C'}{t}][x/u] = t_1$. 
        We reason by cases.

        If $y \notin \fv{v}$, then:
        \[ \begin{array}{llll} 
        \proj(t_0)&=&\proj(v[y/\ctx{C'}{t[x/u]}]) &=\\
	&&\proj(v) [\void/\proj(\ctx{C'}{t[x/u]})]	&\Rewnmod{\New, \unboxed}{\osym} (\ih)\\
        &&\proj(v) [\void/\proj(\ctx{C'}{t})[\void/\proj(u)]] &\Rew{\unboxed}\\
        &&\proj(v) [\void/\proj(\ctx{C'}{t})][\void/\proj(u)] &= \proj(t_1)
        \end{array} \] 

If $y \in \fv{v}$, then:
        \[ \begin{array}{lllll} 
        \proj(t_0)&=&\proj(v[y/\ctx{C'}{t[x/u]}]) &=\\
	&&\proj(v)\isubs{y/\proj(\ctx{C'}{t[x/u]})}	&\Rewnmod{\New, \unboxed}{\osym} (\ih\ \&\ Lem.~\ref{l:newu-pass-to-sub}) &\\
&&\proj(v)\isubs{y/\proj(\ctx{C'}{t})[\void/\proj(u)]} &\Rewnmod{\New, \unboxed}{\osym}(Lem.~\ref{l:out-subs})\\
&&\proj(v)\isubs{y/\proj(\ctx{C'}{t})}[\void/\proj(u)] &=\\
&&\proj(v[y/\ctx{C'}{t}])[\void/\proj(u)] &=\proj(t_1)
        \end{array} \]

\ignore{        
Now suppose $x \in \fv{\ctx{C'}{t}}$.  Again we reason by cases. 

If $y \notin \fv{v}$, then:
        \[ \begin{array}{llll} 
        \proj(t_0)&=&\proj(v[y/\ctx{C'}{t[x/u]}]) &=\\
	&&\proj(v)[y/\proj(\ctx{C'}{t[x/u]})]	&\Rewnmod{\New, \unboxed}{\osym} (\ih)\\
        &&\proj(v)[y/\proj(\ctx{C'}{t})\isubs{x/\proj(u)}] & =\\
        &&\proj(v)[y/\proj(\ctx{C'}{t})]\isubs{x/\proj(u)} &= \proj(t_1)
        \end{array} \] 

If $y \in \fv{v}$, then:
        \[ \begin{array}{lllll} 
        \proj(t_0)&=&\proj(v[y/\ctx{C'}{t[x/u]}]) &=\\
	&&\proj(v)\isubs{y/\proj(\ctx{C'}{t[x/u]})}	&\Rewnmod{\New, \unboxed}{\osym} (\ih\ \&\  Lem.~\ref{l:newu-pass-to-sub}) \\
&&\proj(v)\isubs{y/\proj(\ctx{C'}{t})\isubs{x/\proj(u)}} &=\\
&&\proj(v)\isubs{y/\proj(\ctx{C'}{t})}\isubs{x/\proj(u)} &=\\
&&\proj(v[y/\ctx{C'}{t}])\isubs{x/\proj(u)} &=\proj(t_1)
        \end{array} \] 
}
\item  $t_0 = \ctx{C'}{t[x/u]}[y/v]\Rew{\unboxed}
        \ctx{C'}{t}[y/v][x/u]= t_1$.  Note that $y\notin \fv{u}$, otherwise the rule cannot be applied.         We reason by cases.

        If $y \notin \fv{\ctx{C'}{t}}$, then:
        \[ \begin{array}{lllll} 
           \proj(t_0)&=&\proj(\ctx{C'}{t[x/u]}[y/v]) &= \\
&&\proj(\ctx{C'}{t[x/u]})[\void/\proj(v)]&\Rewnmod{\New, \unboxed}{\osym} (\ih)\\
&&\proj(\ctx{C'}{t})[\void/\proj(u)][\void/\proj(v)]&  \equiv_{\CS} \\
&&\proj(\ctx{C'}{t})[\void/\proj(v)][\void/\proj(u)]&  = \proj(t_1)  
           \end{array} \] 

If $y \in \fv{\ctx{C'}{t}}$, then:
        \[ \begin{array}{lllll} 
           \proj(t_0)&=&\proj(\ctx{C'}{t[x/u]}[y/v]) &= \\
&&\proj(\ctx{C'}{t[x/u]})\isubs{y/\proj(v)}&\Rewnmod{\New, \unboxed}{\osym} (\ih)\\
&&\proj(\ctx{C'}{t})[\void/\proj(u)]\isubs{y/\proj(v)}& =\\
 &&\proj(\ctx{C'}{t})\isubs{y/\proj(v)}[\void/\proj(u)]& = \proj(t_1) 
           \end{array} \] 
           
\ignore{If  $x \in \fv{\ctx{C'}{t}}$, 
Then we conclude by using the \ih\ and 
 Lemma~\ref{l:newu-pass-to-sub} if $y \in \fv{\ctx{C'}{t}}$. }

     \end{enumerate}
\item The inductive cases for the  abstraction, the application 
      and reduction inside substitution are straightforward. 

\ignore{
\item $t_0 = u_0[y/u_1] \Rew{} u_0[y/u'_1] =t_1$, where $u_1
  \Rew{\New} u'_1$ (resp. $u_1 \Rew{\unboxed} u'_1$).
        
        If $y \in \fv{u_0}$, then:
         \[ \begin{array}{ll}
           \proj(t_0) = \proj(u_0)\isubs{y/\proj(u_1)}  &\Rewplusmod{\New, \unboxed}{\osym}\ (resp. \Rewnmod{\New, \unboxed}{\osym})\  (\ih) \\
        \proj(u_0)\isubs{y/\proj(u'_1)}         &  = \proj(t_1)   
        \end{array} \]
       
        If $y \notin \fv{u_0}$, then:

          \[ \begin{array}{ll}
           \proj(t_0) = \proj(u_0)[y/\proj(u_1)]  &\Rewplusmod{\New, \unboxed}{\osym}\ (resp. \Rewnmod{\New, \unboxed}{\osym})\  (\ih) \\
        \proj(u_0)[y/\proj(u'_1)]         &  = \proj(t_1)   
        \end{array} \]

}       

\item $t_0  = u_0[y/u_1] \Rew{} u'_0[y/u_1] =t_1$, where $u_0
  \Rew{\New} u'_0$ (resp. $u_0 \Rew{\unboxed} u'_0$).
       Since $\unboxed$ preserves free variables, then
  $y \in \fv{u_0}\ \&\ y
  \in \fv{u'_0}$ or $y \notin \fv{u_0}\ \&\ y \notin \fv{u'_0}$ 
so that the
  property is straightforward by the \ih\
\qed
\end{enumerate}

\begin{thm}
\label{t:psn-enriched}
The $\modulo{\set{\ldis,
      \unboxed}}{\fsymb}$-calculus enjoys PSN, \ie\ $t \in
\termslambda \cap \SN{\beta}$, then $t \in 
\SN{\modulo{\set{\ldis,\unboxed}}{\fsymb}}$.
\end{thm}

\begin{proof}
We apply Theorem~\ref{t:equational-abstract}, where $\calA = \laux$,
$\calA_1= \set{\Gc, \Var,\DSubs,\unboxed}$, $\calA_2=\\set{\B}$,
$\ttE$ is $\equiv_{\fsymb}$, $\ttF$ is $\equiv_{\osym}$ and
$t\ \R\ \proj(t)$. Property {\bf (P0)} holds by
Lemma~\ref{l:projection}:\ref{ppp-iv}-\ref{ppp-v}, Property {\bf (P1)}
holds by Lemmas~\ref{l:projection}:\ref{ppp-iii}
and~\ref{l:projectionext}, Property {\bf (P2)} holds by
Lemmas~\ref{l:projection}:\ref{ppp-i}. To show Property {\bf (P3)} we
proceed as follows. First of all notice that
$\modulo{\unboxed}{\fsymb}$ is trivially terminating, then show that
$\modulo{\calA_1}{\fsymb}$ is terminating by showing that $t
\Rew{\modulo{\calA_1}{\fsymb}} t'$ implies $\pair{\dm{t}}{t} >_{lex}
\pair{\dm{t'}}{t'}$, where the first component of the pair is compared
with respect to the multiset order, the second with respect to the
terminating relation $\Rew{\modulo{\unboxed}{\fsymb}}$, and the
stability of $\dm{\_}$ by $\eqf$, which is given by Lemma
\ref{l:mul-for-eq}:\ref{p:mul-for-eq-two}.  Now, take $t \in
  \termslambda \cap \SN{\beta}$ so that Corollary~\ref{c:psn-lauxm}
  gives $t \in \SN{\lauxm}$. Since $\proj(t) = t$, then $t \in
  \SN{\modulo{\set{\ldis, \unboxed}}{\fsymb}}$ by application of the
  Theorem. 
\end{proof}

\subsection{Orienting the axioms of $\fsymb$}

Another interesting result concerns a more
traditional form of explicit substitutions calculus, called here the 
\deft{inner structural $\lam$-calculus},  and noted $\ldisin$,  whose rules appear in Figure~\ref{fig:inn-oriented-f}.

\begin{figure}
\[\begin{array}{lll@{\hspace{.5cm}}l}
(\lam x.t)\List\ u & \rRew{\B}&  t[x/u]\List    \\
t[x/u] & \rRew{\Gc} & t &\mbox{if $|t|_x=0$}\\
t[x/u] & \rRew{\Var} & t\set{x/u} &\mbox{if $|t|_x=1$}\\
t[x/u] & \rRew{\DSubs} & t_{[y]_x}[x/u][y/u] &\mbox{if $|t|_x>1$}\\\\

(\lam y. t) [x/u] & \rRew{\inn_1} & \lam y. (t [x/u]) & \\
(t v)[x/u]  & \rRew{\inn_2} & t[x/u] v &\mbox{if }x\notin\fv{v}\\
(t v)[x/u]     & \rRew{\inn_3} & t v[x/u]&\mbox{if }x\notin\fv{t}\ \&\ x\in\fv{v}\\
t[y/v][x/u]    & \rRew{\inn_4} &  t[y/v[x/u]] &\mbox{if }x\notin\fv{t}\ \&\ x\in \fv{v}\\\\
  
  t[x/u][y/v] & \sim_{\CS} & t[y/v][x/u] &\mbox{if }x\notin\fv{v}\ \&\ y\notin\fv{s}
\end{array}\]
\caption{\label{fig:inn-oriented-f} The inner structural $\lam$-calculus $\ldisin$}
\end{figure}

Let $\Rew{\inn}$ be the context closure of the rules $\rRew{\inn_{1,2,3,4}}$ modulo $\eqcs$.

\begin{lem}
\label{l:in-terminates}
The reduction relation $\Rew{\inn}$ is strongly normalising.
\end{lem}

\begin{proof}
Define $M(t)$ to be the sum of all the sizes of the subterms of $t$ directly
affected by jumps. It is easily seen that 
such a measure strictly decreases 
by one-step rewriting and is invariant by 
$\eqcs$.
\end{proof}

\begin{cor}
The inner structural $\lam$-calculus $\ldisin$  enjoys PSN.
\end{cor}

\begin{proof}
By application of Theorem~\ref{t:equational-abstract},
where the required properties 
of the projection of $\ldisin$ into $\ldisf$
are guaranteed by Lemmas~\ref{l:projection}
and~\ref{l:in-terminates}.
\end{proof}

\noindent The inner structural $\lam$-calculus can be seen as a refinement of
Kesner's $\les$~\cite{Kes07}, an explicit
substitution calculus related to Proof-Nets, whose rules are in
Figure~\ref{fig:les-rules}.

\begin{figure}
\begin{center}
 \begin{tabular}{llllllll}
$(\lambda x. t) u $&$\Rew{\shB}$& $t[x/u]$ \\
$ x [x/u]$&$\Rew{\Var'}$& $u$\\
$ t [x/u]$&$\Rew{\Gc}$& $t$  & if $x\notin\fv{t}$ \\
$  (t v) [x/u]$&$\Rew{@_r}$&$ t v [x/u]$ & if $x\notin\fv{t}$ and $x\in\fv{v}$\\
$  (t v) [x/u]$&$\Rew{@_l}$&$ t [x/u] v$ & if $x\in\fv{t}$ and $x\notin\fv{v}$\\
$  (t v) [x/u]$&$\Rew{@}$&$ t [x/u] v [x/u]$& if $x\in\fv{t}$ and $x\in\fv{v}$\\
$ (\lambda y. t) [x/u]$&$\Rew{\lam}$&$ \lambda y. t [x/u]$\\
$  t[x/u][y/v]$    & $\Rew{\comp_1} $& $t[x/u[y/v]]$  & if $y\notin\fv{t}$ and $y\in\fv{u}$\\
$  t[x/u][y/v]$    & $\Rew{\comp_2} $& $t[y/v][x/u[y/v]]$  & if $y\in\fv{t}$ and $y\in\fv{u}$\\\\
$  t[x/u][y/v]$ & $\sim_{\CS}$&$ t[y/v][x/u]$  & if $y\notin\fv{u}$ and $x\notin\fv{v}$ \\ &&&(and $x\neq y$)
\end{tabular}
\end{center}
\caption{The $\les$-calculus\label{fig:les-rules}}
\end{figure}

Indeed, only rules $\set{@,\comp_2}$ are not particular
cases of rules of $\ldisin$, but they can be decomposed 
by using duplication followed by propagations 
as follows:
\[\begin{array}{cccccccccc}
(t v) [x/u]&\Rew{@} &t [x/u] v [x/u]\\\\
\downarrow_\DSubs&& \uparrow^{\inn_3}\\\\
(t v\isubs{x/y}) [x/u][y/u]& \Rew{\inn_2}&(t[x/u] v\isubs{x/y}) [y/u]\eqw{\alpha}(t[x/u] v) [x/u]
\end{array}\]

It is then straightforward to simulate $\les$ inside $\ldisin$, so we get:

\begin{cor}
\label{cor:les-psn}
The $\les$-calculus enjoys PSN.
\end{cor}

The second author shows in~\cite{Kes09}
that from PSN of $\les$ one can infer PSN of a wide range of calculi,
$\lx$, Kesner's $\les$ and $\lesw$~\cite{Kes07},
Milner's calculus $\lsub$~\cite{Milner07}, David's
and Guillaume's $\lam_{\tt ws}$~\cite{DBLP:journals/mscs/DavidG01}, the
calculus with director strings of~\cite{SinotFM03}. Hence PSN for $\ldisf$ encompasses
most results of PSN in the literature of explicit substitutions.

\noindent The interesting feature of $\ldisin$ with respect to $\les$ is that
the propagation subsystem $\Rew{\inn}$ is not needed in order to
compute a normal form. Propagations are rather (linear)
re-arrangements of term constructors which may be used as the basis of
some term transformations used for compilation or
optimisation.\medskip

The strength of a splitting of the whole calculus into a core and 
propagation system lies in the fact that 
the latter can be changed without affecting the former. In
particular, it is possible to orient the axioms $\set{\sig_1, \sig_2,
  \sigt, \sigq}$ in the opposite direction 
by getting the \textit{outer} structural $\lam$-calculus
$\ldisout$, whose rules are in Figure~\ref{fig:out-oriented-f}.

Observe that  in contrast  to the inner  calculus the outer  box rules
act also on  void jumps,  \ie\ they  are not
just an orientation  of the box equations, but  an extension too. This
is  possible because  --- as  showed  earlier (Theorem
\ref{t:psn-enriched})       ---       extending      $\ldisf$       with
unboxing for void jumps is safe (while we
do  not know  whether it  is safe  to extend  $\ldisf$ with
boxing for  void  jumps).    Let
$\Rew{\outm}$  be  the derived  context  closure  of  the outer  rules
$\rRew{\out_{1,2,3,4}}$ modulo $\eqcs$.

\begin{lem}
\label{l:out-terminates}
The reduction relation $\Rew{\outm}$ is strongly normalising.
\end{lem}

\ignore{
\begin{proof}
Informal proof: let $|C|$ denote the size of a context. Then define 
$M(t)=\sum_{t=C[v[x/u]]}|C|$, \ie, the sum of the size of the contexts containing each jump of $t$. It is easily seen that such a measure decreases with any $\Rew{\out}$-step and is invariant by $\eqcs$.
\end{proof}}

\begin{cor}
The outer structural $\lam$-calculus $\ldisout$ enjoys PSN.
\end{cor}

\begin{proof}
By application of Theorem~\ref{t:equational-abstract},
where the required properties 
of the projection of $\ldisout$ into $\ldisf$
are guaranteed by Lemmas~\ref{l:projection}
and~\ref{l:out-terminates}.
\end{proof}

\begin{figure}
\[\begin{array}{lll@{\hspace{.5cm}}l}
(\lam x.t)\List\ u & \rRew{\B}&  t[x/u]\List    \\
t[x/u] & \rRew{\Gc} & t &\mbox{if $|t|_x=0$}\\
t[x/u] & \rRew{\Var} & t\set{x/u} &\mbox{if $|t|_x=1$}\\
t[x/u] & \rRew{\DSubs} & t_{[y]_x}[x/u][y/u] &\mbox{if $|t|_x>1$}\\\\

\lam y. (t [x/u]) & \rRew{\out_1} & (\lam y. t) [x/u] &\mbox{ if } y\notin\fv{u} \\
t[x/u] v  & \rRew{\out_2} & (t v)[x/u] \\
t v[x/u]     & \rRew{\out_3} &  (t v)[x/u]\\
t[y/v[x/u]]    & \rRew{\out_4} &  t[y/v][x/u] &  \\\\

  t[x/u][y/v] & \sim_{\CS} & t[y/v][x/u] & \mbox{ if } x\notin\fv{v}\ \&\ y\notin\fv{s}
\end{array}\]
\caption{\label{fig:out-oriented-f} The outer structural $\lam$-calculus $\ldisout$}
\end{figure}

In fact, it is easily seen that no matter how the axioms $\set{\sig_1,
  \sig_2, \sigt, \sigq}$ are oriented that they get a terminating
rewriting system. As for $\ldisin$ and $\ldisout$,  
PSN can also be proved for the remaining 14
derived calculi, even if it is not clear to what extent
they  would be interesting.

\subsection{Adding equations to $\lam$-terms}
\label{ssec:rene}
We briefly present here the results of~\cite{AKLPAR}, which
extends and complement those of this paper. As discussed in
Section~\ref{s:regnier}, the equations $\eqw{\sig_1}$ and
$\eqw{\sig_2}$ can be seen as a jump reformulation of Regnier's
$\rsig$-equivalence on $\lam$-terms after the elimination of $\B$-redexes. It is
also possible to apply the $\B$-rule in the other sense (\ie\ as a $\B$-expansion)
to the equations $\set{\preeqw{\sigt},\preeqw{\sigq}}$ in order to obtain other
equations  on $\lam$-terms. If $x\notin \fv{t}$
and $x\in \fv{v}$, the equation $\preeqw{\sigt}$ can be $\B$-expanded to the new equation $\rsigt$:
\[ \begin{array}{c@{\sep}c@{\sep}c}
(t v)\grisar{[x/u]}   & \preeqw{\sigt} & t v\grisar{[x/u]} \\\\
\uparrow_\B             & &  	\uparrow_\B \\\\
 \grisar{(\lam x.t v) u} & \preeqw{\rsigt}  & t \grisar{((\lam x. v) u)}
\end{array}\] 
Axiom $\rsigt$ is a more general
instance of the rule called 
${\tt assoc}$~\cite{Moggi89,DBLP:journals/corr/abs-0806-4859,DBLP:journals/tcs/David11}
(which usually is not taken modulo but oriented from right to left). The axiom
$\preeqw{\sigq}$ $\B$-expands to a special case of
$\preeqw{\rsigt}$, and thus it is subsumed 
by it. Indeed:
\[ \begin{array}{cccccc}
t\grisar{[y/v]}[x/u]    & \preeqw{\sigq} & t\grisar{[y/v[x/u]]} \\\\

\uparrow_\B & &  	\uparrow_\B \\\\
(\grisar{(\lam y. t) v})\grisarOscuro{[x/u]}    & \preeqw{\sigt} & 
\grisar{(\lam y. t) v\grisarOscuro{[x/u]}} \\\\
\uparrow_\B & &  	\uparrow_\B \\\\
\grisarOscuro{(\lam x. ((\lam y. t) v))u}    & \preeqw{\rsigt} & 
(\lam y. t) \grisarOscuro{((\lam x. v)u)} \\\\
\end{array}\] 
Last, one can turn the unboxing rule into its $\lam$-calculus form, getting:
\[ \begin{array}{lll@{\hspace{.5cm}}l}
t ((\lam x. v) u)     & \rRew{\runboxed} & (\lam x.t v) u \sep \mbox{ if } x \notin \fv{t}\ \&\ x\in \fv{v} \\
\end{array}\]\vskip2 pt

\noindent Let $\eqP$ be defined as the smallest equivalence relation containing
$\equiv_{\set{\rsig_1,\rsig_2, \rsigt}}$ and $\eqf$. In \cite{AKLPAR} we show
that the $\modulo{\set{\ldis, \unboxed, \runboxed}}{\Psymb}$-calculus
in Figure~\ref{fig:final-calc} enjoys PSN. The proof is obtained via a
simple function which eliminates $\B$-redexes, and that project this
calculus over the $\modulo{\set{\ldis, \unboxed}}{\fsymb}$-calculus,
whose PSN is given by Theorem \ref{t:psn-enriched}. The main result of
\cite{AKLPAR}, however, is that the the $\modulo{\set{\ldis, \unboxed,
    \runboxed}}{\Psymb}$-calculus is also Church-Rosser modulo the
whole equational theory. This is proved via $\tt M$-developments, a
new notion of development taking advantage of jumps. Actually, in
\cite{AKLPAR} we use a macro-steps substitution rule $t[x/u]\Rew{\tt
  sub} t\isubs{x/u}$ instead of our subsystem $\Rew{\dis}$: we do so
because the fine granularity of $\Rew{\dis}$ plays no role in the
proof of these properties, their refinement to $\Rew{\dis}$ is
straightforward.\medskip

Let us call \deft{permutative $\lam$-calculus} (see Figure
\ref{fig:perm-lambda}) the set of $\lam$-terms plus the operational
semantics given by $\modulo{\set{\beta, \runboxed}}{\psymb}$, where
$\eqp$ is the smallest equivalence relation containin $\rsig_1$,
$\rsig_2$, $\rsigt$.  Such a calculus can be (strictly) simulated into
the $\modulo{\set{\ldis, \runboxed, \unboxed}}{\Psymb}$-calculus and
thus it enjoys PSN. This result generalises all known results in the
literature about PSN for $\lam$-calculus extended with permutative
conversion
\cite{DBLP:journals/tcs/David11,DBLP:journals/tcs/Santo11,DBLP:journals/corr/abs-0806-4859}. In
\cite{AKLPAR} we also prove that it is Church-Rosser modulo $\eqp$.

\begin{figure}[ht]
\[\begin{array}{lll@{\hspace{.5cm}}l}
(\lam x.t)\List\ u & \rRew{\B}&  t[x/u]\List    \\
t[x/u] & \rRew{\Gc} & t &\mbox{if $|t|_x=0$}\\
t[x/u] & \rRew{\Var} & t\set{x/u} &\mbox{if $|t|_x=1$}\\
t[x/u] & \rRew{\DSubs} & t_{[y]_x}[x/u][y/u] &\mbox{if $|t|_x>1$}\\\\

\ctx{B}{t[x/u]} &  \rRew{\unboxed}   & \ctx{B}{t}[x/u] & B \mbox{ does not bind } u\\
t ((\lam x. v) u)     & \rRew{\runboxed} & (\lam x.t v) u   & \mbox{if } x \notin \fv{t}\ \&\ x\notin \fv{v}\\\\

(\lam x. \lam y.t) u & \preeqw{\rsig_1} & \lam y. ((\lam x.t) u) & \mbox{if } y\notin\fv{u}\\
(\lam x.t v) u     & \preeqw{\rsig_2} & (\lam x. t) u v    & \mbox{if } x\notin \fv{v}\\
(\lam x.t v) u     & \preeqw{\rsigt} & t ((\lam x. v) u)    & \mbox{if } x \notin \fv{t}\ \&\ x\in \fv{v}\\\\

   t[x/s][y/v] & \sim_{\CS} & t[y/v][x/s] & \mbox{ if } x\notin\fv{v}\ \&\ y\notin\fv{s}  \\
   \lam y. (t [x/s]) & \preeqsigu & (\lam y. t) [x/s]  & \mbox{ if } y\notin \fv{s} \\
   t[x/s] v & \preeqsigt & (t v)[x/s]& \mbox{ if } x\notin\fv{v} \\
(t v)[x/u]     & \preeqw{\sigt} & t v[x/u]    & 
  \mbox{if } x \notin \fv{t}\ \&\  x\in \fv{v}\\
t[y/v][x/u]    & \preeqw{\sigq} & t[y/v[x/u]] & 
  \mbox{if } x\notin \fv{t}\ \&\ x\in \fv{v}\\

\end{array}\]
\caption{\label{fig:final-calc} The structural $\lam$-calculus modulo}
\end{figure}

\begin{figure}
\[\begin{array}{lll@{\hspace{.5cm}}l}
(\lam x.t) u & \rRew{\beta}&  t\isubs{x/u}    \\
t ((\lam x. v) u)     & \rRew{\runboxed} & (\lam x.t v) u   & \mbox{if } x \notin \fv{t}\ \&\ x\notin \fv{v}\\\\

(\lam x. \lam y.t) u & \preeqw{\rsig_1} & \lam y. ((\lam x.t) u) & \mbox{if } y\notin\fv{u}\\
(\lam x.t v) u     & \preeqw{\rsig_2} & (\lam x. t) u v    & \mbox{if } x\notin \fv{v}\\
(\lam x.t v) u     & \preeqw{\rsigt} & t ((\lam x. v) u)    & \mbox{if } x \notin \fv{t}\ \&\ x\in \fv{v}\\

\end{array}\]
\caption{\label{fig:perm-lambda} The permutative $\lam$-calculus}
\end{figure}

\section{Conclusions}

We  have introduced  the  structural $\ldis$-calculus,  a concise  but
expressive   $\lam$-calculus  with  jumps   admitting
graphical interpretations  by means of $\ljdag$s  and Pure Proof-Nets.
Even if $\ldis$  has strong linear logic background,  the calculus can
be understood as a particular reduction system, based on the notion of
multiplicity and reduction at a distance, and being independent from
any  logic or  type system.  We established  different  properties for
$\ldis$ such as confluence  and PSN.  Moreover, full composition holds
without any  need of  structural composition nor  commutation of
jumps.  The $\lj$-calculus  admits a graphical operational equivalence
$\eqo$ allowing to commute jumps with linear constructs.  The relation
$\eqo$ can  be naturally understood  as Regnier's $\sigma$-equivalence
on $\lam$-terms  and turns out  to be a strong  bisimulation.  Moreover,
$\eqo$ can be further  extended to the substitution equivalence $\eqf$
allowing  to  commute  also  jumps  and  non-linear  constructs.   The
resulting  calculus enjoys PSN,  a non-trivial  result from  which one
derives several known PSN results.

PSN  of  $\ldis$ modulo  $\eqf$  is shown  by  means  of an  auxiliary
calculus  $\lauxm$  which  can  be  understood  as  a  \textit{memory}
calculus specified by  means of \textit{void} substitutions.  
A memory calculus due to Klop ~\cite{Klo} is often used for termination
arguments. Its syntax is usually presented as follows:
\[ \sep t, u :: = x \mid \lam x. t \mid t u \mid [t,u] \]
where $x \in \fv{t}$ for every term $\lam x.t$ and the memory construct $[t,u]$ 
is used to collect in $u$ the  arguments of the erasing $\beta$-redexes. The rule associated to this calculus are:
\[\begin{array}{lll}
(\lam x.t) u   & \rRew{\Beta}&  t\isubs{x/u}    \\
{[}t,v{]} u & \rRew{\pi}&  {[}t u, v{]} \\   
\end{array}\]
If       one      interprets       $[t,v]$       as      $t[\void/v]$   then   Klop's
calculus  can be  mapped into  $\lauxm$: $\beta$  maps to  $\beta$ and
$\pi$         becomes        the         reduction        rule
$t[\void/v]  u\Rew{}(t  u)[\void/v]$,  which is  subsumed  by
the equation $\eqw{\sig_2}$ of $\lauxm$. 
Indeed, $\lauxm$ 
is more expressive than  Klop's calculus.  We
claim that  $\lauxm$ is  interesting on  its own and  can be  used for
proving termination results beyond those of this paper.

We do not know whether 
  $\ldisf$ extended with unrestricted \textit{boxing}, in contrast to
    $\ldisf$ extended with unrestricted \textit{unboxing} presented in 
  Section~\ref{s:ldisf+n+u}, enjoys PSN.  The
  point is delicate, indeed from the literature (\cite{Mellies1995a}) we know
  that unrestricted boxing together with the following
  traditional explicit substitution rule (without side condition on
  $x$):
\[\begin{array}{lll@{\hspace{.5cm}}l}
  (t v) [x/u]&\Rew{@}& t [x/u] v [x/u]
\end{array}\]

\noindent break PSN. Now, the rule $\Rew{@}$ cannot be simulated in
$\ldisf$, so it would be interesting to
 understand if $\ldisf$ plus unrestricted boxing
enjoys PSN. 

An  interesting research  direction  would be  to  formalise the  link
between  $\ldis$,  linear  logic  and abstract  machines.  Indeed,  in
contrast  to  explicit substitution  calculi,  $\ldis$
naturally     expresses     the      notion     of     linear     head
reduction~\cite{Danos99opt},  which  relates   in  a  simpler  way  to
Krivine's  Abstract  Machine~\cite{Krivinemachine}. This
is  because  linear head  reduction  performs  the  minimal amount  of
substitutions necessary  to find  which occurrences of  variables will
stand in head positions. While this is not a reduction strategy in the
usual  sense of  $\lam$-calculus, it  can  be seen  as a  clever way  to
implement $\beta$-reduction  by means of  proof-nets technology, which
can be reformulated in the $\ldis$-calculus as a strategy.

The  structural $\lam$-calculus  has   been  used in~\cite{AK10}  to  specify  {\tt
  XL}-developments,  a terminating  notion  of reduction  generalising
those           of           development~\cite{Hindley78}          and
superdevelopment~\cite{KvOvR93}. It would be interesting to better understand {\tt
  XL}-developments.

It would also be interesting to exploit distance and multiplicities in
other frameworks dealing  for example with pattern matching,
continuations  or  differential  features.   A direction  which  seems
particularly  challenging is standardization
for $\lj$. It  would be interesting  in particular to obtain  a
notion    of     standard    reduction   which     is    stable    by
$\eqo$-equivalence (or  at least  $\eqcs$, so that  the result
would    pass   to   $\lj$-dags).   Indeed,
classical  notions as leftmost-outermost  reduction do
not  easily generalise  to $\lj$  modulo $\eqo$, where  jumps can  be
swapped and permuted with linear constructors.

\section*{Acknowledgements}
We would like to thank Stefano Guerrini
for stimulating discussions.

\bibliographystyle{is-abbrv}

\end{document}